\documentclass[preprint, sort&compress]{elsarticle}
\usepackage{graphicx}
\usepackage{graphics}
\usepackage{epsfig}
\usepackage{amsmath}
\usepackage{amsfonts}
\usepackage{amssymb}
\usepackage{dcolumn}
\usepackage{bm}
\usepackage{longtable}
\usepackage{multirow}

\begin{document}

\title{Magnetic interactions in strongly correlated systems: spin and orbital contributions}

\author[RU]{A. Secchi}
\ead{a.secchi@science.ru.nl}
\ead{andrea.secchi@gmail.com}

\author[UH]{A. I. Lichtenstein}

\author[RU]{M. I. Katsnelson}

\address[RU]{Radboud University, Institute for Molecules and Materials, 6525 AJ Nijmegen, The Netherlands}
\address[UH]{Universitat Hamburg, Institut f\"ur Theoretische Physik, Jungiusstra{\ss}e 9, D-20355 Hamburg, Germany}

\date{\today}

\begin{abstract}
We present a technique to map an electronic model with local interactions (a generalized multi-orbital Hubbard model) onto an effective model of interacting \emph{classical} spins, by requiring that the thermodynamic potentials associated to spin rotations in the two systems are equivalent up to second order in the rotation angles, when the electronic system is in a symmetry-broken phase. This allows to determine the parameters of relativistic and non-relativistic magnetic interactions in the effective spin model in terms of equilibrium Green's functions of the electronic model. The Hamiltonian of the electronic system includes, in addition to the non-relativistic part, relativistic single-particle terms such as the Zeeman coupling to an external magnetic fields, spin-orbit coupling, and arbitrary magnetic anisotropies; the orbital degrees of freedom of the electrons are explicitly taken into account. We determine the complete relativistic exchange tensors, accounting for anisotropic exchange, Dzyaloshinskii-Moriya interactions, as well as additional non-diagonal symmetric terms (which may include dipole-dipole interaction). The expressions of all these magnetic interactions are determined in a unified framework, including previously disregarded features such as the vertices of two-particle Green's functions and non-local self-energies. We do not assume any smallness in spin-orbit coupling, so our treatment is in this sense exact. Finally, we show how to distinguish and address separately the spin, orbital and spin-orbital contributions to magnetism, providing expressions that can be computed within a tight-binding Dynamical Mean Field Theory.
\end{abstract}


\maketitle

\section{Introduction}

Establishing a rigorous connection between magnetic and electronic descriptions of condensed matter systems is a challenging problem \cite{Lichtenstein87}, whose formal statement can be formulated as follows: Given a physical system described by means of a completely known electronic Hamiltonian, what is the \emph{spin} Hamiltonian (supposing that it exists) that most closely reproduces the spectral and dynamical features of the system? 

The answer to this important question is, of course, far from being straightforward. It is well known, e.g., that the spectrum of the lowest energy band of the single-orbital Hubbard model at half filling with nearest-neighbour hopping $T$ and strong on-site Coulomb repulsion $U$ can be effectively represented in terms of the antiferromagnetic quantum Heisenberg Hamiltonian, as follows from perturbation theory in small $\left| T \right| / U$. Dynamics of the electronic system, however, may involve hopping transitions via intermediate higher bands, which are not captured in the Heisenberg Hamiltonian alone, as well as real hopping processes become relevant at other electronic fillings (as a first correction, one should consider the $T$-$\mathcal{J}$ model \cite{Auerbach}). A non-Heisenberg character of magnetic interactions in itinerant systems was explicitly demonstrated, e.g., for the narrow-band Hubbard model on the Bethe lattice beyond half filling  \cite{Auslender}. The problem gets much more complicated if one attempts to map more realistic electronic systems to magnetic models: for example, the natural extension of the single-orbital Hubbard model is the \emph{multi}-orbital Hubbard model \cite{Kanamori63, Hubbard65, Kugel82, Lichtenstein98, Georges13}, which includes more than just one orbital per site, being a more appropriate description of relevant systems such as $d$ and $f$ materials. Moreover, when both spin and orbital degrees of freedom of the electrons are taken into account, their interplay gives rise to relativistic interactions such as spin-orbit coupling and anisotropies \cite{White}.

When no smallness in some characteristic energy parameters of the system can be assumed (such as $\left| T  \right| << U$ in the Hubbard model), the parameters describing the magnetic interactions in an electronic system can be \emph{defined} by imposing the equivalence between the response to spin rotations of a quantity characterizing the system and the analogous response computed for a reference \emph{classical} spin model \cite{Lichtenstein87}. In the case of symmetry-broken phases, the quantity which is generally considered is the thermodynamic potential \cite{Katsnelson00, Katsnelson02, Katsnelson10} computed for an \emph{out-of-equilibrium} state or statistical superposition, that is, either a pure state which is \emph{not} an eigenstate of the electronic Hamiltonian, or a statistical superposition of eigenstates whose weights do \emph{not} depend only on their energies (which would be the case for the Boltzmann distribution, with weights $W_n(\beta) = \mathrm{e}^{- \beta E_n} / Z$, where $E_n$ is the eigenenergy of state $n$, $\beta$ is the inverse temperature and $Z$ is the partition function). The idea of using a symmetry-broken state is similar in spirit to the Higgs mechanism: we need first to solve the non-perturbative many-body problem and find the local moments (massive Higgs fields) and then use the information contained in the single-particle Green's functions and the vertex functions to find perturbatively the soft modes related with exchange interactions. It has been shown that the expressions for the exchange parameters obtained by applying this approach in the non-relativistic case, within the framework of time-dependent density functional theory in the adiabatic approximation, provide an accurate expression for the spin-wave stiffness \cite{Katsnelson04}, while the computation of static properties requires the introduction of constraining magnetic fields to equilibrate the non-equilibrium spin configuration \cite{Stocks98, Bruno03}. However, the corresponding corrections to the exchange parameters \cite{Bruno03} are small in the adiabatic approximation, that is, when typical magnon energies are small in comparison with the Stoner splitting \cite{Katsnelson04}. This justifies our approach. In the non-relativistic case, we have recently extended the treatment of Ref.\cite{Katsnelson00} to systems driven explicitly out of equilibrium by time-dependent external electric fields, by considering the potential arising from the non-equilibrium Kadanoff-Baym partition function \cite{Secchi13} (in Refs.\cite{Katsnelson00, Katsnelson02, Katsnelson10, Secchi13} the electronic system was modelled by means of the multi-orbital Hubbard model). 

Making a mapping to a \emph{classical} spin model means that the target of the mapping is a Hamiltonian involving a set of interacting unit vectors $\boldsymbol{e}_i$, which can rotate in space as classical vectors, their components varying in a continuous domain. These vectors can be called \emph{classical spins}, and our goal is to determine the coefficients of their interactions. It should be noted that these effective parameters include information related to the magnitudes of the local spins, since this is not included into the unit vectors $\boldsymbol{e}_i$.

Unfortunately, the most correct spin model to be used for the mapping with a given electronic system may include interactions between up to an arbitrary number of spins, which is not known a priori. In practice, it is generally assumed that the most relevant magnetic parameters are the external magnetic field, which couples linearly with the spins, and the spin-spin pair interactions, quantified by the $3 \times 3$ exchange tensors $\overline{\boldsymbol{\mathcal{H}}}_{i j}$ (one for each pair of spins labelled by $i$ and $j$). In the non-relativistic case, the exchange tensors are proportional to the identity matrix, $\overline{\boldsymbol{\mathcal{H}}}_{i j} \rightarrow \mathcal{J}_{i j} \overline{\boldsymbol{1}}$, where $\mathcal{J}_{i j}$ is the non-relativistic (isotropic) exchange parameter. In the relativistic case, the exchange tensors are general matrices which include anisotropic exchange, Dzyaloshinskii-Moriya, and other (symmetric) pair interactions \cite{White}. It should be remarked that higher-order interactions such as bi-quadratic exchange have been suggested to be relevant for materials such as MnO \cite{Rodbell63}, CuO$ _2$ plaquettes \cite{Honda93}, pnictide superconductors \cite{Wysocki11}, and models such as spin ladders \cite{Brehmer99}.  

However, in this work we will show that the thermodynamic potential under small spin rotations of a rather general relativistic electronic system in a broken-symmetry phase is equivalent to that of a quadratic spin model with a general $3 \times 3$ exchange tensor, if the equivalence is required up to the second order in the rotation angles. We will prove this by determining the \emph{complete} quadratic exchange tensor, which up to now was determined only in the non-relativistic case \cite{Katsnelson00, Katsnelson02} or limited to the Dzyaloshinskii-Moriya interaction in the relativistic case \cite{Katsnelson10}. In addition to that, we will remove two uncontrolled approximations that were previously adopted \cite{Katsnelson00, Katsnelson02}, namely neglecting the vertices of two-particle Green's functions and neglecting non-local components of the self-energies. The electronic model that we will consider has a completely general single-particle Hamiltonian, including terms such as the Zeeman coupling between the magnetic field and the electronic spins, local and non-local spin anisotropies and spin-orbit couplings, in addition to the non-relativistic hopping. We emphasize that \emph{no smallness} will be assumed in \emph{any} of the single-particle terms, so that relativistic terms are accounted for in a non-perturbative way. We will adopt the only simplification of assuming that the interaction Hamiltonian is rotationally invariant (in the same spirit as what was done in the previous literature \cite{Katsnelson00, Katsnelson02, Katsnelson10, Secchi13}). The electronic model is then a \emph{relativistic multi-orbital Hubbard model}. Moreover, in this work we will consider also the contributions to magnetic interactions due to the \emph{orbital} degrees of freedom of the electrons, and we will show how to distinguish them from the usual contributions due to the intrinsic spin-$1/2$.

While the original approach of Ref.\cite{Lichtenstein87}, based on the non-relativistic multi-scattering formalism within density-functional theory, was recently extended to account for relativistic interactions \cite{Udvardi03, Szilva13}, our approach here is different, since we apply a consistent many-body treatment of the multi-band Hubbard model, including both relativistic and non-relativistic effects on equal footing, as well as fully including contributions to magnetism due to orbital and intrinsic spin of the electrons (i.e., we do not assume that the orbital moment is frozen).

This Article is structured as follows. In Section \ref{sec: Hamiltonian} we introduce our reference electronic Hamiltonian, specifying the model and the notation; in Section \ref{sec: Rotation} we explain our procedure to probe the response of the electronic system to rotations of the spin quantization axes; in Sections \ref{sec: Action} and \ref{sec: Effective action} we derive the effective rotational action of the electronic system for small deviations from the initial directions of the quantization axes; in Section \ref{sec: Static potential} we derive the effective potential for static spin rotations applied to the electronic Hamiltonian; in Section \ref{sec: Mapping} we derive the analogous effective potential for a model of classical spins, and we derive a set of equations connecting the parameters of the spin model to Green's functions of the electronic model, which allows to identify the magnetic parameters; in Section \ref{sec: Solution} we solve the system in the general relativistic case (for ease of reference, the resulting formulas are summarized in Section \ref{sec: Summary solution}); in Section \ref{sec: Nonrel} we consider the spin-$1/2$ single-orbital Hubbard model in the non-relativistic regime, establishing correspondence with the previous literature; in Section \ref{sec: Separation} we show how to study separately the spin, orbital, and spin-orbital contributions to magnetism, providing the explicit expressions of the respective contributions to all the magnetic parameters determined in Section \ref{sec: Solution}; in Section \ref{sec: Conclusion} we summarize our results and mention possible future developments. In \ref{app: Invariance} we discuss the condition for the rotational invariance of the interaction, which is the initial hypothesis of this work, while in \ref{app: Analysis} we provide some details related to one of the key quantities contributing to the magnetic parameters.

\section{The electronic Hamiltonian}
\label{sec: Hamiltonian}
The Hamiltonian of a general electronic system is written as
\begin{align}
\hat{H}  \equiv \hat{H}_T^{\phi}  + \hat{H}_V^{\phi}  ,
\end{align}
where $\hat{H}_T^{\phi}$ and $\hat{H}_V^{\phi}$ are, respectively, the single-particle and interaction Hamiltonians.  

\subsection{Single-particle Hamiltonian}

The single-particle Hamiltonian reads:
\begin{align}
\hat{H}_T^{\phi}  = \sum_{1, 2}   \hat{\phi}^{\dagger}_{1}   T^1_{2}   \hat{\phi}^{2}  ,
\label{hopping}
\end{align}
where $1 \equiv ( a_1, n_1, l_1, S_1, M_1)$ is a collective index including all the indexes and quantum numbers that specify an atomically-localized (Wannier) single-electron state. Namely, $a$ is the atomic index, $n$ is the orbital shell quantum number, $l$ is the quantum number of the square orbital angular momentum $\hat{\boldsymbol{l}}^2$, $S$ is the quantum number of the square total angular momentum $\hat{\boldsymbol{S}}^2$, with $\hat{\boldsymbol{S}} = \hat{\boldsymbol{l}} + \hat{\boldsymbol{s}}$ (obviously $s = 1/2$), and $M \in \lbrace -S, -S + 1, \ldots, S - 1, S \rbrace$ is the total third component of the total angular momentum. It is convenient to group the indexes $(a, n, l)$ by introducing the \emph{orbital} index $o_1 \equiv (a_1, n_1, l_1)$ and the \emph{magnetic moment} index $i_1 \equiv (o_1, S_1) \equiv (a_1, n_1, l_1, S_1)$. Angular momenta associated with a single-electron state are measured in a reference frame centered on its respective site, with a $i$-specific quantization axis $\boldsymbol{r}_i \equiv \boldsymbol{u}_z(i)$ (we will employ one of the two notations from case to case). The corresponding reference frame is then specified by the right-handed triple of unit vectors $\left[ \boldsymbol{u}_x(i), \boldsymbol{u}_y(i) , \boldsymbol{u}_z(i) \right]$. For a given orbital index $o_1 \equiv (a_1, n_1, l_1)$ there are of course two possible magnetic moment indexes $i_1^{\pm} \equiv  (a_1, n_1, l_1, l_1 \pm 1/2)$ if $l_1 > 0$, or only one, $i_1   \equiv (a_1, n_1, 0,   1/2)$ if $l = 0$. For $l_1 > 0$, the two fields specified by $i_1^{\pm}$ have the same quantization axis, which defines the common orbital angular momentum and hence their common quantum number $l_1$. In the following, for brevity we will refer to the individual magnetic moments as \emph{spins}, so we will say, e.g., that $i$ is a spin index.

We also emphasize that, in the present formulation, the quantization axes $\boldsymbol{r}_i$ can be chosen arbitrarily, and in particular they may be different for each spin. As we will discuss later, for the purposes of the mapping to the classical spin model one should let $\boldsymbol{r}_i$ coincide with the direction of the expectation value of the $i$-th local magnetic moment in the reference symmetry-broken non-equilibrium state. The fact that the $\boldsymbol{r}_i$ vectors are not restricted to be parallel allows us to treat the \emph{non-collinear} regime.

Equation \eqref{hopping} can be written more explicitly as
\begin{align}
\hat{H}_T^{\phi}  =   \sum_{o_1, S_1, M_1} \sum_{o_2, S_2, M_2} \hat{\phi}^{\dagger}_{o_1, S_1, M_1}   T^{o_1, S_1, M_1}_{o_2, S_2, M_2}   \hat{\phi}^{o_2, S_2, M_2}  .
\end{align}
It is convenient to define the spinors
\begin{align}
\hat{\phi}^{\dagger}_{i_1} \equiv  \hat{\phi}^{\dagger}_{o_1, S_1} \equiv \left( \hat{\phi}^{\dagger}_{o_1, S_1, S_1}, \hat{\phi}^{\dagger}_{o_1, S_1, S_1 - 1}, \ldots, \hat{\phi}^{\dagger}_{o_1, S_1, - S_1 + 1}, \hat{\phi}^{\dagger}_{o_1, S_1, -S_1} \right) ,
\end{align}
and accordingly we write
\begin{align}
\hat{H}_T^{\phi}  =  \sum_{o_1, S_1 } \sum_{o_2,  S_2 }   \hat{\phi}^{\dagger}_{o_1, S_1} \cdot   T^{o_1, S_1 }_{o_2, S_2 } \cdot  \hat{\phi}^{o_2, S_2 } =  \sum_{i_1  } \sum_{i_2  }   \hat{\phi}^{\dagger}_{i_1 } \cdot   T^{i_1  }_{i_2  } \cdot  \hat{\phi}^{i_2   } ,
\label{SP electronic Hamiltonian}
\end{align}
where $T^{o_1, S_1 }_{o_2, S_2 }$ is a matrix in angular momentum space. Note that $\hat{\phi}^{\dagger}_{o_1, S_1}$ has (rows $\times$ columns) dimensions $1 \times (2 S_1 + 1)$, $T^{o_1, S_1 }_{o_2, S_2 } $ has dimensions $(2 S_1 + 1) \times (2 S_2 + 1)$, and $\hat{\phi}^{o_2, S_2 }$ has dimensions $(2 S_2 + 1) \times 1$ in angular momentum space. In particular, it must be noted that the matrix $T^{o_1, S_1 }_{o_2, S_2 }$ is not square but rectangular if $S_1 \neq S_2$.

\subsection*{Example: Zeeman Hamiltonian}

The single-particle Hamiltonian that we are considering is the most general one: by specifying the form of the matrix $T^{o_1, S_1 }_{o_2, S_2 }$ one can obtain any single-electron term, including of course relativistic contributions in addition to the standard kinetic hopping. We give two examples. First, we consider the Zeeman Hamiltonian arising from a position-dependent magnetic field $\boldsymbol{B}_a$ coupling with electronic magnetic moments as
\begin{align}
\hat{H}_{\mathrm{Zeeman}} = \mu_{\mathrm{B}}   \sum_{o,  S} g_{l S} \boldsymbol{B}_a \cdot \hat{\boldsymbol{S}}_{o S}^{\phi}, 
\end{align}
where $o = (a, n, l)$, $\mu_{\mathrm{B}} = \left| e \right| / (2 m c)$ for $\hbar = 1$ (as we assume throughout this Article), $g_{l S}$ is the electron $g$-factor, and
\begin{align}
\hat{\boldsymbol{S}}_{o S}^{\phi} \equiv  \hat{\boldsymbol{S}}_i^{\phi} = \sum_{M, M' = - S}^S \hat{\phi}^{\dagger}_{o, S, M} \left( \boldsymbol{S}_{o S}\right)^M_{M'} \hat{\phi}^{o, S, M'} \equiv \hat{\phi}^{\dagger}_{i} \cdot \boldsymbol{S}_{i} \cdot \hat{\phi}^{i}
\label{local J}
\end{align}
is the angular momentum operator associated with the spin $i$; $\boldsymbol{S}_{o S} = \boldsymbol{S}_{i}$ is the angular momentum matrix of dimension $2 S + 1$ and quantization axis $\boldsymbol{u}_z(i)$. This Zeeman term can be obtained in our formalism by specifying a part of the single-particle matrix $T$ as:
\begin{align}
\left(T_{\mathrm{Zeeman}}\right)^{o_1 , S_1, M_1 }_{o_2 , S_2, M_2 } = \delta^{o_1}_{o_2}    \delta^{S_1}_{S_2}  \mu_{\mathrm{B}}     g_{l_1 S_1} \boldsymbol{B}_{a_1} \cdot  \left( \boldsymbol{S}_{o_1 S_1}\right)^{M_1}_{M_2} .
\label{Zeeman hopping}
\end{align}

\subsection*{Example: atomic spin-orbit Hamiltonian}

The second relativistic example that we consider is the Hamiltonian for atomic spin-orbit coupling,
\begin{align}
\hat{H}_{\mathrm{at. SOC}} & = \sum_{a_1, n_1} \sum_{l_1, S_1, M_1} \sum_{l_2, S_2, M_2}  \hat{\phi}^{\dagger}_{a_1 n_1 l_1 S_1 M_1}   \xi_{a_1 n_1} \left( \boldsymbol{l} \cdot \boldsymbol{s} \right)^{l_1, S_1, M_1}_{l_2, S_2, M_2}  \hat{\phi}^{a_1 n_1 l_2 S_2 M_2} \nonumber \\
& = \sum_{1}    \hat{\phi}^{\dagger}_{1} \frac{1}{2}  \xi_{a_1 n_1} \left[ S_1 (S_1 + 1 ) - \frac{3}{4} - l_1 (l_1 + 1) \right]   \hat{\phi}^{1} ,
\end{align}
where we have used $\hat{\boldsymbol{l}} \cdot \hat{\boldsymbol{s}} = \left( \hat{S}^2 - \hat{s}^2 - \hat{l}^2 \right) / 2$ and $s = 1/2$. The corresponding single-particle Hamiltonian matrix is then
\begin{align}
\left(T_{\mathrm{at. SOC}}\right)^{1 }_{2 }  \equiv \delta^1_2 \frac{1}{2}  \xi_{a_1 n_1} \left[ S_1 (S_1 + 1) - \frac{3}{4} - l_1 (l_1 + 1) \right] . 
\end{align}

\subsection*{Local single-electron Hamiltonian: magnetic field and local magnetic anisotropy}

The general form of the \emph{local} single-electron Hamiltonian can be specified as 
\begin{align}
T^{i M}_{i M'} \equiv E_i \delta^M_{M'}  + \mu_{\mathrm{B}} g_i \boldsymbol{B}_i \cdot \left( \boldsymbol{S}_i \right)^M_{M'}  +   A^{i M}_{i M'}         ,
\label{local T}
\end{align} 
where we have included the $M$-independent energy eigenvalue $E_i$ and a Zeeman term as given in Eq.\eqref{Zeeman hopping}. The last term in Eq.\eqref{local T}, denoted as $A^{i M}_{i M'}$,  originates from crystal field effects and can be considered as a single-site contribution to the magnetic anisotropy. For example, for the case of quadratic spin anisotropy $\left(A_Q\right)^{i M}_{i M'} = \left( \boldsymbol{S}_{i} \cdot  \overline{\boldsymbol{A}}_i \cdot \boldsymbol{S}_{i} \right)^M_{M'}$. We note that here (and in the following) we use the term \emph{local} referring to a fictitious lattice where each site corresponds to one spin, labelled by $i$. This concept of locality may not correspond to locality in position space (i.e., there can be more than one spin associated with a single atom constituting the lattice in position space).

\subsection{The interaction Hamiltonian}
The interaction Hamiltonian is generally written as
\begin{align}
\hat{H}_V^{\phi}  \equiv & \frac{1}{2}   \sum_{1, 2, 3, 4}   \hat{\phi}^{\dagger}_{1}  \hat{\phi}^{\dagger}_{2}   V_{3, 4}^{1, 2} \hat{\phi}^{3}   \hat{\phi}^{4} ,
\end{align}
where
\begin{align}
&  V_{  3  ,  4 }^{ 1 ,  2 } \equiv \int \text{d} v(\boldsymbol{x}) \int \text{d} v(\boldsymbol{x}')  \, \psi^*_{1}(\boldsymbol{x}) \, \psi^*_{2}(\boldsymbol{x}')    V(\boldsymbol{x} - \boldsymbol{x}') \,  \psi_{3}(\boldsymbol{x}') \, \psi_{4}(\boldsymbol{x}) .    
\end{align}
While we have chosen the single-particle Hamiltonian to be the most general one, we assume that the interaction Hamiltonian is invariant under rotation of the electronic quantization axes $\boldsymbol{r}_i$ defined for each orbital. In particular, we consider an intra-site (Hubbard) interaction. A more precise formalization of this requirement will be given in the next Section.

\section{Rotation of the spin quantization axes}
\label{sec: Rotation}

\subsection{Rotation operator}
We define the rotation operator for the quantization axis of the $i$-th individual spin as:
\begin{align}
R_i \equiv   \text{e}^{\text{i} \delta \boldsymbol{\varphi}_i \cdot \boldsymbol{S}_{i}} ,
\label{rotation}
\end{align}
where the individual rotation parameter is
\begin{align}
\delta \boldsymbol{\varphi}_i \equiv \theta_i \boldsymbol{u}_i ,
\label{delta phi def}
\end{align}
where $\boldsymbol{u}_i$ is a unit vector and $\theta_i$ is the azimuthal angle of rotation.

In the initial Hamiltonian we change the basis according to the spinor transformation
\begin{align}
& \hat{\phi}^{\dagger}_{i_1} \equiv \hat{\psi}^{\dagger}_{i_1} \cdot R^{\dagger}_{i_1}, \nonumber \\
& \hat{\phi}^{i_2} \equiv R_{i_2} \cdot \hat{\psi}^{i_2}   .
\label{transformation}
\end{align}
To understand the meaning of the transformation \eqref{transformation}, we note that the expectation value of the angular momentum on a state of one $\phi$ fermion is:
\begin{align}
\left< 0 \left| \hat{\phi}^{i , M} \hat{\boldsymbol{S}}_i^{\phi} \hat{\phi}^{\dagger}_{i , M}    \right| 0 \right> = (\boldsymbol{S}_{i  })^M_M = M \boldsymbol{r}_i = M \boldsymbol{u}_z(i) ,
\end{align} 
where $\hat{\boldsymbol{S}}_i^{\phi}$ is given by Eq.\eqref{local J}; the expectation value of the same operator on a state of one $\psi$ fermion, instead, is given by:
\begin{align}
  \left< 0 \left| \hat{\psi}^{i , M} \hat{\boldsymbol{S}}_i^{\phi} \hat{\psi}^{\dagger}_{i , M}    \right| 0 \right>    =    \left[   R^{\dagger}_i \cdot   \boldsymbol{S}_{i} \cdot R_i \right]^M_M ,
\label{orientation}
\end{align} 
whose direction is therefore specified by the rotation operators, and in particular may be not parallel to $\boldsymbol{u}_z(i)$. If we expand Eq.\eqref{orientation} in powers of small $\theta_i$ up to the second order, we obtain:
\begin{align}
  \left< 0 \left| \hat{\psi}^{i , M} \hat{\boldsymbol{S}}_i^{\phi} \hat{\psi}^{\dagger}_{i , M} \right| 0 \right>  & \approx M \left[ \boldsymbol{r}_i + \boldsymbol{r}_i \times \delta \boldsymbol{\varphi}_i - \frac{1}{2} \boldsymbol{r}_i \left( \delta \boldsymbol{\varphi}_i \right)^2 + \frac{1}{2} \delta \boldsymbol{\varphi}_i \left( \delta \boldsymbol{\varphi}_i \cdot \boldsymbol{r}_i  \right)  \right] \nonumber \\
& \equiv M \boldsymbol{e}_i .
\label{q axes rotated}
\end{align}
The rotated quantization axis $\boldsymbol{e}_i$ satisfies $\boldsymbol{e}_i \cdot \boldsymbol{e}_i = 1$, i.e., it is a unit vector for any $\delta \boldsymbol{\varphi}_i$ (to the specified order). Therefore, we need only two variables $(\theta_i, \varphi_i)$ to specify it completely, which are the polar angles with respect to the initial quantization axis $\boldsymbol{r}_i \equiv \boldsymbol{u}_z(i)$, which means that one of the three components of $\delta \boldsymbol{\varphi}_i$ is redundant. To determine the necessary and sufficient set of variables, we impose
\begin{align}
\boldsymbol{e}_i & \equiv \boldsymbol{u}_x(i) \sin(\theta_i) \cos(\varphi_i)   +   \boldsymbol{u}_y(i) \sin(\theta_i) \sin(\varphi_i)    +   \boldsymbol{u}_z(i) \cos(\theta_i)  \nonumber \\
& \approx \boldsymbol{u}_x(i)  \theta_i \cos(\varphi_i)   +   \boldsymbol{u}_y(i)  \theta_i \sin(\varphi_i)    +   \boldsymbol{r}_i \left( 1 - \frac{(\theta_i)^2}{2} \right)  .
\label{e theta phi}
\end{align}
where in the last line we have kept up to second-order terms in $\theta_i$. Recalling Eq.\eqref{delta phi def} and comparing Eqs.\eqref{q axes rotated} and \eqref{e theta phi}, we see that
\begin{align}
\boldsymbol{r}_i \times \delta \boldsymbol{\varphi}_i   + \frac{1}{2} \delta \boldsymbol{\varphi}_i \left( \delta \boldsymbol{\varphi}_i \cdot \boldsymbol{r}_i  \right) = \boldsymbol{u}_x(i)  \theta_i \cos(\varphi_i)   +   \boldsymbol{u}_y(i)  \theta_i \sin(\varphi_i) .
\label{order discrimination}
\end{align}
The RHS of Eq.\eqref{order discrimination} is of first order in $\theta_i$, while the term $\frac{1}{2} \delta \boldsymbol{\varphi}_i \left( \delta \boldsymbol{\varphi}_i \cdot \boldsymbol{r}_i  \right)$ on the LHS is of second order. Therefore, we need to impose $\delta \boldsymbol{\varphi}_i \cdot \boldsymbol{r}_i = 0$. From Eq.\eqref{order discrimination} we then determine $\delta \boldsymbol{\varphi}_i$ uniquely as:
\begin{align}
\delta \boldsymbol{\varphi}_i \equiv \theta_i \left[    \sin(\varphi_i) \boldsymbol{u}_x(i) - \cos(\varphi_i) \boldsymbol{u}_y(i) \right] .
\end{align}

\subsection*{Example: Spin-$1/2$ rotations}

In the particular case of $S = 1/2$, we have (exactly)
\begin{align}
R_{o, 1/2} & = \cos(\theta_o / 2) + \text{i} \, \sin(\theta_o / 2) \boldsymbol{u}_o \cdot \boldsymbol{\sigma}(o) \nonumber \\
& = \cos(\theta_o / 2) + \text{i} \, \sin(\theta_o / 2) \left[    \sin(\varphi_o) \sigma_x(o) - \cos(\varphi_o) \sigma_y(o) \right]  ,
\end{align}
where $\boldsymbol{\sigma}(o)$ is the vector of Pauli matrices expressed in the reference frame of orbital $o$. This choice is appropriate for studying systems without orbital degrees of freedom, such as the single-band Hubbard model, or if one is interested only in the exchange couplings due to the intrinsic spins of the electrons \cite{Secchi13}. Here we consider the more general case of arbitrary magnetic moments $\hat{\boldsymbol{S}}_i^{\phi} = \hat{\boldsymbol{l}}_i^{\phi} + \hat{\boldsymbol{s}}_i^{\phi}$, accounting for the orbital degrees of freedom.

\subsection{The Hamiltonian after the transformation}

The single-particle Hamiltonian, Eq.\eqref{SP electronic Hamiltonian}, is written in terms of the new $\psi$ fields as:
\begin{align}
\hat{H}_T^{\phi}  = \sum_{i_1} \sum_{i_2}   \hat{\psi}^{\dagger}_{i_1} \cdot \text{e}^{- \text{i} \delta \boldsymbol{\varphi}_{i_1} \cdot \boldsymbol{S}_{i_1}}  \cdot   T^{i_1}_{i_2} \cdot  \text{e}^{\text{i} \delta \boldsymbol{\varphi}_{i_2} \cdot \boldsymbol{S}_{i_2}} \cdot \hat{\psi}^{i_2}  .
\label{H_T rotation}
\end{align}
We consider small deviations from the reference quantization axes of the local total angular momenta. Therefore, we now apply a small-$\theta$ expansion of Eq.\eqref{H_T rotation}, keeping only the terms of orders $\theta^0$, $\theta^1$ and $\theta^2$. The single-particle Hamiltonian is then given, to this order, by
\begin{align}
\hat{H}_T^{\phi} \approx \hat{H}_T^{\psi} +   \hat{H}_T^{\psi, \theta}  +   \hat{H}_T^{\psi, \theta^2}   ,
\end{align}
with 
\begin{align}
  \hat{H}_T^{ \psi, \theta}   \equiv \sum_{i} \delta \boldsymbol{\varphi}_i \cdot \hat{\boldsymbol{\mathcal{V}}}_i
  \label{H1}
\end{align}
and
\begin{align}
  \hat{H}_T^{ \psi, \theta^2  }   \equiv  \frac{1}{2} \sum_{i i'}   \delta \boldsymbol{\varphi}_i \cdot \hat{\overline{\boldsymbol{\mathcal{M}}}}_{i i'} \cdot  \delta \boldsymbol{\varphi}_{i'} ,
  \label{H2}
\end{align} 
where
\begin{align}
\hat{ \mathcal{V} }_{i \alpha}  & \equiv   \mathrm{i} \sum_j \left( \hat{\psi}^{\dagger}_{j} \cdot T^j_i \cdot  S_{i \alpha} \cdot \hat{\psi}^i 
-  \hat{\psi}^{\dagger}_{i} \cdot S_{i \alpha} \cdot T^i_j   \cdot \hat{\psi}^j    \right)   = \mathrm{i} \, \mathrm{Tr}_M \! \left( S_{i \alpha} \cdot \Big[ \hat{\rho} ; T \Big]^i_i    \right) ,
\label{vector V}
\end{align}
\begin{align}
\hat{ \mathcal{M} }_{i \alpha, i' \alpha'}  \equiv   &
- \delta_{i i'} \frac{1}{2} \sum_j \left[   \hat{\psi}^{\dagger}_i \cdot \left( S_{i \alpha} \cdot S_{i \alpha'}   + S_{i \alpha'} \cdot S_{i \alpha} \right) \cdot T^i_{j}  \cdot \hat{\psi}^{j} \right. \nonumber \\
& \left.  \quad \quad \quad \quad \quad + \hat{\psi}^{\dagger}_j \cdot T^j_{i}    \cdot \left( S_{i \alpha} \cdot S_{i \alpha'}  +   S_{i \alpha'} \cdot S_{i \alpha}  \right) \cdot \hat{\psi}^{i}     \right]  \nonumber \\
& + \hat{\psi}^{\dagger}_i \cdot S_{i \alpha} \cdot T^i_{i'} \cdot S_{i' \alpha'} \cdot \hat{\psi}^{i'}   
+ \hat{\psi}^{\dagger}_{i'} \cdot S_{i' \alpha'} \cdot T_i^{i'} \cdot S_{i \alpha} \cdot \hat{\psi}^{i} \nonumber \\
= & \mathrm{Tr}_M  \left( S_{i \alpha} \cdot T^i_{i'} \cdot S_{i' \alpha'} \cdot \hat{\rho}^{i'}_i      
+ S_{i' \alpha'} \cdot T_i^{i'} \cdot S_{i \alpha} \cdot \hat{\rho}_{i'}^i  \right)  \nonumber \\
& - \delta_{i i'} \frac{1}{2} \mathrm{Tr}_M \! \left(     \left(  S_{i \alpha} \cdot S_{i \alpha'}    +   S_{i \alpha'} \cdot S_{i \alpha}   \right) \cdot \Big\{ \hat{\rho} ; T \Big\}^i_i    \right)   ;
\label{matrix M}
\end{align}
in Eqs.\eqref{vector V} and \eqref{matrix M} we have introduced the density matrix operator 
\begin{align}
\hat{\rho}^1_2 \equiv \hat{\psi}^{\dagger}_2 \hat{\psi}^1 ,
\label{density matrix definition}
\end{align}
where $1$ and $2$ are general indexes for the fermionic fields. We have also used the notations $\left[ \hat{A}; \hat{B} \right] \equiv \hat{A}  \hat{B} - \hat{B}  \hat{A}$ and $\left\{ \hat{A}; \hat{B} \right\} \equiv \hat{A}  \hat{B} + \hat{B}  \hat{A}$ for commutators and anti-commutators, respectively (if $\hat{A}$ and $\hat{B}$ are matrices, then matrix products are implied). The matrix operator \eqref{matrix M} has been defined, for later convenience, such that $\hat{ \mathcal{M} }_{i \alpha, i' \alpha'} = \hat{ \mathcal{M} }_{i' \alpha', i \alpha}$.
 
Differently from the single-particle Hamiltonian, the interaction Hamiltonian is \emph{assumed} to be rotationally invariant, i.e.,
\begin{align}
\hat{H}_V^{\phi} = \hat{H}_V^{\psi}.
\end{align}
In \ref{app: Invariance} we discuss the conditions for the fulfilment of this requirement.

\section{Action and partition function}
\label{sec: Action}

\subsection{Action}

The derivation of the effective rotational action for the electronic system proceeds analogously to our previous treatment of the spin-$1/2$ rotations \cite{Secchi13}. We write the Matsubara action as
\begin{align}
\mathcal{S}\left[ \bar{\phi}, \phi  \right] =    \int_{\varepsilon}^{\beta} \text{d} \tau \Bigg\{ \bar{\phi}(\tau) \cdot   \dot{\phi}(\tau - \varepsilon) + K \Big[ \bar{\phi}(\tau) ,  \phi(\tau - \varepsilon) \Big]   \Bigg\} ,
\label{action simple}
\end{align}
where $\bar{\phi}(\tau)$ and $\phi(\tau)$ are contour Grassmann variables \cite{AltlandSimons}, $K = H - \mu N$ is the grand-canonical potential, with $\mu$ being the chemical potential and $N$ the number of electrons. We then apply the rotation transformation to the $\tau$-dependent Grassmann fields, by introducing $i$-dependent rotation fields $\delta \boldsymbol{\varphi}_i(\tau)$ along the contour. Keeping into account that the term $\mu N$ is of course rotationally invariant, we obtain
\begin{align}
\mathcal{S}\left[ \bar{\phi}, \phi \right] = \mathcal{S}\left[ \bar{\psi}, \psi \right] + \mathcal{S}'\left[ \bar{\psi}, \psi, \delta \boldsymbol{\varphi} \right] ,
\end{align}
with 
\begin{align}
  \mathcal{S}'\left[ \bar{\psi}, \psi, \delta \boldsymbol{\varphi} \right] & \equiv \int_{\varepsilon}^{\beta} \text{d} \tau \, \bar{\psi}(\tau) \cdot \Delta(\tau) \cdot \psi(\tau - \varepsilon) \nonumber \\
  & \approx \! \int_{\varepsilon}^{\beta} \! \text{d} \tau \, \bar{\psi}(\tau) \cdot \! R^{\dagger}(\tau) \! \cdot  \! \dot{R}(\tau) \! \cdot \psi(\tau - \varepsilon)  
+ \! \int_{\varepsilon}^{\beta} \! \text{d} \tau \!  \left[   H_T^{ \psi, \theta}\!(\tau) +   H_T^{ \psi, \theta^2}\!(\tau) \right] \!  ,
\end{align}
where $\Delta(\tau)$ introduced in the first line is a kernel which depends on the fields $\delta \boldsymbol{\varphi}(\tau)$ and their derivatives $\delta \dot{\boldsymbol{\varphi}}(\tau)$, in principle to all orders. If we consider the regime of small rotations, up to quadratic order, we obtain the expression in the second line, where $H_T^{ \psi, \theta}\!(\tau)$ and $H_T^{ \psi, \theta^2}\!(\tau)$ correspond, respectively, to the expressions \eqref{H1} and \eqref{H2} with the operators $\hat{\psi}^{\dagger}$ and $\hat{\psi}$ replaced, respectively, by the Grassmann fields $\bar{\psi}(\tau)$ and $\psi(\tau - \varepsilon)$, and we have
\begin{align}
R^{\dagger}(\tau) \cdot \dot{R}(\tau)     \approx  \text{i} \left[ \delta \dot{\boldsymbol{\varphi}}(\tau) \cdot \boldsymbol{S} \, \right]  +  \frac{\mathrm{i}}{2}   \left[ \delta  \boldsymbol{\varphi}(\tau) \times \delta \dot{\boldsymbol{\varphi}}(\tau)  \right] \cdot \boldsymbol{S} ,
\label{R dagger R dot}
\end{align}
where we have used the commutation relations of the spin matrices $\left[ S_{\alpha}, S_{\beta} \right] = \mathrm{i} \sum_{\gamma} \varepsilon^{\alpha \beta \gamma} S_{\gamma}$.

We then distinguish the terms which are of the same order in product combinations of the $\delta \boldsymbol{\varphi}$ and the $\delta \dot{\boldsymbol{\varphi}}$ fields, and accordingly we put
\begin{align}
\mathcal{S}' \equiv \sum_{n = 1}^{\infty} \mathcal{S}^{(n)} , \quad \mathcal{S}^{(n)}  &  \equiv \int_{\varepsilon}^{\beta} \text{d} \tau \, \bar{\psi}(\tau) \cdot \Delta^{(n)}(\tau) \cdot \psi(\tau - \varepsilon)  .
\label{S^n}
\end{align}

\subsection{Partition function}
\label{subsec: partition}

The grand-canonical partition function is written as a path integral over Grassmann variables as:
\begin{align}
Z \equiv  \mathrm{Tr} \left\{  \mathrm{e}^{- \beta \left( \hat{H} - \mu \hat{N} \right)  } \right\}  \equiv \int \mathrm{D}\!\left[ \bar{\phi}, \phi \right] \text{e}^{- \mathcal{S}\left[ \bar{\phi}, \phi \right] }  ,
\end{align}
where the trace is taken over the complete set of many-body eigenstates of the Hamiltonian. It should be emphasized that the states are weighted by Boltzmann factors, which depend only on the energy \emph{and therefore cannot distinguish between degenerate broken-symmetry states}. However, the mapping to a classical spin model makes sense only if the reference electronic state is not symmetric with respect to rotations of the spins, since this is an essential property of classical spin configurations. We therefore define a ``broken-symmetry'' partition function, which we label as $Z^*$, as
\begin{align}
Z^* \equiv  \left< \lbrace \boldsymbol{r}_i \rbrace \right|   \mathrm{e}^{- \beta \left( \hat{H} - \mu \hat{N} \right)  } \left| \lbrace \boldsymbol{r}_i \rbrace \right>   \equiv \int \mathrm{D}^*\!\left[ \bar{\phi}, \phi \right] \text{e}^{- \mathcal{S}\left[ \bar{\phi}, \phi \right] }  ,
\end{align}
where $\left| \lbrace \boldsymbol{r}_i \rbrace \right>$ is a reference electronic state which realizes the spin configuration specified by the set of unit vectors $\lbrace \boldsymbol{r}_i \rbrace$, and the measure $\mathrm{D}^*$ of the path integral in the last passage is defined formally. 
 
We implement the rotations of the spin quantization axes by applying the transformation discussed above from the $\left[ \bar{\phi}, \phi \right]$ to the $\left[ \bar{\psi}, \psi \right]$ fermions, and we define the functional
\begin{align}
\mathcal{Z}\left[ \delta \boldsymbol{\varphi}_i(\tau) \right] \equiv    \int \mathcal{D}\left[ \bar{\psi}, \psi \right] \text{e}^{- \mathcal{S}\left[ \bar{\psi}, \psi \right] }    
 \text{e}^{- \mathcal{S}'\left[ \bar{\psi}, \psi, \delta \boldsymbol{\varphi} \right] } ,
\label{partition}
\end{align} 
where $\mathcal{D}\left[ \bar{\phi}, \phi \right] \equiv \mathrm{D}^*\!\left[ \bar{\phi}, \phi \right] / Z^*$. We expand Eq.\eqref{partition} as
\begin{align}
\mathcal{Z}\left[ \delta \boldsymbol{\varphi}_i(\tau)  \right] & \equiv   \int \mathcal{D}\left[ \bar{\psi}, \psi \right] \text{e}^{- \mathcal{S}\left[ \bar{\psi}, \psi \right] }    
  \left\{   1 + \sum_{m = 1}^{\infty} \frac{ \left( - \mathcal{S}'\left[ \bar{\psi}, \psi, \delta \boldsymbol{\varphi} \right]  \right)^m  }{m!}   \right\} \nonumber \\
  & \equiv \sum_{n = 0}^{\infty} \mathcal{Z}^{(n)}\left[ \delta \boldsymbol{\varphi}_i(\tau)  \right] ,
\label{partition expanded}
\end{align} 
where $\mathcal{Z}^{(n)}$ includes all the terms that require the product of $n$ fields $\delta \varphi$ or $\delta \dot{\varphi}$. Here we focus on the contributions to the broken-symmetry partition function coming from trajectories $\delta \boldsymbol{\varphi}_i(\tau)$ close to $\delta \boldsymbol{\varphi}_i(\tau) = \boldsymbol{0}$, i.e., we study the regime of small rotations of the spin quantization axes from their initial configuration. Specifically, we consider $\mathcal{Z}^{(0)}$, $\mathcal{Z}^{(1)}$ and $\mathcal{Z}^{(2)}$; with reference to Eq.\eqref{S^n}, these are given by 
\begin{align}
& \mathcal{Z}^{(0)}   \equiv \int \mathcal{D}\left[ \bar{\psi}, \psi \right] \text{e}^{- \mathcal{S}\left[ \bar{\psi}, \psi \right] }    
  =  1 , \nonumber \\
& \mathcal{Z}^{(1)}\left[ \delta \boldsymbol{\varphi}_i(\tau)  \right]   \equiv \int \mathcal{D}\left[ \bar{\psi}, \psi \right] \text{e}^{- \mathcal{S}\left[ \bar{\psi}, \psi \right] }    
    \left\{ - \mathcal{S}^{(1)}\left[ \bar{\psi}, \psi, \delta \boldsymbol{\varphi}  \right]  \right\}  ,   \nonumber \\
& \mathcal{Z}^{(2)}\left[ \delta \boldsymbol{\varphi}_i(\tau)  \right]    \equiv \int     \mathcal{D}\!\left[ \bar{\psi}, \psi \right] \text{e}^{- \mathcal{S}\left[ \bar{\psi}, \psi \right] }    
  \left\{     - \mathcal{S}^{(2)}\!\left[ \bar{\psi}, \psi, \delta \boldsymbol{\varphi}  \right]  + \frac{1}{2} \Big( \mathcal{S}^{(1)}\!\left[ \bar{\psi}, \psi, \delta \boldsymbol{\varphi}  \right]  \Big)^2   \right\} 
\label{before fermion integration}
\end{align}
(the definitions used here are slightly different from Eqs.(43) of Ref.\cite{Secchi13}).

\section{Effective rotational action for small spin deviations}
\label{sec: Effective action}

We now derive an effective action for the fields $\delta \boldsymbol{\varphi}_i(\tau)$ in the regime of small $\delta \boldsymbol{\varphi}_i(\tau)$ by integrating out the fermionic fields $\bar{\psi}_i(\tau)$ and $\psi_i(\tau)$.

\subsection{Fermionic integration}
To integrate out the fermionic fields, we use the identities
\begin{align}
&  \text{i} \int \mathcal{D}\left[ \bar{\psi}, \psi \right] \text{e}^{- \mathcal{S}\left[ \bar{\psi}, \psi \right] }    \bar{\psi}_{2}(\tau') \, \psi_{1}(\tau) \equiv G^{1}_{2}(\tau, \tau') = - \mathrm{i} \left< \mathcal{T}_{\gamma} \hat{\psi}_1(\tau) \hat{\psi}_2^{\dagger}(\tau') \right> , \nonumber \\
& \text{i}^2 \int \mathcal{D}\!\left[ \bar{\psi}, \psi \right] \text{e}^{- \mathcal{S}\left[ \bar{\psi}, \psi \right] }    \bar{\psi}_{2}(\tau_2) \, \psi_{1}(\tau_1) \bar{\psi}_{4}(\tau_4) \, \psi_{3}(\tau_3) \equiv \chi^{1, 3}_{2, 4}(\tau_1, \tau_2, \tau_3, \tau_4) \nonumber \\
& \quad \quad \quad =  \mathrm{i}^2  \left< \mathcal{T}_{\gamma} \hat{\psi}_1(\tau_1) \, \hat{\psi}_2^{\dagger}(\tau_2) \, \hat{\psi}_3(\tau_3) \, \hat{\psi}_4^{\dagger}(\tau_4)  \right>  ,
\end{align}
where $G$ and $\chi$ respectively denote single-particle and two-particle Matsubara Green's functions, $\hat{\psi}(\tau) \equiv \mathrm{e}^{\tau \hat{K}} \hat{\psi} \mathrm{e}^{- \tau \hat{K}}$, $\hat{K} = \hat{H} - \mu \hat{N}$, and $\mathcal{T}_{\gamma}$ is the time-ordering operator along the Matsubara axis $\gamma$. We recall that the Green's functions should be computed for the electronic state $\left| \lbrace \boldsymbol{r}_i \rbrace \right>$, as follows from the formal definition of the measure $\mathcal{D}\!\left[ \bar{\psi}, \psi \right]$ given in Section \ref{subsec: partition}. The two-particle Green's function can be written as
\begin{align}
\chi^{1, 3}_{2, 4}(\tau_1, \tau_2, \tau_3, \tau_4) \equiv & G^1_{2}(\tau_1, \tau_2) G^3_{4}(\tau_3, \tau_4)  - G^1_{4}(\tau_1, \tau_4) G^3_{2}(\tau_3, \tau_2) \nonumber \\
& + \sum_{1' 2' 3' 4'} \int \text{d} \tau_1' \int \text{d} \tau_2' \int \text{d} \tau_3' \int \text{d} \tau_4' G^1_{1'}(\tau_1, \tau_1') G^3_{3'}(\tau_3, \tau_3') \nonumber \\
& \quad \quad \quad \quad   \times \Gamma^{1' , 3'}_{2' , 4'}(\tau_1', \tau_2', \tau_3', \tau_4') G^{4'}_{4}(\tau_4', \tau_4) G^{2'}_{2}(\tau_2', \tau_2) \nonumber \\
\equiv & \left(\chi^0\right)^{1 , 3}_{2, 4}(\tau_1, \tau_2, \tau_3, \tau_4) + \left( \chi^{\Gamma} \right)^{1, 3}_{2, 4}(\tau_1, \tau_2, \tau_3, \tau_4)  , 
\label{2p GF}
\end{align}
where $\chi^{\Gamma}$ is the sum of connected Feynman diagrams, depending on the vertex $\Gamma$. While a number of previous works on magnetic interactions \cite{Lichtenstein87, Katsnelson00, Katsnelson02, Secchi13} have employed the simplifying approximation of neglecting vertices in two-particle Green's functions ($\Gamma = 0$), we will here remove this assumption and carry on the derivation with the full two-particle Green's functions. In fact, while it has been shown that neglecting vertices leads nevertheless to the correct expression for the spin-wave stiffness \emph{if the self-energy is local} \cite{LKvolume}, there is no guaranty that other magnetic properties will be unaffected by that approximation, which is, as a matter of fact, uncontrolled. Moreover, we consider here the general case of a non-local self-energy. 

After the fermionic integration, we obtain:
\begin{align}
& \mathcal{Z}^{(1)}\left[ \delta \boldsymbol{\varphi}_i(\tau)  \right] =  \int \mathcal{D}\left[ \bar{\psi}, \psi \right] \text{e}^{- \mathcal{S}\left[ \bar{\psi}, \psi \right] }    
   \left\{ -  \int_{\varepsilon}^{\beta} \text{d} \tau \, \bar{\psi}(\tau) \cdot \Delta^{(1)}(\tau) \cdot \psi(\tau - \varepsilon)  \right\} \nonumber \\
& =        \sum_1 \sum_2     \int_{\varepsilon}^{\beta} \text{d} \tau      \left( \Delta^{(1)}(\tau) \right)^2_1  \mathrm{i} \, G^1_2(\tau - \varepsilon, \tau )         \equiv       - \mathrm{Tr}_{\lbrace i,   M \rbrace}  \left[    \rho  \cdot \int_{0}^{\beta} \text{d} \tau   \Delta^{(1)}(\tau)   \right]     \nonumber \\
& \equiv       -   \mathcal{S}_1\!\left[  \delta \boldsymbol{\varphi}_i(\tau)  \right]   ,
\label{Z^1}
\end{align}
\begin{align}
 &  \mathcal{Z}^{(2)}\left[ \delta \boldsymbol{\varphi}_i(\tau)  \right]  \equiv \int \mathcal{D}\left[ \bar{\psi}, \psi \right] \text{e}^{- \mathcal{S}\left[ \bar{\psi}, \psi \right] }    
   \Bigg\{   -    \int_{\varepsilon}^{\beta} \text{d} \tau \, \bar{\psi}(\tau) \cdot \Delta^{(2)}(\tau) \cdot \psi(\tau^-)    \nonumber \\
& \quad \quad \quad \quad \quad \quad  + \frac{1}{2}       \int_{\varepsilon}^{\beta}  \! \text{d} \tau \, \bar{\psi}(\tau) \cdot \Delta^{\!(1)}\!(\tau) \cdot \psi(\tau^-)          \int_{\varepsilon'}^{\beta} \!  \text{d} \tau' \, \bar{\psi}(\tau') \cdot \Delta^{(1)}(\tau') \cdot \psi(\tau'^- )  \Bigg\}    \nonumber \\
& \equiv        - \mathrm{Tr}_{\lbrace i,  M \rbrace}  \left[    \rho  \cdot \int_{0}^{\beta} \text{d} \tau   \Delta^{(2)}(\tau)   \right]  \nonumber \\
&    \quad     -    \frac{1}{2}     \sum_{1 2 3 4}   \int_{0}^{\beta}   \text{d} \tau    \int_{0}^{\beta}  \text{d} \tau'     \left( \Delta^{(1)}(\tau) \right)^2_1  \, \chi^{1, 3}_{2 , 4}(\tau, \tau^+, \tau', \tau'^+)  \, \left( \Delta^{(1)}(\tau') \right)^4_3       \nonumber \\
& \equiv     -   \mathcal{S}_2\!\left[  \delta \boldsymbol{\varphi}_i(\tau)  \right] ;
\label{Z^2}
\end{align}
for the sake of brevity, we do not report here the expressions for $\Delta^{(1)}(\tau)$ and $\Delta^{(2)}(\tau)$. In Eqs.\eqref{Z^1} and \eqref{Z^2} we have introduced the symbol for the single-particle density matrix,
\begin{align}
G(\tau - \varepsilon, \tau) \equiv \mathrm{i} \, \rho .
\end{align}
It should be noted that the only two-particle Green's function appearing in Eq.\eqref{Z^2} is of the form
\begin{align}
\chi^{1, 3}_{2 , 4}(\tau, \tau^+, \tau', \tau'^+) = - \left< \mathcal{T}_{\gamma} \hat{\rho}^1_2(\tau) \, \hat{\rho}^3_4(\tau') \right> ,
\label{only two}
\end{align}
which is a correlator between two density-matrix operators [see Eq.\eqref{density matrix definition}] taken at different imaginary times. The result depends on imaginary time only via the combination $(\tau - \tau')$.

\subsection{Effective action}

The contributions to the rotation functional up to the quadratic order in the $\delta \boldsymbol{\varphi}$ fields can be written as
\begin{align}
\mathcal{Z}^{(0)} + \mathcal{Z}^{(1)}\!\left[ \delta \boldsymbol{\varphi}_i(\tau)  \right] + \mathcal{Z}^{(2)}\!\left[ \delta \boldsymbol{\varphi}_i(\tau)  \right] =   1 - \mathcal{S}_1\!\left[  \delta \boldsymbol{\varphi}_i(\tau)  \right] - \mathcal{S}_2\!\left[  \delta \boldsymbol{\varphi}_i(\tau)  \right]  \approx 
 \text{e}^{ -  \mathcal{S}\left[  \delta \boldsymbol{\varphi}_i(\tau)  \right]} ,
\label{Z0 + Z1 + Z2}
\end{align}
where the effective action is defined as $\mathcal{S}\!\left[  \delta \boldsymbol{\varphi}_i(\tau)  \right] =  \mathcal{S}_1\!\left[  \delta \boldsymbol{\varphi}_i(\tau)  \right] +  \mathcal{S}_2\!\left[  \delta \boldsymbol{\varphi}_i(\tau)  \right] + \frac{1}{2} \left( \mathcal{S}_1\!\left[  \delta \boldsymbol{\varphi}_i(\tau)  \right] \right)^2$. Namely,
\begin{align}
& \mathcal{S}\!\left[  \delta \boldsymbol{\varphi}_i(\tau)  \right] =     \mathrm{Tr}_{\lbrace i,   J, M \rbrace} \left\{ \rho \cdot   \int_{0}^{\beta} \text{d} \tau  \left[ \Delta^{(1)}(\tau) + \Delta^{(2)}(\tau) \right] \right\} \nonumber \\
 &   + \frac{1}{2}     \sum_{1 2 3 4}   \int_{0}^{\beta}   \text{d} \tau    \int_{0}^{\beta}  \text{d} \tau'    \left( \Delta^{(1)}(\tau)   \right)^2_1       \left[ \chi^{1, 3}_{2, 4}(\tau, \tau^+, \tau', \tau'^+) + \rho^1_2   \rho^3_4   \right] \left(  \Delta^{(1)}(\tau') \right)^4_3 .
 \label{action before EoM}
\end{align}
We now write the expression in Eq.\eqref{action before EoM} explicitly. Using the property 
\begin{align}
\chi^{1, 3}_{2, 4}(\tau, \tau^+, \tau', \tau'^+) =   \chi^{3, 1}_{4, 2}(\tau', \tau'^+, \tau, \tau^+) ,
\label{symmetry chi}
\end{align}
and defining 
\begin{align}
\hat{\mathcal{V}}_{i \alpha}(\tau) \equiv \mathrm{i} \, \mathrm{Tr}_{M}\!\left( S_{i \alpha} \cdot \Big[ \hat{\rho}(\tau) ; T \Big]^i_i \right) ,
\label{time dependent V}
\end{align}
\begin{align}
&   \mathcal{V}_{i \alpha}   \equiv   \left< \hat{\mathcal{V}}_{i \alpha}  \right>, \nonumber \\
&   \mathcal{M}^{i \alpha}_{i' \alpha'}  \equiv \left< \hat{\mathcal{M}}^{i \alpha}_{i' \alpha'}  \right> =  \mathcal{M}_{i \alpha}^{i' \alpha'}   ,
\label{average V and M}
\end{align}
we can rewrite the action of the electronic model (for small spin rotations) as:
\begin{align}
 \mathcal{S}\left[ \delta \boldsymbol{\varphi}_i(\tau) \right] \equiv &      \int_{0}^{\beta} \text{d} \tau  \Bigg\{ \mathrm{i}  \sum_{i}    \left[  \delta \dot{\boldsymbol{\varphi}}_{i}(\tau) + \frac{1}{2} \Big( \delta \boldsymbol{\varphi}_{i}(\tau) \times \delta \dot{\boldsymbol{\varphi}}_{i}(\tau) \Big)      \right] \cdot \left<   \hat{\boldsymbol{S}}_{i} \right>        \nonumber \\
 & \quad \quad +    \sum_{i}  \sum_{\alpha} \delta \varphi_{i \alpha}(\tau) \mathcal{V}_{i \alpha}   + \frac{1}{2}   \sum_{i, i'} \sum_{\alpha, \alpha'} \delta  \varphi_{i \alpha}(\tau) \, \delta \varphi_{i' \alpha'}(\tau)   \mathcal{M}^{i \alpha}_{i' \alpha'} \Bigg\}  \nonumber \\
     &  + \frac{1}{2}         \int_{0}^{\beta}   \text{d} \tau    \int_{0}^{\beta}  \text{d} \tau'   \sum_{i, i' } \sum_{\alpha, \alpha'} \Bigg\{   \nonumber \\
     & \quad \quad        \delta \dot{\varphi}_{i \alpha}(\tau) \,   \delta \dot{ \varphi}_{i' \alpha'}(\tau')  \, \left[    \left<   \mathcal{T}_{\gamma} \,   \hat{S}_{i \alpha}(\tau) \,       \hat{S}_{i' \alpha'}(\tau')     \right>      -    \left<  \hat{S}_{i \alpha}  \right>   \left<    \hat{S}_{i' \alpha'}      \right>       \right] \nonumber \\
    &   \quad \quad -   2 \mathrm{i} \, \delta \varphi_{i \alpha}(\tau) \, \delta \dot{\varphi}_{i' \alpha'}(\tau')   \,   \left[          \left< \mathcal{T}_{\gamma} \, \hat{\mathcal{V}}_{i \alpha}(\tau) \, \hat{S}_{i' \alpha'}( \tau')      \right>   -  \mathcal{V}_{i \alpha} \left<  \hat{S}_{i' \alpha'}     \right>   \right]  \nonumber \\
        &  \quad \quad -               \delta  \varphi_{i \alpha}(\tau)  \, \delta  \varphi_{i' \alpha'}(\tau')      \left[  \left< \mathcal{T}_{\gamma} \, \hat{\mathcal{V}}_{i \alpha}(\tau)   \, \hat{\mathcal{V}}_{i' \alpha'}(\tau')   \right>     - \mathcal{V}_{i \alpha} \mathcal{V}_{i' \alpha'}    \right] \Bigg\}      .   
\label{Hubbard EQ action with J's}
\end{align}
It must be stressed that, after the integration of the fermionic variables, the rotational fields $\delta \boldsymbol{\varphi}_i(\tau)$ are the \emph{only} dynamical variables left. All the features of the electronic dynamical processes are accounted for by the electronic Green's functions.

\section{Static potential for the spin rotations}
\label{sec: Static potential}

We now consider the action \eqref{Hubbard EQ action with J's} in the particular case of static rotations, $\delta  \varphi_{i \alpha}(\tau)  \rightarrow \delta  \varphi_{i \alpha}$:
\begin{align}
 \mathcal{S}\left[ \delta \boldsymbol{\varphi}_i \right]  
        \equiv &        \beta  \left( \sum_{i}  \sum_{\alpha} \delta \varphi_{i \alpha}  \mathcal{V}_{i \alpha}   +  \frac{1}{2}    \sum_{i, i'} \sum_{\alpha, \alpha'} \delta  \varphi_{i \alpha}  \, \delta \varphi_{i' \alpha'} \widetilde{\mathcal{M}}^{i \alpha}_{i' \alpha'}       \right)
\label{Hubbard static action}
\end{align}
where
\begin{align}
 \widetilde{\mathcal{M}}^{i \alpha}_{i' \alpha'} \equiv  \mathcal{M}^{i \alpha}_{i' \alpha'}     - \frac{1}{\beta}\int_{0}^{\beta}   \text{d} \tau    \int_{0}^{\beta}  \text{d} \tau' \left< \mathcal{T}_{\gamma} \, \hat{\mathcal{V}}_{i \alpha}(\tau)   \, \hat{\mathcal{V}}_{i' \alpha'}(\tau')   \right>     + \beta  \mathcal{V}_{i \alpha} \mathcal{V}_{i' \alpha'}    . 
 \label{tilde M}
\end{align}
In this case, we can define the effective broken-symmetry partition function in the presence of a spin rotation as
\begin{align}
Z^* \left[ \delta \boldsymbol{\varphi}_i  \right] \equiv Z^* \mathcal{Z}\left[ \delta \boldsymbol{\varphi}_i  \right] \equiv \mathrm{e}^{- \beta \left( \Omega_0^* +  \Omega\left[ \delta \boldsymbol{\varphi}_i \right] \right) } ,
\end{align}
where $\Omega_0^* = - \beta^{-1} \mathrm{ln}(Z^*)$, and $\Omega\left[ \delta \boldsymbol{\varphi}_i \right] \equiv \beta^{-1} \mathcal{S}\left[ \delta \boldsymbol{\varphi}_i \right]$ is the effective thermodynamic potential associated with spin rotations from the broken-symmetry non-equilibrium configuration $\left| \lbrace \boldsymbol{r}_i \rbrace \right>$.

\section{Mapping to a classical spin model}
\label{sec: Mapping}

\subsection{Classical spin model}
\label{sec: Classical}

The Hamiltonian of a classical quadratic spin model is given by 
\begin{align}
H\left[   \boldsymbol{e}_i \right] \equiv  \sum_{i}      \boldsymbol{e}_{i}  \cdot  \boldsymbol{\mathcal{B}}_{i} + \frac{1}{2} \sum_{i, i'}   \sum_{\alpha, \alpha'}     e_{i, \alpha}       e_{i', \alpha'}  \mathcal{H}_{i i' }^{\alpha \alpha'}  ,
\label{general quadratic model}
\end{align}
where the $\boldsymbol{e}_i$'s are unit vectors representing the directions of the classical magnetic moments, and $\alpha, \alpha' \in \lbrace x, y, z \rbrace$. The vector $\boldsymbol{\mathcal{B}}_{i}$ is a local magnetic field, while $\mathcal{H}_{i i' }^{\alpha \alpha'}$ is the exchange tensor, which can be chosen, without loss of generality, to satisfy the symmetry property
\begin{align}
\mathcal{H}_{i j}^{\alpha \beta} = \mathcal{H}_{j i}^{\beta \alpha} .
\label{symmetry second order tensor}
\end{align}
Furthermore, the tensor can be decomposed into three vectors $\boldsymbol{\mathcal{J}}_{i j}$, $\boldsymbol{\mathcal{D}}_{i j}$, and $\boldsymbol{\mathcal{C}}_{i j}$ as follows:
\begin{align}
\mathcal{H}_{i j}^{\alpha \beta} \equiv  \delta^{\alpha \beta} \mathcal{J}_{i j}^{\alpha} + \sum_{\gamma} \left( \varepsilon^{\alpha \beta \gamma} \mathcal{D}_{i j}^{\gamma} + \left| \varepsilon^{\alpha \beta \gamma} \right| \mathcal{C}_{i j}^{\gamma} \right)  ,
\end{align}
where the parameters
\begin{align}
\mathcal{J}_{i j}^{\alpha} \equiv \mathcal{H}_{i j}^{\alpha \alpha} , \quad  \mathcal{D}^{\gamma}_{i j} \equiv \frac{1}{2} \sum_{\alpha , \beta} \varepsilon^{\alpha \beta \gamma} \mathcal{H}_{i j}^{\alpha \beta}  , \quad \text{and} \quad \mathcal{C}^{\gamma}_{i j} \equiv \frac{1}{2} \sum_{\alpha , \beta} \left| \varepsilon^{\alpha \beta \gamma} \right| \mathcal{H}_{i j}^{\alpha \beta}  
\end{align}
account, respectively, for anisotropic exchange, Dzyaloshinskii-Moriya interaction, and an additional traceless symmetric interaction (which includes, e.g., dipole-dipole\footnote{For example, the second part of a dipole-dipole interaction term \cite{White, Akhiezer} of the form  \begin{align*} \frac{1}{R_{i j}^3} \left[  \boldsymbol{e}_i \cdot  \boldsymbol{e}_j - 3 \left( \frac{ \boldsymbol{e}_i \cdot \boldsymbol{R}_{ij}}{R_{ij}} \right) \left(   \frac{ \boldsymbol{e}_j \cdot \boldsymbol{R}_{ij}}{R_{ij}}   \right)    \right], \end{align*} where $\boldsymbol{R}_{ij}$ is the vector connecting the positions of magnetic moments $i$ and $j$, may be included in $\boldsymbol{\mathcal{C}}_{i j}$. }). One can easily verify that $\boldsymbol{\mathcal{J}}_{i j} = \boldsymbol{\mathcal{J}}_{j i}$, $\boldsymbol{\mathcal{D}}_{i j} = - \boldsymbol{\mathcal{D}}_{j i}$, and $\boldsymbol{\mathcal{C}}_{i j} = \boldsymbol{\mathcal{C}}_{j i}$, as a consequence of the symmetry given by Eq.\eqref{symmetry second order tensor}. We can write the total magnetic Hamiltonian as
\begin{align}
H\left[   \boldsymbol{e}_i \right] = & \sum_i   \boldsymbol{e}_{i} \cdot \boldsymbol{\mathcal{B}}_i  + \frac{1}{2} \sum_{i, i'} \left[ \boldsymbol{e}_{i} \cdot  \boldsymbol{\overline{ \mathcal{J}}}_{i i'} \cdot \boldsymbol{e}_{i'}   +     \boldsymbol{\mathcal{D}}_{i i'} \cdot \left( \boldsymbol{e}_{i} \times \boldsymbol{e}_{i'} \right)     +   \boldsymbol{\mathcal{C}}_{i i'} \cdot \left( \boldsymbol{e}_{i} \otimes \boldsymbol{e}_{i'} \right) \right] ,
\label{most general eff H}
\end{align}
where $\boldsymbol{\overline{\mathcal{J}}}_{i i'}$ is the diagonal matrix with elements $\mathcal{J}_{i i'}^{\alpha}$ on the diagonal, and we have put
\begin{align}
\boldsymbol{e}_{i} \otimes \boldsymbol{e}_{i'} \equiv \sum_{\alpha \beta \gamma} \left| \varepsilon^{\alpha \beta \gamma} \right| e_{i \alpha} e_{i' \beta} \boldsymbol{u}_{\gamma} .
\end{align}
It should be noted that the exchange tensor has also single-spin terms corresponding to $i = i'$. If the index $i$ labelling the magnetic moments can be identified with a space coordinate, such as an atomic index, then these \emph{local} terms should be identified with the \emph{local anisotropy tensor}, having 6 independent components: $\mathcal{J}_{i i}^{\alpha}$ and $\mathcal{C}_{i i}^{\alpha}$, for $\alpha \in \lbrace x, y, z \rbrace$. We note that the energy contribution from the diagonal part of this tensor can be written as
\begin{align}
\frac{1}{2} \sum_i \sum_{\alpha = x, y, z} \left( e_{i \alpha} \right)^2 \mathcal{J}_{i i}^{\alpha} = \frac{1}{2} \sum_i   \left[ e^2_{i x}  \left(  \mathcal{J}_{i i}^{x} - \mathcal{J}_{i i}^{z}  \right)    
+ e^2_{i y}  \left(  \mathcal{J}_{i i}^{y} - \mathcal{J}_{i i}^{z}  \right)  + \mathcal{J}_{i i}^z \right] ,  
\label{diagonal anisotropy}
\end{align}   
where we have used the constraint $\boldsymbol{e}_i^2 = 1$. The last term of Eq.\eqref{diagonal anisotropy} is obviously rotationally invariant, so that the response of the system to spin rotations will not depend individually on the three parameters $ \mathcal{J}_{i i}^{\alpha}$, but only on their relative differences as expressed, for example, in the combinations $ \left(  \mathcal{J}_{i i}^{x} - \mathcal{J}_{i i}^{z}  \right)$ and $ \left(  \mathcal{J}_{i i}^{y} - \mathcal{J}_{i i}^{z}  \right)$.

\subsection{Static potential for the spin rotations}

The effective potential for the classical spin model, expressing the energy change when the unit vectors are rotated from a given configuration $\lbrace \boldsymbol{r}_i \rbrace$, is obtained from Eq.\eqref{most general eff H} by replacing
\begin{align}
\boldsymbol{e}_i    & \approx    \boldsymbol{r}_i + \boldsymbol{r}_i \times \delta \boldsymbol{\varphi}_i - \frac{1}{2} \boldsymbol{r}_i \left( \delta \boldsymbol{\varphi}_i \right)^2      ,
\label{unit vector}
\end{align}
which follows from Eq.\eqref{q axes rotated} and the requirement that $\delta \boldsymbol{\varphi}_i \cdot \boldsymbol{r}_i = 0$. In order to map the effective potential for spin rotations relative to the classical spin model onto the potential derived for the electronic model, we need in fact to require that the $\delta \boldsymbol{\varphi}_i$ fields have the same meaning in the two cases. Therefore, the vectors $\lbrace \boldsymbol{r}_i \rbrace$ which specify the classical spin configuration of reference must be the same as the vectors specifying the spin configuration of the electronic state of reference $\left| \lbrace \boldsymbol{r}_i \rbrace \right>$, introduced in Section \ref{subsec: partition}. 

The effective potential for the classical spin model (to second order in the rotation angles) is:
\begin{align}
\Omega_{\mathrm{cl}}\left[ \delta \boldsymbol{\varphi}_i \right] = \Omega^{(1)}_{\mathrm{cl}}\left[ \delta \boldsymbol{\varphi}_i \right] + \Omega^{(2)}_{\mathrm{cl}}\left[ \delta \boldsymbol{\varphi}_i \right] ,
\end{align}
where
\begin{align}
\Omega^{(1)}_{\mathrm{cl}}\left[ \delta \boldsymbol{\varphi}_i \right] = \sum_i \left\{ \delta \varphi_{i x} \! \left[ \mathcal{B}_i^y + \sum_{i'} \left( \mathcal{C}_{i i'}^{x}  + \mathcal{D}_{i i'}^{x}  \right)     \right] -  \delta \varphi_{i y} \! \left[  \mathcal{B}_i^x + \sum_{i'} \left( \mathcal{C}_{i i'}^{y} - \mathcal{D}_{i i'}^{y} \right) \right] \!  \right\} ,
\label{Omega1}
\end{align}
\begin{align}
& \Omega^{(2)}_{\mathrm{cl}}\left[ \delta \boldsymbol{\varphi}_i \right] =     \frac{1}{2} \sum_{i, i' \neq i} \left( \begin{matrix} \delta \varphi_{i x} & \delta \varphi_{i y}\end{matrix}\right) \left(   \begin{matrix}    \mathcal{J}_{i i'}^y & - \mathcal{C}_{i i'}^z + \mathcal{D}_{i i'}^z \\ - \mathcal{C}_{i i'}^z - \mathcal{D}_{i i'}^z  &    \mathcal{J}_{i i'}^x \end{matrix}  \right)   \left( \begin{matrix}   \delta \varphi_{i' x} \\ \delta \varphi_{i' y} \end{matrix} \right)  \nonumber \\
& + \frac{1}{2} \sum_{i  } \left( \begin{matrix} \delta \varphi_{i x} & \delta \varphi_{i y}\end{matrix}\right) \left(   \begin{matrix}   -    \mathcal{B}_i^z - \sum_{j} \mathcal{J}_{i j}^z   + \mathcal{J}_{i i}^y & - \mathcal{C}_{i i}^z  \\ - \mathcal{C}_{i i}^z     &  -    \mathcal{B}_i^z - \sum_{j} \mathcal{J}_{i j}^z   + \mathcal{J}_{i i}^x \end{matrix}  \right)   \left( \begin{matrix}   \delta \varphi_{i x} \\ \delta \varphi_{i y} \end{matrix} \right)  .
\label{Omega2}
\end{align}
The matrices appearing in Eq.\eqref{Omega2} are maximally symmetrized.

\subsection{Equations for the effective magnetic parameters}

We determine the effective magnetic parameters of the classical spin model by putting $\Omega\left[ \delta \boldsymbol{\varphi}_i \right] \equiv \Omega_{\mathrm{cl}}\left[ \delta \boldsymbol{\varphi}_i \right]$ and identifying the terms which depend on the same orders and same combinations of the $\delta \varphi_{i \alpha}$ fields. We obtain the following three sets of equations:
\begin{align}
&   \mathcal{B}_i^y + \sum_{i'} \left( \mathcal{C}_{i i'}^{x}  + \mathcal{D}_{i i'}^{x}  \right)  = \mathcal{V}_{i x} , \nonumber \\
&   - \mathcal{B}_i^x - \sum_{i'} \left( \mathcal{C}_{i i'}^{y} - \mathcal{D}_{i i'}^{y} \right)  = \mathcal{V}_{i y} , 
\label{first set}
\end{align}
\begin{align}
& -    \mathcal{B}_i^z - \sum_{j} \mathcal{J}_{i j}^z   + \mathcal{J}_{i i}^y   = \widetilde{\mathcal{M}}^{i x}_{i x},   \nonumber \\
& -    \mathcal{B}_i^z - \sum_{j} \mathcal{J}_{i j}^z   + \mathcal{J}_{i i}^x   = \widetilde{\mathcal{M}}^{i y}_{i y},   \nonumber \\
&   \mathcal{C}_{i i}^z  = - \widetilde{\mathcal{M}}^{i x}_{i y} ,
\label{third set}
\end{align}
\begin{align}
&  \mathcal{J}_{i i'}^y = \widetilde{\mathcal{M}}^{i x}_{i' x}, \quad i \neq i', \nonumber \\
&  \mathcal{D}_{i i'}^z =  \frac{1}{2} \left(  \widetilde{\mathcal{M}}^{i x}_{i' y}  -  \widetilde{\mathcal{M}}^{i y}_{i' x}   \right)  , \quad i \neq i', \nonumber \\ 
&  \mathcal{C}_{i i'}^z =  -  \frac{1}{2} \left(   \widetilde{\mathcal{M}}^{i x}_{i' y}  +  \widetilde{\mathcal{M}}^{i y}_{i' x} \right) , \quad i \neq i', \nonumber \\  
&  \mathcal{J}_{i i'}^x = \widetilde{\mathcal{M}}^{i y}_{i' y}, \quad i \neq i'  .  
\label{second set}
\end{align}
The first set, Eqs.\eqref{first set}, is obtained from the identity of the terms of the effective potentials which are of the first order in the rotation angles. The second set, \eqref{third set}, is obtained from second-order rotations of a single spin, while the third set, \eqref{second set}, is obtained from second-order rotations involving different spins $i, i'$. While the equations of the third set are already explicitly solved, we need to solve the first set, Eqs.\eqref{first set}, and two equations of the second set, Eqs.\eqref{third set}. As it can be seen, the equations arising from single-spin rotations do not have a unique solution. To obtain the identity of the thermodynamic potentials up to the second order in the rotations, however, we just need \emph{one} solution, and in the following we will determine it following general principles of physical reasonability and respecting the symmetries of the parameters. For example, the three components of the effective magnetic field $\boldsymbol{\mathcal{B}}_i$ must be proportional to the respective components of the field $\boldsymbol{B}_i$ entering the electronic Hamiltonian, although one could in principle set $\boldsymbol{\mathcal{B}}_i = \boldsymbol{0}$ and absorb all the terms depending on $\boldsymbol{\mathcal{B}}_i$ into the exchange tensor. Obviously this last solution would make no sense. 

We also wish to note that an extension of this analysis to the identity of the third-order terms in the rotation angles (leaving the spin model unchanged) may provide additional constraints on the quantities which are not completely determined from the sets \eqref{first set} and \eqref{third set}. This analysis, however, is beyond the scope of the present work.

\section{Solution in the general relativistic regime}
\label{sec: Solution}

In this Section we solve the equations needed to determine the remaining parameters of the effective spin model. We will report all the relevant details, so that the derivation can be followed in its entirety. For the sake of clarity, in Section \ref{sec: Summary solution} we will summarize and list all the results.

\subsection{First set - Equations \eqref{first set}}
\label{subsec: First}

From the definitions, Eqs.\eqref{vector V} and \eqref{average V and M}, we write the right-hand-sides of Eqs.\eqref{first set}, for $\alpha \in \lbrace x, y \rbrace$, as  
\begin{align}
\mathcal{V}_{i \alpha}  & =    \mathrm{i}     \,  \mathrm{Tr}_M \Big[ S_{i \alpha} \cdot \left( \rho \cdot T  -  T \cdot \rho  \right)^{i}_{i}  \Big]  =    \mathrm{i}     \,  \mathrm{Tr}_M \sum_j \Big[ S_{i \alpha} \cdot \left( \rho^i_j \cdot T^j_i - T^i_j \cdot \rho^j_i     \right)   \Big] \nonumber \\
& =    \mathrm{i}     \,  \mathrm{Tr}_M  \Big[ \rho^i_i \cdot \left( T^i_i \cdot S_{i \alpha} -  S_{i \alpha} \cdot T^i_i          \right)   \Big] +  \mathrm{i}     \,  \mathrm{Tr}_M \sum_{j \neq i} \Big[ S_{i \alpha} \cdot \left( \rho^i_j \cdot T^j_i -  T^i_j \cdot \rho^j_i      \right)   \Big] .
\label{eq 1 RHS}
\end{align}
Identifying Eqs.\eqref{eq 1 RHS} with the corresponding left-hand-sides given in Eqs.\eqref{first set}, we are naturally led to separate local $(j = i)$ and non-local $(j \neq i)$ terms. We obtain the following equations:
\begin{align}
&    \mathrm{i}     \,  \mathrm{Tr}_M  \Big[ \rho^i_i \cdot \left( T^i_i \cdot S_{i x} -  S_{i x} \cdot T^i_i      \right)   \Big]  =  \mathcal{B}_i^y + \mathcal{C}_{ii}^x  , \nonumber \\
&    \mathrm{i}     \,  \mathrm{Tr}_M  \Big[ \rho^i_i \cdot \left( T^i_i \cdot S_{i y} -  S_{i y} \cdot T^i_i      \right)   \Big]  =   - \mathcal{B}_i^x - \mathcal{C}_{ii}^y ,
\label{separated 1}
\end{align}
\begin{align}
&   \mathrm{i}     \,  \mathrm{Tr}_M \sum_{j \neq i} \Big[ S_{i x} \cdot \left( \rho^i_j \cdot T^j_i -  T^i_j \cdot \rho^j_i      \right)   \Big]    =  \sum_{j \neq i} \left( \mathcal{C}_{i j}^{x}  + \mathcal{D}_{i j}^{x}  \right)    , \nonumber \\
&   \mathrm{i}     \,  \mathrm{Tr}_M \sum_{j \neq i} \Big[ S_{i y} \cdot \left( \rho^i_j \cdot T^j_i -  T^i_j \cdot \rho^j_i      \right)   \Big]    =   - \sum_{j \neq i} \left( \mathcal{C}_{i j}^{y} - \mathcal{D}_{i j}^{y} \right)   .
\label{separated 2}
\end{align}
Except for the presence of the magnetic field $\boldsymbol{\mathcal{B}}_i$, Eqs.\eqref{first set} were also obtained in Ref.\cite{Katsnelson10}, which focused on the determination of the Dzyaloshinskii-Moriya interactions. Here we will also discuss the other terms of the exchange tensor that can be determined from this set, namely the parameters $\mathcal{C}_{i j}^x$ and $\mathcal{C}_{i j}^y$. Moreover, we note that the parameters $\mathcal{C}_{i j}^z$ and $\mathcal{D}_{i j}^z$ cannot be determined from the first-order term of the rotational potential, due to the fact that $\delta \varphi_{i z} = 0$, which was not discussed in Ref.\cite{Katsnelson10}. These terms are indeed obtained from the second-order terms of the potential, see Eqs.\eqref{second set} and the last among Eqs.\eqref{third set}.

We now consider Eqs.\eqref{separated 1}. Using Eq.\eqref{local T}, we compute the LHSs of Eqs.\eqref{separated 1}, from which we can separate the contributions due to the magnetic field and to the anisotropy as
\begin{align}
& \mathcal{B}_i^x  =  - \mu_{\mathrm{B}} g_i \left(   \boldsymbol{B}_i \times \left< \hat{\boldsymbol{S}}_i \right>  \right)   \cdot \boldsymbol{u}_i^y  , \nonumber \\
& \mathcal{B}_i^y  =  \mu_{\mathrm{B}} g_i \left(   \boldsymbol{B}_i \times \left< \hat{\boldsymbol{S}}_i \right>  \right)   \cdot \boldsymbol{u}_i^x  , \nonumber \\
& \mathcal{C}_{i i}^x =   \mathrm{i}     \,  \mathrm{Tr}_M  \Big[ \rho^i_i \cdot \left( A^i_i \cdot S_{i x} -  S_{i x} \cdot A^i_i      \right)   \Big] , \nonumber \\
& \mathcal{C}_{i i}^y = - \mathrm{i}     \,  \mathrm{Tr}_M  \Big[ \rho^i_i \cdot \left( A^i_i \cdot S_{i y} -  S_{i y} \cdot A^i_i      \right)   \Big]  .
\label{solution first set local}
\end{align}
Since the local quantization axes are chosen to be parallel to the local magnetic moments, i.e., $\left< \hat{\boldsymbol{S}}_i \right> \equiv S_i \boldsymbol{u}_i^z$, then it follows that $\mathcal{B}_i^x = \mu_{\mathrm{B}} g_i S_i B_i^x$ and $\mathcal{B}_i^y = \mu_{\mathrm{B}} g_i S_i B_i^y$.

To solve Eqs.\eqref{separated 2}, analogously to Ref.\cite{Katsnelson10} we use the symmetries of the magnetic parameters under exchange of the indexes $i \leftrightarrow j$. In general, one can always write
\begin{align}
\sum_{j \neq i} f_{i j} = \frac{1}{2} \sum_{j \neq i} \left[ \left( f_{i j} + f_{j i} \right) + \left( f_{i j} - f_{j i} \right) \right] ,
\label{S AS separation}
\end{align} 
where we have distinguished the functions $\left( f_{i j} + f_{j i} \right) / 2$ and $\left( f_{i j} - f_{j i} \right) / 2$, which are respectively symmetric and antisymmetric under the permutation of $i$ and $j$. After applying \eqref{S AS separation} to the LHSs of Eqs.\eqref{separated 2}, one could think of identifying the symmetric and antisymmetric functions of $(i, j)$, respectively, with the corresponding components of the vectors $\boldsymbol{\mathcal{C}}_{i j}$ and $\boldsymbol{\mathcal{D}}_{i j}$ appearing on the RHSs. We want to note that this procedure is not unique. Indeed, if we have an identity of the form \begin{align*} \sum_{j \neq i} a_{i j} = \sum_{j \neq i} b_{i j}, \end{align*} where $a_{i j}$ and $b_{i j}$ have the same symmetry under $i \leftrightarrow j$, say $a_{i j} = \pm a_{j i}$ and $b_{i j} = \pm b_{j i}$, we can in general only assume that $a_{i j} = b_{i j} + x_{i j}$, where $x_{i j} = \pm x_{j i}$ and $\sum_{j} x_{i j} = 0$. The undetermined quantity $x_{i j}$ may be relevant if some unknowns appearing in $a_{i j}$ or $b_{i j}$ must satisfy constraints imposed also by other equations. In other words, in the most general case we cannot split the equations of the first set into pairs of equations for symmetric and antisymmetric components, since doing so would require introducing additional unknowns (the $x_{i j}$ quantities of the previous example) that cannot be determined uniquely. However, the quantities $\mathcal{C}_{i j}^x$, $\mathcal{C}_{i j}^y$, $\mathcal{D}_{i j}^x$ and $\mathcal{D}_{i j}^y$ do not appear in any of the other equations that must be satisfied for the mapping to hold. Therefore, we can just take any solution of Eqs.\eqref{separated 2} with the correct symmetry properties. We therefore obtain
\begin{align}
              \mathcal{D}_{i j}^{x}         = & \frac{ \mathrm{i} }{2 }       \mathrm{Tr}_M  \Big[ S_{i x} \cdot \left( \rho^i_j \cdot T^j_i - T^i_j \cdot \rho^j_i    \right)  - S_{j x} \cdot \left( \rho^j_i \cdot T^i_j -  T^j_i \cdot \rho^i_j     \right)  \! \Big]  ,    \nonumber \\
        \mathcal{D}_{i j}^{y}                = &  \frac{ \mathrm{i} }{2    }      \mathrm{Tr}_M  \Big[ S_{i y} \cdot \left( \rho^i_j \cdot T^j_i - T^i_j \cdot \rho^j_i     \right)  - S_{j y} \cdot \left( \rho^j_i \cdot T^i_j  -  T^j_i \cdot \rho^i_j   \right) \!  \Big]  ,    \nonumber \\
               \mathcal{C}_{i j}^{x}                 = & \frac{ \mathrm{i} }{2 }       \mathrm{Tr}_M  \Big[ S_{i x} \cdot \left( \rho^i_j \cdot T^j_i - T^i_j \cdot \rho^j_i    \right)  + S_{j x} \cdot \left( \rho^j_i \cdot T^i_j -  T^j_i \cdot \rho^i_j     \right)  \! \Big]    , \nonumber \\
       \mathcal{C}_{i j}^{y}                = & -  \frac{ \mathrm{i} }{2    }      \mathrm{Tr}_M  \Big[ S_{i y} \cdot \left( \rho^i_j \cdot T^j_i - T^i_j \cdot \rho^j_i     \right)  + S_{j y} \cdot \left( \rho^j_i \cdot T^i_j  -  T^j_i \cdot \rho^i_j   \right) \!  \Big]  .
\label{sr C and D parameters}
\end{align}
This completes the determination of the $\boldsymbol{\mathcal{C}}_{i j}$ and $\boldsymbol{\mathcal{D}}_{i j}$ vectors. The components of the Dzyaloshinskii-Moriya vector $ \mathcal{D}_{i j}^{x}$ and $ \mathcal{D}_{i j}^{y}$ are in agreement with Ref.\cite{Katsnelson10}, except that here the parameters are defined in such a way that they exhibit definite symmetry under the permutation of the indexes $i, j$. Moreover, here we have allowed $\boldsymbol{S}_i$ to be in general different from $\boldsymbol{S}_j$, so Eqs.\eqref{sr C and D parameters} are valid also in the \emph{non-collinear} regime. By comparing Eqs.\eqref{second set} and \eqref{sr C and D parameters}, we observe that the expressions for the $z$ components of the vectors $\boldsymbol{\mathcal{C}}$ and $\boldsymbol{\mathcal{D}}$ look formally different from those that give the $x$ and $y$ components of the same vectors. Recalling that the reference frames $\lbrace \boldsymbol{u}^x_i, \boldsymbol{u}^y_i, \boldsymbol{u}^z_i \rbrace$ are defined locally (they depend on $i$), this different status of the direction $\boldsymbol{u}^z_i$ with respect to the plane defined by the directions $\lbrace \boldsymbol{u}^x_i, \boldsymbol{u}^y_i \rbrace$ reflects the fact that $\boldsymbol{u}^z_i$ is the direction of the $i$-th magnetic moment, and our procedure involves rotations in spin space, whose definition is not insensitive the choice of $\boldsymbol{u}^z_i$. In other terms, the differences in the formulas are due to the fact that the definition of the Green's functions is influenced by the choice of the quantization axis, so the expressions based on Green's functions exhibit ``special'' directions, which coincide with the vectors $\boldsymbol{u}^z_i \equiv \boldsymbol{r}_i$.

\subsection{Second set - Equations \eqref{third set}}
\label{subsec: Second}

We now solve Eqs.\eqref{third set}. The first step is to identify $\mathcal{B}_i^z$ with a corresponding term proportional to $B^z_i$ that is included in the RHSs of both the first and the second among Eqs.\eqref{third set}. Thus, we have to find such a term in the expression for $\widetilde{\mathcal{M}}^{i \alpha}_{i \alpha}$. For brevity, from Eq.\eqref{tilde M} we put
\begin{align}
 \widetilde{\mathcal{M}}^{i \alpha}_{i' \alpha'} \equiv  \mathcal{M}^{i \alpha}_{i' \alpha'}     - \mathcal{W}^{i \alpha}_{i' \alpha'}     + \beta  \mathcal{V}_{i \alpha} \mathcal{V}_{i' \alpha'}   , 
 \label{tilde M brevity}
\end{align}
where we have defined the quantity
\begin{align}
\mathcal{W}^{i \alpha}_{i' \alpha'} \equiv \frac{1}{\beta} \int_0^{\beta} \mathrm{d} \tau \int_0^{\beta} \mathrm{d} \tau' \left< \mathcal{T}_{\gamma} \hat{\mathcal{V}}_{i \alpha}(\tau) \, \hat{\mathcal{V}}_{i' \alpha'}(\tau') \right> ,
\label{A tilde}
\end{align}
which can be written explicitly as
\begin{align}
& \mathcal{W}^{i \alpha}_{i' \alpha'}    \equiv \sum_{j , j'}  \sum_{M_1 M_2 M_3 M_4}  \!   \left[ \left( S_{i \alpha}  \cdot T^{i}_{j}   \right)^{M_2}_{M_1}   \left( S_{i' \alpha'}  \cdot T^{i'}_{j'}   \right)^{M_4}_{M_3} \widetilde{\chi}^{(j M_1) (j' M_3) }_{ (i M_2)  (i' M_4)}    \right.  \nonumber \\
        & \quad \quad  \quad \quad \quad \quad -           \left( S_{i \alpha}  \cdot T^{i}_{j}   \right)^{M_2}_{M_1}     \left( T^{j'}_{i'} \cdot   S_{i' \alpha'} \right)^{M_4}_{M_3}  \widetilde{\chi}^{(j M_1) (i' M_3)  }_{ (i M_2) (j' M_4)}       \nonumber \\
    &   \quad \quad \quad \quad  \quad \quad    -        \left( T^{j}_{i} \cdot   S_{i \alpha} \right)^{M_2}_{M_1}   \left( S_{i' \alpha'}  \cdot T^{i'}_{j'}   \right)^{M_4}_{M_3} \widetilde{\chi}^{(i M_1)  (j' M_3) }_{ (j M_2)  (i' M_4)}    \nonumber \\
    & \quad \quad \quad \quad \quad \quad +      \left. \left( T^{j}_{i} \cdot   S_{i \alpha} \right)^{M_2}_{M_1}      \left( T^{j'}_{i'} \cdot   S_{i' \alpha'} \right)^{M_4}_{M_3} \widetilde{\chi}^{(i M_1) (i' M_3)  }_{ (j M_2) (j' M_4)}     \right] ,
    \label{A tilde ii'}
\end{align}
where we have put
\begin{align}
\widetilde{\chi}^{1, 3 }_{ 2,  4} \equiv \frac{1}{\beta} \int_0^{\beta} \mathrm{d} \tau \int_0^{\beta} \mathrm{d} \tau' \chi^{1, 3 }_{ 2,  4}(\tau, \tau^+, \tau', \tau'^+) , 
\label{chi integrated} 
\end{align}
which has the property $\widetilde{\chi}^{1, 3 }_{ 2,  4} = \widetilde{\chi}^{3, 1}_{4, 2}$, as follows from Eq.\eqref{symmetry chi}. Consequently, 
\begin{align}
\mathcal{W}^{i \alpha}_{i' \alpha'} = \mathcal{W}_{i \alpha}^{i' \alpha'} .
\label{symmetry A tilde}
\end{align}
All the three terms appearing in the RHS of Eq.\eqref{tilde M brevity} depend explicitly on the magnetic field, both via the single- and two- electron Green's functions entering their definitions, and via the explicit dependence of the hopping parameters. While we will comment more in detail about this in \ref{app: Analysis}, for the purpose of solving Eqs.\eqref{third set} we just need to find the term that should be identified with $\mathcal{B}_i^z$. To this end, we note that
\begin{align}
\left[ S_{i \alpha}, T^i_i \right] =  \left[ S_{i \alpha}, A^i_i \right]  +  \mathrm{i} \mu_{\mathrm{B}} g_i \left( \boldsymbol{B}_i \times \boldsymbol{S}_i \right) \cdot \boldsymbol{u}_i^{\alpha},
\label{local commutator}
\end{align}
from which it follows that
\begin{align}
& \mathcal{V}_{i \alpha} = \mu_{\mathrm{B}} g_i \left( \boldsymbol{B}_i \times \left< \hat{\boldsymbol{S}}_i \right> \right)  \cdot \boldsymbol{u}^{\alpha}_i + \left. \mathcal{V}_{i \alpha} \right|_{\boldsymbol{B}_i = \boldsymbol{0}}  , \nonumber \\
& \mathcal{M}_{i \alpha}^{i \alpha} = - \mu_{\mathrm{B}} g_i \left(  \boldsymbol{B}_i \cdot \left< \hat{\boldsymbol{S}}_i \right> - B_{i}^{\alpha} \left< \hat{S}_{i \alpha} \right> \right)   +   \left. \mathcal{M}_{i \alpha}^{i \alpha} \right|_{\boldsymbol{B}_i = \boldsymbol{0}}  ,
\end{align}  
where with the notation $ \left. x   \right|_{\boldsymbol{B}_i = \boldsymbol{0}} $, here and in the following, we \emph{do not} mean that $x$ does not depend explicitly on $\boldsymbol{B}_i$, we mean instead that $x$ does not vanish when $\boldsymbol{B}_i = \boldsymbol{0}$. In the presence of $\boldsymbol{B}_i \neq \boldsymbol{0}$, also $\left. x   \right|_{\boldsymbol{B}_i = \boldsymbol{0}}$ will depend on $\boldsymbol{B}_i$ via the Green's functions, as well as via the Peierls phases due to the electromagnetic field, which have to be included into the hopping parameters.

Using the fact that $\left< \hat{\boldsymbol{S}}_i \right> = S_i \boldsymbol{u}_i^z$, a term in the expression of $\mathcal{M}_{i \alpha}^{i \alpha}$ becomes 
\begin{align}
   \boldsymbol{B}_i \cdot \left< \hat{\boldsymbol{S}}_i \right> - B_{i}^{\alpha} \left< \hat{S}_{i \alpha} \right> \rightarrow B_i^z S_i ,
\end{align}
and the identification is obvious,
\begin{align}
\mathcal{B}_i^z = \mu_{\mathrm{B}} g_i S_i B_i^z,
\end{align}
analogously to the other components of the magnetic field. It should be noted that the whole procedure relies on the existence of non-zero magnetic moments $S_i$, which is therefore a requirement for the mapping of the electronic model to the classical spin model via the equivalence of the thermodynamic potentials for spin rotations.

The final task is to determine $\boldsymbol{\mathcal{J}}_{i i}$ and $\mathcal{J}_{i j}^z$. Let us define
\begin{align}
\widetilde{\mathcal{M}}^{i \alpha}_{i \alpha} \equiv \widetilde{\mathcal{N}}^{i \alpha}_{i \alpha} - \mu_{\mathrm{B}} g_i \left( \boldsymbol{B}_i \cdot \left< \hat{\boldsymbol{S}}_i\right>  -  B_i^{\alpha} \left< \hat{S}_{i \alpha} \right>\right) ,
\label{definition N}
\end{align}
so that the first and the second among Eqs.\eqref{third set} become:
\begin{align}
&   - \sum_{j \neq i} \mathcal{J}_{i j}^z   + \mathcal{J}_{i i}^y -  \mathcal{J}_{i i}^z = \widetilde{\mathcal{N}}^{i x}_{i x},   \nonumber \\
&   - \sum_{j \neq i} \mathcal{J}_{i j}^z   + \mathcal{J}_{i i}^x -  \mathcal{J}_{i i}^z = \widetilde{\mathcal{N}}^{i y}_{i y} ,
\end{align}
or, more compactly,
\begin{align}
&   - \sum_{j \neq i} \mathcal{J}_{i j}^z   + \mathcal{J}_{i i}^{\bar{\alpha}} -  \mathcal{J}_{i i}^z = \widetilde{\mathcal{N}}^{i \alpha}_{i \alpha}, 
\label{third set remaining}
\end{align}
where $\alpha, \bar{\alpha} \in \lbrace x, y \rbrace$ and $\alpha  \neq \bar{\alpha}$. In the general relativistic case, $\widetilde{\mathcal{N}}^{i x}_{i x} \neq \widetilde{\mathcal{N}}^{i y}_{i y}$, and there seems to be no uniquely defined way of separating the various unknown terms. However, we note that one should definitely have $\widetilde{\mathcal{N}}^{i x}_{i x} = \widetilde{\mathcal{N}}^{i y}_{i y}$  in the non-relativistic regime (more on this in the following Section \ref{sec: Nonrel}), and consistently $\mathcal{J}_{i i}^x = \mathcal{J}_{i i}^y =  \mathcal{J}_{i i}^z \equiv \mathcal{J}_{i i}$. In this way, in fact, the corresponding Hamiltonian term becomes $\sum_i \mathcal{J}_{i i}  \left| \boldsymbol{e}_i \right|^2 = \sum_i \mathcal{J}_{i i}$, which is just a constant term, expressing the fact that there is no on-site anisotropy. Therefore, in the relativistic regime the quantity $- \sum_{j \neq i} \mathcal{J}_{i j}^z$ must be obtained from Eqs.\eqref{third set remaining}, it must be independent of $\alpha$, and it must reduce to $\left( \widetilde{\mathcal{N}}^{i \alpha}_{i \alpha} \right)_{\mathrm{nrel}}$ in the non-relativistic regime. Additionally, the quantities $\mathcal{J}_{i j}^z$ must be symmetric under $i \leftrightarrow j$. We can keep into account all these requirements by putting
\begin{align}
   \mathcal{J}_{i j}^z \stackrel{!}{=} \frac{1}{2} \left( \widetilde{\mathcal{M}}^{i x}_{j x} + \widetilde{\mathcal{M}}^{i y}_{j y}   \right) ,  \quad i \neq j
\label{ansatz}
\end{align}
from which it follows that
\begin{align}
&  \mathcal{J}_{i i}^{y} -  \mathcal{J}_{i i}^z = \widetilde{\mathcal{N}}^{i x}_{i x} + \frac{1}{2} \sum_{j \neq i} \left(   \widetilde{\mathcal{M}}^{i x}_{j x} + \widetilde{\mathcal{M}}^{i y}_{j y}  \right) , \nonumber \\
&  \mathcal{J}_{i i}^{x} -  \mathcal{J}_{i i}^z =  \widetilde{\mathcal{N}}^{i y}_{i y} + \frac{1}{2} \sum_{j \neq i} \left(   \widetilde{\mathcal{M}}^{i x}_{j x} + \widetilde{\mathcal{M}}^{i y}_{j y}  \right)  .
\label{rot inv anisotropy}
\end{align}
The equations only allow to determine two of the parameters $\lbrace \mathcal{J}_{i i}^{x} , \mathcal{J}_{i i}^{y}, \mathcal{J}_{i i}^{z} \rbrace $ as functions of the third one, which is of course understandable since, as discussed in Section \ref{sec: Classical}, the diagonal part of the local anisotropy tensor provides energy contributions under spin rotations only depending on the relative differences between the three elements, becoming rotationally invariant when these differences disappear. Hence, the rotationally invariant part is undetermined. 

We note that in the non-relativistic regime, as we will discuss in Section \ref{sec: Nonrel}, we have $\widetilde{\mathcal{M}}^{i x}_{j x} \stackrel{\mathrm{nrel}}{=} \widetilde{\mathcal{M}}^{i y}_{j y}$, and the sum rule $\sum_j \widetilde{\mathcal{M}}^{i \alpha}_{j \alpha} \stackrel{\mathrm{nrel}}{=} 0$. Noting that $\widetilde{\mathcal{N}}^{i \alpha}_{j \alpha} \stackrel{\mathrm{nrel}}{=} \widetilde{\mathcal{M}}^{i \alpha}_{j \alpha}$, we conclude that, in the non-relativistic regime, our solutions \eqref{ansatz} and \eqref{rot inv anisotropy} correctly give $\mathcal{J}_{i j}^z  \stackrel{\mathrm{nrel}}{=}  \widetilde{\mathcal{M}}^{i \alpha}_{j \alpha}  \stackrel{\mathrm{nrel}}{=}  \mathcal{J}_{i j}^x   \stackrel{\mathrm{nrel}}{=} \mathcal{J}_{i j}^y$, as well as $ \mathcal{J}_{i i}^{x} \stackrel{\mathrm{nrel}}{=}   \mathcal{J}_{i i}^{y} \stackrel{\mathrm{nrel}}{=}  \mathcal{J}_{i i}^{z} $. In the relativistic case, instead, Eqs.\eqref{rot inv anisotropy} account for the non-equivalence of the spatial directions.

\subsection{Summary of the formulas for the effective interactions}
\label{sec: Summary solution}

For the convenience of the reader, we here summarize the resulting formulas, which completely establish the mapping from the multi-orbital Hubbard model with rotationally invariant interaction (see the related discussion in \ref{app: Invariance}) to the general quadratic Hamiltonian of \emph{classical} spins given by Eq.\eqref{most general eff H}, under the requirement that the thermodynamic potentials for spin rotations of the two models are the same up to second order in the rotation angles. The results are:
\begin{align}
\boldsymbol{\mathcal{B}}_i = \mu_{\mathrm{B}} g_i S_i \boldsymbol{B}_i \,  ;
\label{B}
\end{align}
\begin{align}
      &        \mathcal{D}_{i j}^{x}         =   \frac{ \mathrm{i} }{2 }       \mathrm{Tr}_M  \Big[ S_{i x} \cdot \left( \rho^i_j \cdot T^j_i - T^i_j \cdot \rho^j_i    \right)  - S_{j x} \cdot \left( \rho^j_i \cdot T^i_j -  T^j_i \cdot \rho^i_j     \right)  \! \Big]    ,  \nonumber \\
     &   \mathcal{D}_{i j}^{y}                =    \frac{ \mathrm{i} }{2    }      \mathrm{Tr}_M  \Big[ S_{i y} \cdot \left( \rho^i_j \cdot T^j_i - T^i_j \cdot \rho^j_i     \right)  - S_{j y} \cdot \left( \rho^j_i \cdot T^i_j  -  T^j_i \cdot \rho^i_j   \right) \!  \Big] ,     \nonumber \\
&  \mathcal{D}_{i j}^z =  \frac{1}{2}  \left(  \widetilde{\mathcal{M}}^{i x}_{j y}  -  \widetilde{\mathcal{M}}^{i y}_{j x} \right) \,  ;  
\label{D}
\end{align}
for $i \neq j$,
\begin{align}
    &           \mathcal{C}_{i j}^{x}                 =  \frac{ \mathrm{i} }{2 }       \mathrm{Tr}_M  \Big[ S_{i x} \cdot \left( \rho^i_j \cdot T^j_i - T^i_j \cdot \rho^j_i    \right)  + S_{j x} \cdot \left( \rho^j_i \cdot T^i_j -  T^j_i \cdot \rho^i_j     \right)  \! \Big]    , \nonumber \\
 &      \mathcal{C}_{i j}^{y}                =  -  \frac{ \mathrm{i} }{2    }      \mathrm{Tr}_M  \Big[ S_{i y} \cdot \left( \rho^i_j \cdot T^j_i - T^i_j \cdot \rho^j_i     \right)  + S_{j y} \cdot \left( \rho^j_i \cdot T^i_j  -  T^j_i \cdot \rho^i_j   \right) \!  \Big]  , \nonumber \\
       &  \mathcal{C}_{i j}^z =  - \frac{1}{2} \left(   \widetilde{\mathcal{M}}^{i x}_{j y}  +  \widetilde{\mathcal{M}}^{i y}_{j x} \right)  \, ;  
\label{C nonlocal}
\end{align}
for $i = j$,
\begin{align}
& \mathcal{C}_{i i}^x =   \mathrm{i}     \,  \mathrm{Tr}_M  \Big[ \rho^i_i \cdot \left( A^i_i \cdot S_{i x} -  S_{i x} \cdot A^i_i      \right)   \Big] , \nonumber \\
& \mathcal{C}_{i i}^y = - \mathrm{i}     \,  \mathrm{Tr}_M  \Big[ \rho^i_i \cdot \left( A^i_i \cdot S_{i y} -  S_{i y} \cdot A^i_i      \right)   \Big]  , \nonumber \\
& \mathcal{C}_{i i}^z  = - \widetilde{\mathcal{M}}^{i x}_{i y}  \, ;
\label{C local}
\end{align}
for $i \neq j$,
\begin{align}
&  \mathcal{J}_{i j}^x = \widetilde{\mathcal{M}}^{i y}_{j y},  \nonumber \\ 
&  \mathcal{J}_{i j}^y = \widetilde{\mathcal{M}}^{i x}_{j x},  \nonumber \\  
&  \mathcal{J}_{i j}^z = \frac{1}{2} \left( \widetilde{\mathcal{M}}^{i x}_{j x} + \widetilde{\mathcal{M}}^{i y}_{j y}    \right)  \, ; 
\label{J nonlocal}
\end{align}
for $i = j$,
\begin{align}
&  \mathcal{J}_{i i}^{y} -  \mathcal{J}_{i i}^z = \widetilde{\mathcal{N}}^{i x}_{i x} + \frac{1}{2} \sum_{j \neq i} \left(   \widetilde{\mathcal{M}}^{i x}_{j x} + \widetilde{\mathcal{M}}^{i y}_{j y}  \right) , \nonumber \\
&  \mathcal{J}_{i i}^{x} -  \mathcal{J}_{i i}^z =  \widetilde{\mathcal{N}}^{i y}_{i y} + \frac{1}{2} \sum_{j \neq i} \left(   \widetilde{\mathcal{M}}^{i x}_{j x} + \widetilde{\mathcal{M}}^{i y}_{j y}  \right)  .
\label{J local}
\end{align}
We recall that the quantities $\widetilde{\mathcal{M}}^{i \alpha}_{i' \alpha'}$ are defined in Eq.\eqref{tilde M}, in terms of Eqs.\eqref{vector V}, \eqref{matrix M}, \eqref{time dependent V} and \eqref{average V and M}. The quantities $\widetilde{\mathcal{N}}^{i \alpha}_{i \alpha}$ are defined in Eq.\eqref{definition N}. 

We also recall that, for $i \neq j$, the quantities $\mathcal{J}^x_{i j}$, $\mathcal{J}^y_{i j}$, $\mathcal{C}^z_{i j}$ and $\mathcal{D}^z_{i j}$ are uniquely determined in our procedure, as well as $\mathcal{C}^z_{i i}$. The other magnetic parameters were determined, instead, as particular (physically reasonable) solutions of a set of equations whose number is far smaller than the number of parameters. It may be that requiring the equivalence of the thermodynamic potentials for spin rotations to higher orders in the rotation angles will lead to modifications of the formulas related to these latter parameters, however any modification must satisfy Eqs.\eqref{first set} and \eqref{second set}.

\section{Spin $1/2$ in the non-relativistic regime (single-orbital Hubbard model)}
\label{sec: Nonrel}

A particular case is obtained for the single-orbital Hubbard model, with $S = 1/2$ (no orbital exchange, or $l = 0$), in the non-relativistic regime and in the absence of external magnetic fields. We will label this particular case as ``soH'' in the following. In that case, the magnetic moment indexes $(i, i')$ coincide with the atomic indexes, the single-particle Hamiltonian is $T^{i M}_{i' M'} = \delta^M_{M'} T^i_{i'}$, as well as $T^i_{i'} = T_i^{i'} \equiv T_{i i'}$. Analogously, $\rho^{i M}_{i' M'} = \delta^M_{M'} \rho^M_{i i'} =  \delta^M_{M'} \rho^M_{i' i}$. The index $M$ assumes the values $M \in \lbrace + 1/2, - 1/2 \rbrace \equiv \lbrace \uparrow, \downarrow \rbrace$. Moreover, $\boldsymbol{S}_i = \boldsymbol{s} = \boldsymbol{\sigma} / 2$, where $\boldsymbol{\sigma}$ is the vector of Pauli matrices, and we have removed the subscript $i$ because we consider a collinear spin configuration. As a consequence,
\begin{align}
\Big[ S_{i \alpha}, T^i_i \Big] \stackrel{\mathrm{soH}}{=} 0 . 
\end{align}
In the soH case, it is easy to compute all the traces and sums over $M$. Using Eq.\eqref{tilde M}, and recalling that $\alpha, \alpha' \in \lbrace x, y \rbrace$, we obtain:
\begin{align}
     \mathcal{V}_{i \alpha} \stackrel{\mathrm{soH}}{=} 0 , 
\end{align} 
\begin{align}
   \widetilde{\mathcal{M}}^{i \alpha}_{i' \alpha'} & \stackrel{\mathrm{soH}}{=} \left( \mathcal{M}^{i \alpha}_{i' \alpha'} \right)^{\mathrm{nrel}}    - \left(  \widetilde{\mathbb{A}}^{i \alpha}_{i' \alpha'}  \right)_{\mathrm{nloc}}^{\mathrm{nrel}}    \nonumber \\
&   \stackrel{(1)}{=} \frac{1}{2} \delta_{\alpha \alpha'} T_{i i'}       \left(    \rho^{\uparrow}_{i i'} +    \rho^{\downarrow}_{i i'}      \right)   - \delta_{i i'} \delta_{\alpha \alpha'} \frac{1}{2} \sum_j  T_{i j} \left(          \rho_{i j}^{\uparrow}   +  \rho_{i j}^{\downarrow}    \right)  \nonumber \\
&  \quad    - \frac{1}{4} \sum_{j \neq i} \sum_{ j' \neq i'} T_{i j} T_{i' j'}  \sum_{M  M'}    \left( \sigma_{  \alpha}     \right)^{\bar{M}}_{M}   \left( \sigma_{  \alpha'}      \right)^{\bar{M}'}_{M'}  \Bigg[  \widetilde{\chi}^{(j M) (j' M') }_{ (i \bar{M})  (i' \bar{M}')}          \nonumber \\
    &   \quad \quad \quad -  \widetilde{\chi}^{(j M) (i' M')  }_{ (i \bar{M}) (j' \bar{M}')}       -  \widetilde{\chi}^{(i M)  (j' M') }_{ (j \bar{M}) (i' \bar{M}')}   +  \widetilde{\chi}^{(i M) (i' M')  }_{ (j \bar{M}) (j' \bar{M}')}     \Bigg] \nonumber \\
    &   \stackrel{(2)}{=} \frac{1}{2} \delta_{\alpha \alpha'} T_{i i'}       \left(    \rho^{\uparrow}_{i i'} +    \rho^{\downarrow}_{i i'}      \right)   - \delta_{i i'} \delta_{\alpha \alpha'} \frac{1}{2} \sum_j  T_{i j} \left(          \rho_{i j}^{\uparrow}   +  \rho_{i j}^{\downarrow}    \right)  \nonumber \\
&  \quad    - \frac{1}{4} \sum_{j \neq i} \sum_{ j' \neq i'} T_{i j} T_{i' j'}  \sum_{M}    \left( \sigma_{  \alpha}     \right)^{\bar{M}}_{M}   \left( \sigma_{  \alpha'}      \right)^{M}_{\bar{M}}  \Bigg[  \widetilde{\chi}^{(j M) (j' \bar{M}) }_{ (i \bar{M})  (i' M)}          \nonumber \\
    &   \quad \quad \quad -  \widetilde{\chi}^{(j M) (i' \bar{M})  }_{ (i \bar{M}) (j' M)}       -  \widetilde{\chi}^{(i M)  (j' \bar{M}) }_{ (j \bar{M}) (i' M)}   +  \widetilde{\chi}^{(i M) (i' \bar{M})  }_{ (j \bar{M}) (j' M)}     \Bigg] \nonumber \\
    &   \stackrel{(3)}{=} \frac{1}{2} \delta_{\alpha \alpha'} \Bigg\{ T_{i i'}       \left(    \rho^{\uparrow}_{i i'} +    \rho^{\downarrow}_{i i'}      \right)   - \delta_{i i'}   \sum_j  T_{i j} \left(          \rho_{i j}^{\uparrow}   +  \rho_{i j}^{\downarrow}    \right)  \nonumber \\
&  \quad \quad    - \frac{1}{2} \sum_{j \neq i} \sum_{ j' \neq i'} T_{i j} T_{i' j'}  \sum_{M = \uparrow, \downarrow}      \Bigg[  \widetilde{\chi}^{(j M) (j' \bar{M}) }_{ (i \bar{M})  (i' M)}          \nonumber \\
    &   \quad \quad \quad \quad -  \widetilde{\chi}^{(j M) (i' \bar{M})  }_{ (i \bar{M}) (j' M)}       -  \widetilde{\chi}^{(i M)  (j' \bar{M}) }_{ (j \bar{M}) (i' M)}   +  \widetilde{\chi}^{(i M) (i' \bar{M})  }_{ (j \bar{M}) (j' M)}     \Bigg] \Bigg\} ,
    \label{M nonrel}
\end{align}
where we have used the fact that in the non-relativistic regime the Hamiltonian cannot alter the total number of electrons with a given spin projection $\uparrow$ or $\downarrow$, therefore the only non-vanishing terms of $\widetilde{\chi}^{(i_1 M_1) (i_3 M_3)}_{(i_2 M_2) (i_4 M_4)}$ are those with $M_3 = M_4$ \emph{and} $M_1 = M_2$, or those with $M_1 = M_4$ \emph{and} $M_2 = M_3$. This selects the terms with $M = - M'$ in going from passage $(1)$ to passage $(2)$ in the previous equation. Then, one observes that $\widetilde{\chi}^{ 1, 3}_{2, 4} = \left( \widetilde{\chi}_{ 1, 3}^{2, 4} \right)^*$, and in the non-relativistic case this quantity is real. Thus, the terms with $\alpha \neq \alpha'$ vanish, and we obtain the last passage $(3)$. 

We see immediately that in this case Eqs.\eqref{J nonlocal} give (for $i \neq i'$) 
\begin{align}
& \mathcal{J}_{i i'}^{x} \stackrel{\mathrm{soH}}{=} \mathcal{J}_{i i'}^{y} \stackrel{\mathrm{soH}}{=} \mathcal{J}_{i i'}^{z} \stackrel{\mathrm{soH}}{\equiv} \mathcal{J}_{i i'}  \stackrel{\mathrm{soH}}{=} \widetilde{\mathcal{M}}^{i x}_{i' x} \stackrel{\mathrm{soH}}{=} \widetilde{\mathcal{M}}^{i y}_{i' y}  \nonumber \\
& =  \frac{1}{2}    T_{i i'}       \left(    \rho^{\uparrow}_{i i'} +    \rho^{\downarrow}_{i i'}      \right)    \nonumber \\
&  \quad \!\!     - \frac{1}{4} \sum_{j \neq i} \sum_{ j' \neq i'} T_{i j} T_{i' j'} \! \! \! \sum_{M = \uparrow  \downarrow}   \! \!  \Bigg[  \widetilde{\chi}^{(j M) (j' \bar{M}) }_{ (i \bar{M})  (i' M)}   \!   -  \widetilde{\chi}^{(j M) (i' \bar{M})  }_{ (i \bar{M}) (j' M)}       -  \widetilde{\chi}^{(i M)  (j' \bar{M}) }_{ (j \bar{M}) (i' M)}   +  \widetilde{\chi}^{(i M) (i' \bar{M})  }_{ (j \bar{M}) (j' M)}     \Bigg]  \! ,
\label{isotropy of exchange}
\end{align}
that is, exchange is isotropic. All the other magnetic parameters, determined for the general relativistic regime, vanish in the soH case. The mapping equations, \eqref{first set}, \eqref{third set} and \eqref{second set}, reduce to:
\begin{align}
&  \mathcal{J}_{i i'}  \stackrel{\mathrm{soH}}{=} \widetilde{\mathcal{M}}^{i \alpha}_{i' \alpha}, \quad i \neq i', \nonumber \\
&  - \sum_{j \neq i} \mathcal{J}_{i j}      \stackrel{\mathrm{soH}}{=}  \widetilde{\mathcal{M}}^{i \alpha}_{i \alpha}  .
\label{third set 2}
\end{align}
The sum rule given by the second among Eqs.\eqref{third set 2}, combined with the first equation, becomes
\begin{align}
  \sum_{i'} \widetilde{\mathcal{M}}^{i \alpha}_{i' \alpha} \stackrel{\mathrm{soH}}{=}  0 .
\label{sum rule nonrel}
\end{align}
To check whether this is valid, we use Eq.\eqref{M nonrel}, obtaining:
\begin{align}
 &   \sum_{i'} \sum_{j \neq i} \sum_{ j' \neq i'} T_{i j} T_{i' j'}  \sum_{M = \uparrow, \downarrow}      \Bigg[  \widetilde{\chi}^{(j M) (j' \bar{M}) }_{ (i \bar{M})  (i' M)}   -  \widetilde{\chi}^{(j M) (i' \bar{M})  }_{ (i \bar{M}) (j' M)}       -  \widetilde{\chi}^{(i M)  (j' \bar{M}) }_{ (j \bar{M}) (i' M)}   +  \widetilde{\chi}^{(i M) (i' \bar{M})  }_{ (j \bar{M}) (j' M)}     \Bigg]  \nonumber \\
        & = 0 ,
\end{align}
which is identically true, as follows from the interchange of the dummy indexes $i'$ and $j'$ in the second and fourth term on the LHS. Therefore, the sum rule \eqref{sum rule nonrel} is satisfied by the expression for the exchange parameters given in the first among Eqs.\eqref{third set 2}.

We compare our results with the previous literature on non-relativistic exchange. Most of the previous works on this subject \cite{Lichtenstein87, Katsnelson00, Katsnelson02, Secchi13} neglected the vertices in the two-electron Green's functions. This amounts to putting $\Gamma = 0$ in Eq.\eqref{2p GF}, replacing $\chi$ with $\chi^0$. We will now show what we obtain in the present case when such approximation is performed. We will denote all the equations derived under this approximation with the equality symbol $\stackrel{\Gamma = 0}{=}$. The only two-particle Green's function that we need is given by Eq.\eqref{only two}, which becomes
\begin{align}
 \chi^{1, 3}_{2 , 4}(\tau, \tau^+, \tau', \tau'^+) & \stackrel{\Gamma = 0}{=} \left( \chi^0 \right)^{1, 3}_{2 , 4}(\tau, \tau^+, \tau', \tau'^+) \nonumber \\
 & = -    \rho^1_2 \, \rho^3_4  - G^1_4(\tau - \tau' - \varepsilon ) \, \,  G^3_2(\tau' - \tau - \varepsilon) .
\end{align}
Using the Matsubara-frequency representation, $G(\tau) \equiv \frac{1}{\beta} \sum_{\omega} G(  \mathrm{i} \omega) \, \mathrm{e}^{- \mathrm{i} \omega \tau}$, we re-write Eq.\eqref{chi integrated} for $\Gamma = 0$, after integrating over $\mathrm{d}\tau$ and $\mathrm{d}\tau'$, as:
\begin{align}
\left(\widetilde{\chi}^0\right)^{1, 3 }_{ 2,  4} = - \beta \rho^1_2 \rho^3_4 - \frac{1}{\beta} \sum_{\omega} \mathrm{e}^{\mathrm{i} \omega 0^+} G^1_4( \mathrm{i} \omega ) \, G^3_2(  \mathrm{i} \omega ) .
\label{integration Matsubara}
\end{align}
In the soH case, we have $G^{i \sigma}_{i' \sigma'} \equiv \delta^{\sigma}_{\sigma'} G_{i i'}^{\sigma} = \delta^{\sigma}_{\sigma'} G_{i' i}^{\sigma}$, where $G_{i i'}^{\uparrow} \neq G_{i i'}^{\downarrow}$ for a symmetry-broken configuration of the system. We obtain
\begin{align}
&   \widetilde{\mathcal{M}}^{i \alpha}_{i' \alpha}      \stackrel{\mathrm{soH}, \Gamma = 0}{=} \frac{1}{2}   \Bigg[ T_{i i'}       \left(    \rho^{\uparrow}_{i i'} +    \rho^{\downarrow}_{i i'}      \right)   - \delta_{i i'}   \sum_j  T_{i j} \left(          \rho_{i j}^{\uparrow}   +  \rho_{i j}^{\downarrow}    \right) \Bigg] \nonumber \\
   & +  \frac{1}{4 \beta} \sum_{\sigma = \uparrow, \downarrow} \sum_{\omega} \mathrm{e}^{  \mathrm{i} \omega 0^+} \! \Bigg\{ \left[ T \cdot G^{\sigma}(\mathrm{i} \omega) \right]^i_{i'} \left[ T \cdot G^{\bar{\sigma}}(\mathrm{i} \omega) \right]_i^{i'} - \left[ T \cdot G^{\sigma}(\mathrm{i} \omega) \cdot T \right]^i_{i'} \! \left[   G^{\bar{\sigma}}(\mathrm{i} \omega) \right]_i^{i'} \nonumber \\
   & \quad \quad \quad \quad \quad \quad   \quad     - \left[   G^{\sigma}(\mathrm{i} \omega) \right]^i_{i'} \left[ T \cdot G^{\bar{\sigma}}(\mathrm{i} \omega) \cdot T \right]_i^{i'} 
   + \left[   G^{\sigma}(\mathrm{i} \omega) \cdot T \right]^i_{i'} \left[   G^{\bar{\sigma}}(\mathrm{i} \omega) \cdot T \right]_i^{i'}    \Bigg\} .
\label{J SSB no gamma}
\end{align}
The formulas for exchange parameters are often expressed in terms of self-energies \cite{Katsnelson00, Katsnelson02, Secchi13}, since these are the key quantities for numerical evaluation within the framework of Dynamical Mean Field Theory (DMFT) \cite{Metzner89, Georges96, Kotliar06}. To do so, we use the equations of motion for Matsubara Green's functions, which we write in general matrix notation as
\begin{align}
& \left( \omega - \mathrm{i} \mu \right) G^{\sigma}(\mathrm{i} \omega) + \mathrm{i} T \cdot G^{\sigma}(\mathrm{i} \omega) = 1  -   \Sigma^{\sigma}(\mathrm{i} \omega) \cdot G^{\sigma}(\mathrm{i} \omega) , \nonumber \\
& \left( \omega - \mathrm{i} \mu \right) G^{\sigma}(\mathrm{i} \omega) + \mathrm{i}   G^{\sigma}(\mathrm{i} \omega)  \cdot T  = 1 -  G^{\sigma}(\mathrm{i} \omega) \cdot  \Sigma^{\sigma}(\mathrm{i} \omega)   
\end{align} 
(units have been chosen so that $\Sigma$ has the dimensions of an energy). These equations allow to express Eq.\eqref{J SSB no gamma} in terms of single-particle Green's functions and self-energies $\Sigma$, removing the hopping parameters $T$. After some algebra, for $i \neq i'$ we obtain 
\begin{align}
 &  \mathcal{J}_{i i'} \stackrel{\mathrm{soH}, \Gamma = 0}{=} -   \frac{1}{4 \beta} \sum_{\sigma = \pm 1} \sum_{\omega} \mathrm{e}^{ \mathrm{i} \omega 0^+} \Bigg\{ \left[ \Sigma^{\sigma}(\mathrm{i}  \omega) \cdot G^{\sigma}(\mathrm{i}  \omega) \right]^i_{i'} \left[ \Sigma^{\bar{\sigma}}(\mathrm{i}  \omega) \cdot G^{\bar{\sigma}}(\mathrm{i}  \omega) \right]_i^{i'} \nonumber \\
 & - \left[ \Sigma^{\sigma}(\mathrm{i}  \omega) \cdot G^{\sigma}(\mathrm{i}  \omega) \cdot \Sigma^{\sigma}(\mathrm{i}  \omega) \right]^i_{i'} \left[   G^{\bar{\sigma}}(\mathrm{i}  \omega) \right]_i^{i'}       - \left[   G^{\sigma}\!(\mathrm{i}  \omega) \right]^i_{i'} \left[ \Sigma^{\bar{\sigma}}(\mathrm{i}  \omega)   \cdot   G^{\bar{\sigma}}\!(\mathrm{i}  \omega)   \cdot   \Sigma^{\bar{\sigma}}\!(\mathrm{i}  \omega) \right]_i^{i'}  \nonumber \\
 & + \left[   G^{\sigma}(\mathrm{i}  \omega) \cdot \Sigma^{\sigma}(\mathrm{i}  \omega) \right]^i_{i'} \left[   G^{\bar{\sigma}}(\mathrm{i}  \omega) \cdot \Sigma^{\bar{\sigma}}(\mathrm{i}  \omega) \right]_i^{i'}      
 +     \left[   \Sigma^{\sigma}(\mathrm{i}  \omega) \right]_{i'}^i \left[   G^{\bar{\sigma}}(\mathrm{i}  \omega) \right]^{i'}_i   \nonumber \\
& +    \left[   G^{\sigma}(\mathrm{i}  \omega) \right]_{i'}^i \left[   \Sigma^{\bar{\sigma}}(\mathrm{i}  \omega) \right]^{i'}_i  \Bigg\} .
\label{J SSB no gamma Sigma}
\end{align}
Equation \eqref{J SSB no gamma Sigma} is in agreement with Eqs.(185) and (155) from Ref.\cite{Secchi13} (as it can be seen by using the symmetries of the Green's functions). The different pre-factor $- 2$ is due to the different definition of the exchange parameters in the Hamiltonian \eqref{general quadratic model}.   

Finally, if we assume the self-energy to be local (``LsoH'' assumption), which is a requirement for the direct application of DMFT, by putting $\Sigma_{i i'}^{\sigma} \stackrel{\mathrm{LsoH}}{=} \delta_{i i'} \Sigma_{i}^{\sigma}$ we obtain the simple formula
\begin{align}
\mathcal{J}_{i i'} \stackrel{\mathrm{LsoH}, \Gamma = 0}{=}  \frac{2 }{ \beta}  \sum_{\omega} \mathrm{e}^{ \mathrm{i} \omega 0^+} \left[ G_{i i'}^{\uparrow}(\mathrm{i} \omega) \, \Sigma^{\mathrm{S}}_{i'}(\mathrm{i} \omega) \, G_{i' i}^{\downarrow}(\mathrm{i} \omega) \, \Sigma^{\mathrm{S}}_i(\mathrm{i} \omega) \right] ,
\label{J Lsb}
\end{align}
where $\Sigma^{\mathrm{S}}_i(\mathrm{i} \omega) \equiv \left[ \Sigma^{\uparrow}_i(\mathrm{i} \omega) - \Sigma^{\downarrow}_i(\mathrm{i}\omega) \right] / 2$. Equation \eqref{J Lsb} is in agreement with Eq.(21) from Ref.\cite{Katsnelson00}, Eq.(19) from Ref.\cite{Katsnelson02} and Eq.(191) from Ref.\cite{Secchi13}, again up to a factor $- 2$ due to the different definition mentioned above. We have thus recovered the results of the previous literature as particular cases of our present formulation.

\section{Spin, orbital, and spin-orbital contributions to magnetism}
\label{sec: Separation}

We now go back to the relativistic regime and to the results summarized in Section \ref{sec: Summary solution}. In this work, as stated in the introduction, we are considering as dynamical variables some effective classical ``spins'' which are represented by the unit vectors $\boldsymbol{e}_i$. The coefficients of the interactions, that we have determined, are related to the response of the system under rotations of the total local magnetic moments expressed by the operators $\hat{\boldsymbol{S}}_i = \hat{\boldsymbol{l}}_i + \hat{\boldsymbol{s}}_i$. It is interesting to compare the response of the system under this rotation to the response that is obtained when only the spin-$1/2$ ($\hat{\boldsymbol{s}}_i$) or the orbital ($\hat{\boldsymbol{l}}_i$) components of the magnetic moments are rotated. In order to address this question, we need to separate in the effective magnetic parameters the contributions coming from the rotation of $\hat{\boldsymbol{s}}_i$ from the contributions coming from the rotation of $\hat{\boldsymbol{l}}_i$. If these contributions could be decoupled, we could identify them individually as distinct contributions to magnetism.

To perform this decoupling, we need to switch from the initial basis for the electronic fields, where the single-electron wave functions were characterized by the quantum numbers $(a, n, l, S, M)$, to the basis characterized by $(a, n, l, m, \sigma)$, where $m$ and $\sigma = \pm 1/2$ are the quantum numbers, respectively, of the operators $\hat{l}^z$ and $\hat{s}^z$. The change of basis goes via the Clebsch-Gordan transformation,
\begin{align}
\hat{\psi}^{\dagger}_{a, n, l, S, M} \equiv  \sum_{m = - l}^l  \, \, \sum_{\sigma = \pm 1/2} \,  C^{m \sigma}_{S M}(l) \, \, \hat{\psi}^{\dagger}_{a, n, l, m, \sigma} ,
\end{align}
where $C_{S M}^{m \sigma}(l)$ is a Clebsch-Gordan coefficient. We note that, up to now, we have considered unit vectors $\boldsymbol{e}_i$ depending on the index $i \equiv (a, n, l, S)$, meaning that we have defined in principle a different spin for each value of $S$ corresponding to a given orbital set $o \equiv (a, n, l)$. It is not possible to separate spin-$1/2$ from orbital contributions in this situation. To achieve this separation, we need the spins to be independent of $S$, i.e., $\boldsymbol{e}_i \rightarrow \boldsymbol{e}_o$. In this way our effective spin Hamiltonian, from Eq.\eqref{general quadratic model}, becomes:
\begin{align}
H\left[   \boldsymbol{e}_i \right] & =  \sum_{i}      \boldsymbol{e}_{i}  \cdot  \boldsymbol{\mathcal{B}}_{i} + \frac{1}{2} \sum_{i, i'}   \sum_{\alpha, \alpha'}     e_{i, \alpha}       e_{i', \alpha'}  \mathcal{H}_{i i' }^{\alpha \alpha'}  \nonumber \\
& \equiv \sum_{o}      \boldsymbol{e}_{o}  \cdot  \boldsymbol{\mathcal{B}}_{o}  + \frac{1}{2} \sum_{o, o'}   \sum_{\alpha, \alpha'}     e_{o, \alpha}       e_{o', \alpha'}\mathcal{H}_{ o    o'   }^{\alpha \alpha'}  ,
\label{general quadratic model orbital}
\end{align}
where
\begin{align}
\boldsymbol{\mathcal{B}}_{o} \equiv \sum_{S} \boldsymbol{\mathcal{B}}_{o S} , \quad \mathcal{H}_{ o    o'   }^{\alpha \alpha'} \equiv   \sum_{S, S'} \mathcal{H}_{(o S) (o' S') }^{\alpha \alpha'} .
\label{redefined parameters}
\end{align}
Note that this is a \emph{particular case} of the general procedure that we have followed up to now, corresponding to the less general case of the rotations depending only on $o$ rather than on $(o, S)$. Therefore, we simply have to re-define the parameters according to Eqs.\eqref{redefined parameters}, without altering the spin Hamiltonian.

Summing over the $S$ quantum numbers now allows to separate orbital and spin contributions. The most compact way to show how it works is to consider the quantities $\mathcal{V}_{i \alpha}$ and $\widetilde{\mathcal{M}}^{i \alpha}_{i' \alpha'}$, since all the magnetic parameters are obtained as linear combinations of these quantities (or parts of them). First, let us consider $\mathcal{V}_{i \alpha}$, which in the new Hamiltonian given by the second passage of Eq.\eqref{general quadratic model orbital} will be replaced by [cfr. Eq.\eqref{eq 1 RHS}] 
\begin{align}
\mathcal{V}_{o \alpha} & = \sum_S \mathcal{V}_{(o S) \alpha} = \sum_S \mathrm{i}     \,  \mathrm{Tr}_M \sum_{ o' S' } \Big[ S_{(o S) \alpha} \cdot \left( \rho^{ o S }_{ o' S' } \cdot T^{ o' S' }_{ o S } - T^{ o S }_{ o' S' } \cdot \rho^{ o' S' }_{ o S }     \right)   \Big] \nonumber \\
& =   \mathrm{i}     \,  \mathrm{Tr}_{S, M}   \Big[ S_{ o   \alpha} \cdot \left( \rho  \cdot T - T  \cdot \rho     \right)^o_o   \Big]   \stackrel{!}{=} \mathrm{i}     \,  \mathrm{Tr}_{m, \sigma}   \Big[ \left( s_{ o   \alpha} + l_{ o   \alpha}\right) \cdot \left( \rho  \cdot T  - T  \cdot \rho     \right)^o_o   \Big] \nonumber \\
& \equiv \mathcal{V}_{o \alpha}^{\mathrm{spin}} + \mathcal{V}_{o \alpha}^{\mathrm{orb}} ,
\label{separation V spin orbital}
\end{align}
where we have defined the separate spin-$1/2$ and orbital contributions, respectively, as
\begin{align}
&  \mathcal{V}_{o \alpha}^{\mathrm{spin}} \equiv \mathrm{i}     \,  \mathrm{Tr}_{ \sigma}   \Big[   s_{ o   \alpha}   \cdot \mathrm{Tr}_{m } \left( \rho  \cdot T  - T  \cdot \rho     \right)^o_o   \Big] , \nonumber \\
&  \mathcal{V}_{o \alpha}^{\mathrm{orb}} \equiv \mathrm{i}     \,  \mathrm{Tr}_{m }   \Big[   l_{ o   \alpha}  \cdot \mathrm{Tr}_{ \sigma} \left( \rho  \cdot T  - T  \cdot \rho     \right)^o_o   \Big] .
\end{align}
The passage marked as $\stackrel{!}{=}$ in Eq.\eqref{separation V spin orbital} is the step that allows to go from the representation in the $(a, n, l, S, M)$ basis to the representation in the $(a, n, l, m, \sigma)$ basis, where it is possible to split the total spin matrix into the spin-$1/2$ and the orbital contribution. The matrices defined in the new basis, such as the density matrix $\rho$ and the hopping parameters $T$, are obtained from the previous representation via the Clebsch-Gordan transformation:
\begin{align}
\rho^{a n l S M}_{a' n' l' S' M'} = \sum_{m \sigma} \sum_{m' \sigma'} \rho^{a n l m \sigma}_{a' n' l' m' \sigma'} C^{S M}_{m \sigma}(l) \,  C_{S' M'}^{m' \sigma'}(l') .
\end{align}
The equivalence of the traces is a consequence of the completeness relation
\begin{align}
\sum_{S M} C^{S M}_{m \sigma}(l) \,  C_{S M}^{m' \sigma'}(l) = \delta^{m'}_m \delta^{\sigma'}_{\sigma} . 
\end{align}
It is then clear why we need the sum over $S$ to split the spin and orbital contributions, and for this reason our rotation parameters must depend only on $(a, n, l)$.

We then consider the term $\widetilde{\mathcal{M}}^{i \alpha}_{i' \alpha'}$, which in the new Hamiltonian given by the second passage of Eq.\eqref{general quadratic model orbital} will be replaced by [cfr. Eq.\eqref{tilde M brevity}]
\begin{align}
\widetilde{\mathcal{M}}^{o \alpha}_{o' \alpha'} & = \sum_{S, S'} \widetilde{\mathcal{M}}^{(o S) \alpha}_{(o' S') \alpha'} = \mathrm{Tr}_{S, M} \left( S_{o \alpha} \cdot T^o_{o'} \cdot S_{o' \alpha'} \cdot \rho^{o'}_o    +   S_{o' \alpha'} \cdot T_o^{o'} \cdot  S_{o \alpha} \cdot \rho_{o'}^o  \right)  \nonumber \\
& \quad - \delta^{o  }_{o'  } \frac{1}{2}  \mathrm{Tr}_{S, M} \Big( \left( S_{o \alpha} \cdot S_{o \alpha'} +   S_{o \alpha'} \cdot S_{o \alpha}    \right) \cdot \left\{ \rho ; T \right\}^o_o\Big)   + \beta \mathcal{V}_{o \alpha} \mathcal{V}_{o' \alpha'} \nonumber \\
& \quad - \frac{1}{\beta} \int_0^{\beta} \mathrm{d} \tau \int_0^{\beta} \mathrm{d} \tau'  \left< \mathcal{T}_{\gamma} \hat{\mathcal{V}}_{o \alpha}(\tau) \, \hat{\mathcal{V}}_{o' \alpha'}(\tau') \right> ,
\label{orbital tilde M}
\end{align}
where 
\begin{align}
\hat{\mathcal{V}}_{o \alpha}(\tau)  & = \mathrm{i}  \, \mathrm{Tr}_{S, M} \Big( S_{ o   \alpha} \cdot \left[ \hat{\rho}(\tau) ; T \right]^{o }_{o } \Big) \nonumber \\
&  = \mathrm{i}  \, \mathrm{Tr}_{ \sigma} \Big(   s_{ o   \alpha}   \cdot \mathrm{Tr}_{m }  \left[ \hat{\rho}(\tau) ; T \right]^{o }_{o } \Big) 
+ \mathrm{i}  \, \mathrm{Tr}_{m } \Big(   l_{ o   \alpha}   \cdot \mathrm{Tr}_{ \sigma} \left[ \hat{\rho}(\tau) ; T \right]^{o }_{o } \Big) \nonumber \\
&  \equiv \hat{\mathcal{V}}^{\mathrm{spin}}_{o \alpha}(\tau) + \hat{\mathcal{V}}^{\mathrm{orb}}_{o \alpha}(\tau) .
\end{align}
We can then apply the change of basis, replace $\mathrm{Tr}_{S, M}$ with $\mathrm{Tr}_{m, \sigma}$, and separate the spin-$1/2$ and the orbital contributions by splitting the $S_{o \alpha}$ matrix into $s_{o \alpha} + l_{o \alpha}$. In the case of the quantity $\widetilde{\mathcal{M}}^{o \alpha}_{o' \alpha'}$ given by Eq.\eqref{orbital tilde M}, which depends on quadratic combinations of the spin matrices, we can distinguish spin-spin, orbital-orbital and spin-orbital contributions:
\begin{align}
\widetilde{\mathcal{M}}^{o \alpha}_{o' \alpha'} \equiv \left( \widetilde{\mathcal{M}}^{o \alpha}_{o' \alpha'} \right)^{\mathrm{spin-spin}} + \left( \widetilde{\mathcal{M}}^{o \alpha}_{o' \alpha'} \right)^{\mathrm{spin-orb}} + \left( \widetilde{\mathcal{M}}^{o \alpha}_{o' \alpha'} \right)^{\mathrm{orb-orb}}  ,
\end{align} 
where
\begin{align}
\left( \widetilde{\mathcal{M}}^{o \alpha}_{o' \alpha'} \right)^{\mathrm{spin-spin}}  \equiv & \, \mathrm{Tr}_{m, \sigma} \left( s_{o \alpha} \cdot T^o_{o'} \cdot s_{o' \alpha'} \cdot \rho^{o'}_o    +   s_{o' \alpha'} \cdot T_o^{o'} \cdot  s_{o \alpha} \cdot \rho_{o'}^o  \right)  \nonumber \\
&   - \delta^{o  }_{o'  } \delta^{\alpha}_{\alpha'} \frac{1}{4}  \mathrm{Tr}_{m, \sigma}    \left\{ \rho ; T \right\}^o_o   \nonumber \\
&  + \beta \mathcal{V}^{\mathrm{spin}}_{o \alpha} \mathcal{V}^{\mathrm{spin}}_{o' \alpha'}    - \frac{1}{\beta} \int_0^{\beta} \mathrm{d} \tau \int_0^{\beta} \mathrm{d} \tau'  \left< \mathcal{T}_{\gamma} \hat{\mathcal{V}}^{\mathrm{spin}}_{o \alpha}(\tau) \, \hat{\mathcal{V}}^{\mathrm{spin}}_{o' \alpha'}(\tau') \right> ,
\end{align}
where in the second line we have used the fact that $s_{o \alpha} \cdot s_{o \alpha'} +   s_{o \alpha'} \cdot s_{o \alpha} = \frac{1}{2} \delta_{\alpha \alpha'} 1$, which is a property of the Pauli matrices,
\begin{align}
\left( \widetilde{\mathcal{M}}^{o \alpha}_{o' \alpha'} \right)^{\mathrm{orb-orb}}  \equiv & \, \mathrm{Tr}_{m, \sigma} \left( l_{o \alpha} \cdot T^o_{o'} \cdot l_{o' \alpha'} \cdot \rho^{o'}_o    +   l_{o' \alpha'} \cdot T_o^{o'} \cdot  l_{o \alpha} \cdot \rho_{o'}^o  \right)  \nonumber \\
&   - \delta^{o  }_{o'  } \frac{1}{2}  \mathrm{Tr}_{m } \Big( \left( l_{o \alpha} \cdot l_{o \alpha'} +   l_{o \alpha'} \cdot l_{o \alpha}    \right) \cdot \mathrm{Tr}_{ \sigma} \left\{ \rho ; T \right\}^o_o\Big)   \nonumber \\
& + \beta \mathcal{V}^{\mathrm{orb}}_{o \alpha} \mathcal{V}^{\mathrm{orb}}_{o' \alpha'}   - \frac{1}{\beta} \int_0^{\beta} \mathrm{d} \tau \int_0^{\beta} \mathrm{d} \tau'  \left< \mathcal{T}_{\gamma} \hat{\mathcal{V}}^{\mathrm{orb}}_{o \alpha}(\tau) \, \hat{\mathcal{V}}^{\mathrm{orb}}_{o' \alpha'}(\tau') \right> ,
\end{align}
\begin{align}
& \left( \widetilde{\mathcal{M}}^{o \alpha}_{o' \alpha'} \right)^{\mathrm{spin-orb}} \nonumber \\
& \equiv  \, \mathrm{Tr}_{m, \sigma} \left( s_{o \alpha} \cdot T^o_{o'} \cdot l_{o' \alpha'} \cdot \rho^{o'}_o    +   s_{o' \alpha'} \cdot T_o^{o'} \cdot  l_{o \alpha} \cdot \rho_{o'}^o  \right.  \nonumber \\
& \quad \quad \quad  \quad +   \left. l_{o \alpha} \cdot T^o_{o'} \cdot s_{o' \alpha'} \cdot \rho^{o'}_o    +   l_{o' \alpha'} \cdot T_o^{o'} \cdot  s_{o \alpha} \cdot \rho_{o'}^o  \right)  \nonumber \\
& \quad   - \delta^{o  }_{o'  } \frac{1}{2}  \mathrm{Tr}_{m, \sigma} \Big( \left( s_{o \alpha} \cdot l_{o \alpha'} +   s_{o \alpha'} \cdot l_{o \alpha} + l_{o \alpha} \cdot s_{o \alpha'} +   l_{o \alpha'} \cdot s_{o \alpha}    \right) \cdot \left\{ \rho ; T \right\}^o_o\Big)    \nonumber \\
& \quad  + \beta \left( \mathcal{V}^{\mathrm{spin}}_{o \alpha} \mathcal{V}^{\mathrm{orb}}_{o' \alpha'}  +   \mathcal{V}^{\mathrm{orb}}_{o \alpha} \mathcal{V}^{\mathrm{spin}}_{o' \alpha'} \right)     \nonumber \\
& \quad  - \frac{1}{\beta} \int_0^{\beta} \mathrm{d} \tau \int_0^{\beta} \mathrm{d} \tau'  \left< \mathcal{T}_{\gamma} \left[  \hat{\mathcal{V}}^{\mathrm{spin}}_{o \alpha}(\tau) \, \hat{\mathcal{V}}^{\mathrm{orb}}_{o' \alpha'}(\tau')   + \hat{\mathcal{V}}^{\mathrm{orb}}_{o \alpha}(\tau) \, \hat{\mathcal{V}}^{\mathrm{spin}}_{o' \alpha'}(\tau')   \right] \right> .
\end{align}

This separation, which we have applied to the quantities $\mathcal{V}_{o \alpha}$ and $\widetilde{\mathcal{M}}^{o \alpha}_{o' \alpha'}$, has then to be transferred to the effective magnetic parameters $\boldsymbol{\mathcal{B}}_{o}$ and $\mathcal{H}_{o o'}^{\alpha \alpha'}$ of Eq.\eqref{general quadratic model orbital}, via the solutions of the mapping equations summarized in Section \ref{sec: Summary solution}. We notice that the magnetic parameters can be separated into two groups: the \emph{first group} given by $ \left\{ \boldsymbol{\mathcal{B}}_{o},  \mathcal{D}^x_{o o'}, \mathcal{D}^y_{o o'}, \mathcal{C}^x_{o o'}, \mathcal{C}^y_{o o'} \right\}$, and the \emph{second group} given by $\left\{ \mathcal{D}^z_{o o'}, \mathcal{C}^z_{o o'}, \boldsymbol{\mathcal{J}}_{o o'} \right\}$. The terms of the first group have the following features:
\begin{itemize}
\item they are expressed in terms of the quantities $\mathcal{V}_{o \alpha}$ or parts of them;
\item their evaluation requires computation of single-particle Green's functions (of the density-matrix form);
\item they can be split into spin and orbital contributions.
\end{itemize}
The terms of the second group have the following features:
\begin{itemize}
\item they are expressed in terms of the quantities $\mathcal{V}_{o \alpha}$ and $\widetilde{\mathcal{M}}^{o \alpha}_{o' \alpha'}$;
\item their evaluation requires computation of single-particle and two-particle Green's functions;
\item they can be split into spin-spin, spin-orbital and orbital-orbital contributions.
\end{itemize}
In the latter case, while the spin-spin and orbital-orbital parts obviously arise from rotations involving only one of the two contributions to the total local magnetic moments, respectively, the spin-orbital term does not arise in such individual rotations, appearing only when the whole magnetic moments are rotated.

In the next Sections we list the explicit formulas for all the parameters of the magnetic interactions, separated into spin, orbital and (when applicable) spin-orbital parts. For the magnetic parameters of the second group, we show not only the complete formulas with the full two-particle Green's functions, but also the formulas obtained when the vertices are neglected. This approximation, which produces formulas depending only on single-particle Green's functions, has been routinely applied for computations of isotropic exchange parameters within the framework of the Hubbard model with quenched orbital moments \cite{Katsnelson00}. The formulas that we list below are the natural extension to the unquenched case. The approximated expressions are listed after the exact ones, separated from them by the symbol $\stackrel{\Gamma = 0}{=}$, analogously to the convention used in Section \ref{sec: Nonrel}. The approximation is achieved by applying Eq.\eqref{integration Matsubara}. In particular, we have
\begin{align}
&  \beta \mathcal{V}^{\mathrm{X}}_{o \alpha} \mathcal{V}^{\mathrm{Y}}_{o' \alpha'} - \frac{1}{\beta} \int_0^{\beta} \mathrm{d} \tau   \int_0^{\beta} \mathrm{d} \tau' \left< \mathcal{T}_{\gamma}   \hat{\mathcal{V}}^{\mathrm{X}}_{o \alpha}(\tau)  \,   \hat{\mathcal{V}}^{\mathrm{Y}}_{o' \alpha'}(\tau')   \right> \nonumber \\
&  \stackrel{\Gamma = 0}{=}   \frac{1}{\beta}  \sum_{\omega} \mathrm{e}^{\mathrm{i} \omega 0^+} \mathrm{Tr}_{m, \sigma} \Big\{ S^{\mathrm{X}}_{o \alpha}  \cdot \left[ G(\mathrm{i} \omega ) \cdot T \right]^o_{o'} \cdot S^{\mathrm{Y}}_{o' \alpha'} \cdot \left[ G(\mathrm{i} \omega ) \cdot T \right]_o^{o'}  \nonumber \\
& \quad \quad \quad \quad   \quad \quad  \quad \quad   \quad \quad  -   S^{\mathrm{X}}_{o \alpha}  \cdot   G(\mathrm{i} \omega )^o_{o'} \cdot S^{\mathrm{Y}}_{o' \alpha'} \cdot \left[ T \cdot G(\mathrm{i} \omega ) \cdot T \right]_o^{o'}   \nonumber \\
& \quad \quad \quad \quad   \quad \quad  \quad \quad   \quad \quad  -   S^{\mathrm{X}}_{o \alpha}  \cdot  \left[ T \cdot G(\mathrm{i} \omega ) \cdot T \right]^o_{o'} \cdot S^{\mathrm{Y}}_{o' \alpha'} \cdot  G(\mathrm{i} \omega )_o^{o'}  \nonumber \\
& \quad \quad \quad \quad   \quad \quad  \quad \quad   \quad \quad + S^{\mathrm{X}}_{o \alpha}  \cdot \left[ T \cdot G(\mathrm{i} \omega )  \right]^o_{o'} \cdot S^{\mathrm{Y}}_{o' \alpha'} \cdot \left[ T \cdot G(\mathrm{i} \omega )  \right]_o^{o'}   \Big\} ,
\end{align}
where X and Y refer to either spin- or orbital- related terms. Since the resulting expressions for the magnetic parameters become very long and involved, for the sake of readability we make use of the permutation symbol $\overrightarrow{\mathcal{P}}_{o \leftrightarrow o'}$, which switches the indexes $o$ and $o'$ of any tensor placed on its right side, that is,
\begin{align}
f_{o o'} \overrightarrow{\mathcal{P}}_{o \leftrightarrow o'} g_{o o'} = f_{o o'} g_{o' o} .
\end{align}
Obviously, the combination $1 + \overrightarrow{\mathcal{P}}_{o \leftrightarrow o'}$ is then the symmetrization symbol, while $1 - \overrightarrow{\mathcal{P}}_{o \leftrightarrow o'}$ is the anti-symmetrization symbol.

We stress that the magnetic parameters of the first group do not require the evaluation of two-particle Green's functions, so their expressions in terms of single-particle density matrices are exact.  

Here follows the list of all the explicit expressions.

\subsection{Dzyaloshinskii-Moriya interaction}
 
From Eqs.\eqref{D}, we see that $\mathcal{D}_{o o'}^{\alpha} \equiv \left( \mathcal{D}_{o o'}^{\alpha} \right)^{\mathrm{spin}}   +   \left( \mathcal{D}_{o o'}^{\alpha} \right)^{\mathrm{orb}}$ for $\alpha = x$ or $y$, while $\mathcal{D}_{o o'}^{z} \equiv \left( \mathcal{D}_{o o'}^{z} \right)^{\mathrm{spin-spin}}   +   \left( \mathcal{D}_{o o'}^{z} \right)^{\mathrm{orb-orb}} + \left( \mathcal{D}_{o o'}^{z} \right)^{\mathrm{spin-orb}}$, where
\begin{align}
            \left(  \mathcal{D}_{o o'}^{x}  \right)^{\mathrm{spin}}         =  &  \frac{ \mathrm{i} }{2 }  \left( 1 - \overrightarrow{\mathcal{P}}_{o \leftrightarrow o'} \right)     \mathrm{Tr}_{\sigma}  \Big[ s_{o x} \cdot  \mathrm{Tr}_{m}  \left( \rho^o_{o'} \cdot T^{o'}_o - T^o_{o'} \cdot \rho^{o'}_o    \right)    \Big]    ,  
\end{align}
\begin{align}
            \left(  \mathcal{D}_{o o'}^{x}  \right)^{\mathrm{orb}}         =  &  \frac{ \mathrm{i} }{2 }    \left( 1 - \overrightarrow{\mathcal{P}}_{o \leftrightarrow o'} \right)      \mathrm{Tr}_{m}  \Big[ l_{o x} \cdot  \mathrm{Tr}_{\sigma}  \left( \rho^o_{o'} \cdot T^{o'}_o - T^o_{o'} \cdot \rho^{o'}_o    \right)     \Big]    ,  
\end{align}
\begin{align}
            \left(  \mathcal{D}_{o o'}^{y}  \right)^{\mathrm{spin}}         =  &  \frac{ \mathrm{i} }{2 }  \left( 1 - \overrightarrow{\mathcal{P}}_{o \leftrightarrow o'} \right)        \mathrm{Tr}_{\sigma}  \Big[ s_{o y} \cdot  \mathrm{Tr}_{m}  \left( \rho^o_{o'} \cdot T^{o'}_o - T^o_{o'} \cdot \rho^{o'}_o    \right)      \Big]    ,  
\end{align}
\begin{align}
            \left(  \mathcal{D}_{o o'}^{y}  \right)^{\mathrm{orb}}         =  &  \frac{ \mathrm{i} }{2 }   \left( 1 - \overrightarrow{\mathcal{P}}_{o \leftrightarrow o'} \right)       \mathrm{Tr}_{m}  \Big[ l_{o y} \cdot  \mathrm{Tr}_{\sigma}  \left( \rho^o_{o'} \cdot T^{o'}_o - T^o_{o'} \cdot \rho^{o'}_o    \right)   \Big]    ,  
\end{align}
\begin{align}
&  \left( \mathcal{D}_{o o'}^z \right)^{\mathrm{spin-spin}} \nonumber \\
&  =     \frac{1}{2}  \left( 1 - \overrightarrow{\mathcal{P}}_{o \leftrightarrow o'} \right) \Bigg\{  \mathrm{Tr}_{m, \sigma} \Big( s_{o x} \cdot T^o_{o'} \cdot s_{o' y} \cdot \rho^{o'}_o    +   s_{o' y} \cdot T_o^{o'} \cdot  s_{o x} \cdot \rho_{o'}^o    \Big) \nonumber \\
& \quad  +    \beta \mathcal{V}^{\mathrm{spin}}_{o x} \mathcal{V}^{\mathrm{spin}}_{o' y}          - \frac{1}{  \beta} \int_0^{\beta} \mathrm{d} \tau \int_0^{\beta} \mathrm{d} \tau'  \left< \mathcal{T}_{\gamma}  \hat{\mathcal{V}}^{\mathrm{spin}}_{o x}(\tau) \, \hat{\mathcal{V}}^{\mathrm{spin}}_{o' y}(\tau')       \right>  \Bigg\}   \nonumber \\
&  \stackrel{\Gamma = 0}{=}     \frac{1 }{2}   \left( 1 - \overrightarrow{\mathcal{P}}_{o \leftrightarrow o'} \right) \Bigg\{ \mathrm{Tr}_{m, \sigma} \Big( s_{o x} \cdot T^o_{o'} \cdot s_{o' y} \cdot \rho^{o'}_o    +   s_{o' y} \cdot T_o^{o'} \cdot  s_{o x} \cdot \rho_{o'}^o    \Big) \nonumber \\
& \quad  + \frac{1  }{  \beta}  \sum_{\omega} \mathrm{e}^{\mathrm{i} \omega 0^+} \mathrm{Tr}_{m, \sigma} \Bigg[ s_{o x}  \cdot \left[ G(\mathrm{i} \omega ) \cdot T \right]^o_{o'} \cdot s_{o' y} \cdot \left[ G(\mathrm{i} \omega ) \cdot T \right]_o^{o'}  \nonumber \\
&  \quad    -   s_{o x}  \cdot   G(\mathrm{i} \omega )^o_{o'} \cdot s_{o' y} \cdot \left[ T \cdot G(\mathrm{i} \omega ) \cdot T \right]_o^{o'}      -   s_{o x}  \cdot  \left[ T \cdot G(\mathrm{i} \omega ) \cdot T \right]^o_{o'} \cdot s_{o' y} \cdot  G(\mathrm{i} \omega )_o^{o'}  \nonumber \\
&  \quad   + s_{o x}  \cdot \left[ T \cdot G(\mathrm{i} \omega )  \right]^o_{o'} \cdot s_{o' y} \cdot \left[ T \cdot G(\mathrm{i} \omega )  \right]_o^{o'}     \Bigg] \Bigg\} ,
\end{align}
\begin{align}
&  \left( \mathcal{D}_{o o'}^z \right)^{\mathrm{orb-orb}}  \nonumber \\
& =  \frac{1}{2} \left( 1 - \overrightarrow{\mathcal{P}}_{o \leftrightarrow o'} \right)  \Bigg\{ \mathrm{Tr}_{m, \sigma} \Big( l_{o x} \cdot T^o_{o'} \cdot l_{o' y} \cdot \rho^{o'}_o    +   l_{o' y} \cdot T_o^{o'} \cdot  l_{o x} \cdot \rho_{o'}^o     \Big)  \nonumber \\
& \quad + \beta  \mathcal{V}^{\mathrm{orb}}_{o x} \mathcal{V}^{\mathrm{orb}}_{o' y}     - \frac{1}{  \beta} \int_0^{\beta} \mathrm{d} \tau \int_0^{\beta} \mathrm{d} \tau'  \left< \mathcal{T}_{\gamma}  \hat{\mathcal{V}}^{\mathrm{orb}}_{o x}(\tau) \, \hat{\mathcal{V}}^{\mathrm{orb}}_{o' y}(\tau')          \right>   \Bigg\} \nonumber \\
&  \stackrel{\Gamma = 0}{=}     \frac{1 }{2} \left(  1 - \overrightarrow{\mathcal{P}}_{o \leftrightarrow o'} \right)   \Bigg\{  \mathrm{Tr}_{m, \sigma} \Big( l_{o x} \cdot T^o_{o'} \cdot l_{o' y} \cdot \rho^{o'}_o    +   l_{o' y} \cdot T_o^{o'} \cdot  l_{o x} \cdot \rho_{o'}^o    \Big) \nonumber \\
& \quad  + \frac{1  }{  \beta}  \sum_{\omega} \mathrm{e}^{\mathrm{i} \omega 0^+} \mathrm{Tr}_{m, \sigma} \Bigg[  l_{o x}  \cdot \left[ G(\mathrm{i} \omega ) \cdot T \right]^o_{o'} \cdot l_{o' y} \cdot \left[ G(\mathrm{i} \omega ) \cdot T \right]_o^{o'}  \nonumber \\
&  \quad    -   l_{o x}  \cdot   G(\mathrm{i} \omega )^o_{o'} \cdot l_{o' y} \cdot \left[ T \cdot G(\mathrm{i} \omega ) \cdot T \right]_o^{o'}      -   l_{o x}  \cdot  \left[ T \cdot G(\mathrm{i} \omega ) \cdot T \right]^o_{o'} \cdot l_{o' y} \cdot  G(\mathrm{i} \omega )_o^{o'}  \nonumber \\
&  \quad  + l_{o x}  \cdot \left[ T \cdot G(\mathrm{i} \omega )  \right]^o_{o'} \cdot l_{o' y} \cdot \left[ T \cdot G(\mathrm{i} \omega )  \right]_o^{o'}     \Bigg] \Bigg\} ,
\end{align}
\begin{align}
&  \left( \mathcal{D}_{o o'}^z \right)^{\mathrm{spin-orb}}  \nonumber \\
& =  \frac{1}{2}  \left(  1 - \overrightarrow{\mathcal{P}}_{o \leftrightarrow o'} \right) \Bigg\{    \mathrm{Tr}_{m, \sigma} \Big( s_{o x} \cdot T^o_{o'} \cdot l_{o' y} \cdot \rho^{o'}_o    +   s_{o' y} \cdot T_o^{o'} \cdot  l_{o x} \cdot \rho_{o'}^o     \nonumber \\
& \quad \quad \quad \quad \quad \quad \quad \quad \quad \quad  \quad  \quad +     l_{o x} \cdot T^o_{o'} \cdot s_{o' y} \cdot \rho^{o'}_o    +   l_{o' y} \cdot T_o^{o'} \cdot  s_{o x} \cdot \rho_{o'}^o        \Big)  \nonumber \\
& \quad  +     \beta \mathcal{V}^{\mathrm{spin}}_{o x} \mathcal{V}^{\mathrm{orb}}_{o' y}     + \beta   \mathcal{V}^{\mathrm{orb}}_{o x} \mathcal{V}^{\mathrm{spin}}_{o' y}     \nonumber \\
& \quad  -     \frac{1}{  \beta} \int_0^{\beta} \mathrm{d} \tau \int_0^{\beta} \mathrm{d} \tau'  \Big< \mathcal{T}_{\gamma} \Big[  \hat{\mathcal{V}}^{\mathrm{spin}}_{o x}(\tau) \, \hat{\mathcal{V}}^{\mathrm{orb}}_{o' y}(\tau')   + \hat{\mathcal{V}}^{\mathrm{orb}}_{o x}(\tau) \, \hat{\mathcal{V}}^{\mathrm{spin}}_{o' y}(\tau')          \Big] \Big>  \Bigg\} \nonumber \\
 &  \stackrel{\Gamma = 0}{=}  \frac{1}{2}  \left(  1 - \overrightarrow{\mathcal{P}}_{o \leftrightarrow o'} \right) \Bigg\{    \mathrm{Tr}_{m, \sigma} \Big( s_{o x} \cdot T^o_{o'} \cdot l_{o' y} \cdot \rho^{o'}_o    +   s_{o' y} \cdot T_o^{o'} \cdot  l_{o x} \cdot \rho_{o'}^o     \nonumber \\
& \quad \quad \quad \quad \quad \quad \quad \quad \quad \quad  \quad  \quad +     l_{o x} \cdot T^o_{o'} \cdot s_{o' y} \cdot \rho^{o'}_o    +   l_{o' y} \cdot T_o^{o'} \cdot  s_{o x} \cdot \rho_{o'}^o        \Big)  \nonumber \\
& \quad  +     \frac{1}{\beta}  \sum_{\omega} \mathrm{e}^{\mathrm{i} \omega 0^+} \mathrm{Tr}_{m, \sigma} \Bigg[ s_{o x}  \cdot \left[ G(\mathrm{i} \omega ) \cdot T \right]^o_{o'} \cdot l_{o' y} \cdot \left[ G(\mathrm{i} \omega ) \cdot T \right]_o^{o'}  \nonumber \\
& \quad  -   s_{o x}  \cdot   G(\mathrm{i} \omega )^o_{o'} \cdot l_{o' y} \cdot \left[ T \cdot G(\mathrm{i} \omega ) \cdot T \right]_o^{o'}      -  s_{o  x}  \cdot  \left[ T \cdot G(\mathrm{i} \omega ) \cdot T \right]^o_{o'} \cdot l_{o' y} \cdot  G(\mathrm{i} \omega )_o^{o'}  \nonumber \\
&   \quad + s_{o x}  \cdot \left[ T \cdot G(\mathrm{i} \omega )  \right]^o_{o'} \cdot   l_{o' y} \cdot \left[ T \cdot G(\mathrm{i} \omega )  \right]_o^{o'}    
   +      l_{o x}  \cdot \left[ G(\mathrm{i} \omega ) \cdot T \right]^o_{o'} \cdot s_{o' y} \cdot \left[ G(\mathrm{i} \omega ) \cdot T \right]_o^{o'}  \nonumber \\
&   \quad  -   l_{o x}  \cdot   G(\mathrm{i} \omega )^o_{o'} \cdot s_{o' y} \cdot \left[ T \cdot G(\mathrm{i} \omega ) \cdot T \right]_o^{o'}      -   l_{o  x}  \cdot  \left[ T \cdot G(\mathrm{i} \omega ) \cdot T \right]^o_{o'} \cdot s_{o' y} \cdot  G(\mathrm{i} \omega )_o^{o'}  \nonumber \\
&   \quad + l_{o x}  \cdot \left[ T \cdot G(\mathrm{i} \omega )  \right]^o_{o'} \cdot s_{o' y} \cdot \left[ T \cdot G(\mathrm{i} \omega )  \right]_o^{o'}   \Bigg]   \Bigg\} .
\end{align}

\subsection{Symmetric out-of-diagonal interactions}

From Eqs.\eqref{C nonlocal} and \eqref{C local}, we see that $\mathcal{C}_{o o'}^{\alpha} \equiv \left( \mathcal{C}_{o o'}^{\alpha} \right)^{\mathrm{spin}}   +   \left( \mathcal{C}_{o o'}^{\alpha} \right)^{\mathrm{orb}}$ for $\alpha = x$ or $y$, while $\mathcal{C}_{o o'}^{z} \equiv \left( \mathcal{C}_{o o'}^{z} \right)^{\mathrm{spin-spin}}   +   \left( \mathcal{C}_{o o'}^{z} \right)^{\mathrm{orb-orb}} + \left( \mathcal{C}_{o o'}^{z} \right)^{\mathrm{spin-orb}}$, where, in the case of $o \neq o'$,
\begin{align}
            \left(  \mathcal{C}_{o o'}^{x}  \right)^{\mathrm{spin}}         =  &  \frac{ \mathrm{i} }{2 }    \left(  1 + \overrightarrow{\mathcal{P}}_{o \leftrightarrow o'} \right)    \mathrm{Tr}_{\sigma}  \Big[ s_{o x} \cdot  \mathrm{Tr}_{m}  \left( \rho^o_{o'} \cdot T^{o'}_o - T^o_{o'} \cdot \rho^{o'}_o    \right)    \Big]    ,  
\end{align}
\begin{align}
            \left(  \mathcal{C}_{o o'}^{x}  \right)^{\mathrm{orb}}         =  &  \frac{ \mathrm{i} }{2 }   \left(  1 + \overrightarrow{\mathcal{P}}_{o \leftrightarrow o'} \right)     \mathrm{Tr}_{m}  \Big[ l_{o x} \cdot  \mathrm{Tr}_{\sigma}  \left( \rho^o_{o'} \cdot T^{o'}_o - T^o_{o'} \cdot \rho^{o'}_o    \right)    \Big]    ,  
\end{align}
\begin{align}
            \left(  \mathcal{C}_{o o'}^{y}  \right)^{\mathrm{spin}}         =  & - \frac{ \mathrm{i} }{2 }  \left(  1 + \overrightarrow{\mathcal{P}}_{o \leftrightarrow o'} \right)      \mathrm{Tr}_{\sigma}  \Big[ s_{o y} \cdot  \mathrm{Tr}_{m}  \left( \rho^o_{o'} \cdot T^{o'}_o - T^o_{o'} \cdot \rho^{o'}_o    \right)      \Big]    ,  
\end{align}
\begin{align}
            \left(  \mathcal{C}_{o o'}^{y}  \right)^{\mathrm{orb}}         =  &  - \frac{ \mathrm{i} }{2 }   \left(  1 + \overrightarrow{\mathcal{P}}_{o \leftrightarrow o'} \right)    \mathrm{Tr}_{m}  \Big[ l_{o y} \cdot  \mathrm{Tr}_{\sigma}  \left( \rho^o_{o'} \cdot T^{o'}_o - T^o_{o'} \cdot \rho^{o'}_o    \right)     \Big]    ,  
\end{align}
\begin{align}
&  \left( \mathcal{C}_{o o'}^z \right)^{\mathrm{spin-spin}} \nonumber \\
&  =   -  \frac{1}{2}  \left( 1 + \overrightarrow{\mathcal{P}}_{o \leftrightarrow o'} \right) \Bigg\{  \mathrm{Tr}_{m, \sigma} \Big( s_{o x} \cdot T^o_{o'} \cdot s_{o' y} \cdot \rho^{o'}_o    +   s_{o' y} \cdot T_o^{o'} \cdot  s_{o x} \cdot \rho_{o'}^o    \Big) \nonumber \\
& \quad  +    \beta \mathcal{V}^{\mathrm{spin}}_{o x} \mathcal{V}^{\mathrm{spin}}_{o' y}          - \frac{1}{  \beta} \int_0^{\beta} \mathrm{d} \tau \int_0^{\beta} \mathrm{d} \tau'  \left< \mathcal{T}_{\gamma}  \hat{\mathcal{V}}^{\mathrm{spin}}_{o x}(\tau) \, \hat{\mathcal{V}}^{\mathrm{spin}}_{o' y}(\tau')       \right>  \Bigg\}   \nonumber \\
&  \stackrel{\Gamma = 0}{=}   -  \frac{1 }{2}   \left( 1 + \overrightarrow{\mathcal{P}}_{o \leftrightarrow o'} \right) \Bigg\{ \mathrm{Tr}_{m, \sigma} \Big( s_{o x} \cdot T^o_{o'} \cdot s_{o' y} \cdot \rho^{o'}_o    +   s_{o' y} \cdot T_o^{o'} \cdot  s_{o x} \cdot \rho_{o'}^o    \Big) \nonumber \\
& \quad  + \frac{1  }{  \beta}  \sum_{\omega} \mathrm{e}^{\mathrm{i} \omega 0^+} \mathrm{Tr}_{m, \sigma} \Bigg[ s_{o x}  \cdot \left[ G(\mathrm{i} \omega ) \cdot T \right]^o_{o'} \cdot s_{o' y} \cdot \left[ G(\mathrm{i} \omega ) \cdot T \right]_o^{o'}  \nonumber \\
&  \quad    -   s_{o x}  \cdot   G(\mathrm{i} \omega )^o_{o'} \cdot s_{o' y} \cdot \left[ T \cdot G(\mathrm{i} \omega ) \cdot T \right]_o^{o'}      -   s_{o x}  \cdot  \left[ T \cdot G(\mathrm{i} \omega ) \cdot T \right]^o_{o'} \cdot s_{o' y} \cdot  G(\mathrm{i} \omega )_o^{o'}  \nonumber \\
&  \quad   + s_{o x}  \cdot \left[ T \cdot G(\mathrm{i} \omega )  \right]^o_{o'} \cdot s_{o' y} \cdot \left[ T \cdot G(\mathrm{i} \omega )  \right]_o^{o'}     \Bigg] \Bigg\} ,
\end{align}
\begin{align}
&  \left( \mathcal{C}_{o o'}^z \right)^{\mathrm{orb-orb}}  \nonumber \\
& =  - \frac{1}{2} \left( 1 + \overrightarrow{\mathcal{P}}_{o \leftrightarrow o'} \right)  \Bigg\{ \mathrm{Tr}_{m, \sigma} \Big( l_{o x} \cdot T^o_{o'} \cdot l_{o' y} \cdot \rho^{o'}_o    +   l_{o' y} \cdot T_o^{o'} \cdot  l_{o x} \cdot \rho_{o'}^o     \Big)  \nonumber \\
& \quad + \beta  \mathcal{V}^{\mathrm{orb}}_{o x} \mathcal{V}^{\mathrm{orb}}_{o' y}     - \frac{1}{  \beta} \int_0^{\beta} \mathrm{d} \tau \int_0^{\beta} \mathrm{d} \tau'  \left< \mathcal{T}_{\gamma}  \hat{\mathcal{V}}^{\mathrm{orb}}_{o x}(\tau) \, \hat{\mathcal{V}}^{\mathrm{orb}}_{o' y}(\tau')          \right>   \Bigg\} \nonumber \\
&  \stackrel{\Gamma = 0}{=}   -  \frac{1 }{2} \left(  1 + \overrightarrow{\mathcal{P}}_{o \leftrightarrow o'} \right)   \Bigg\{  \mathrm{Tr}_{m, \sigma} \Big( l_{o x} \cdot T^o_{o'} \cdot l_{o' y} \cdot \rho^{o'}_o    +   l_{o' y} \cdot T_o^{o'} \cdot  l_{o x} \cdot \rho_{o'}^o    \Big) \nonumber \\
& \quad  + \frac{1  }{  \beta}  \sum_{\omega} \mathrm{e}^{\mathrm{i} \omega 0^+} \mathrm{Tr}_{m, \sigma} \Bigg[  l_{o x}  \cdot \left[ G(\mathrm{i} \omega ) \cdot T \right]^o_{o'} \cdot l_{o' y} \cdot \left[ G(\mathrm{i} \omega ) \cdot T \right]_o^{o'}  \nonumber \\
&  \quad    -   l_{o x}  \cdot   G(\mathrm{i} \omega )^o_{o'} \cdot l_{o' y} \cdot \left[ T \cdot G(\mathrm{i} \omega ) \cdot T \right]_o^{o'}      -   l_{o x}  \cdot  \left[ T \cdot G(\mathrm{i} \omega ) \cdot T \right]^o_{o'} \cdot l_{o' y} \cdot  G(\mathrm{i} \omega )_o^{o'}  \nonumber \\
&  \quad  + l_{o x}  \cdot \left[ T \cdot G(\mathrm{i} \omega )  \right]^o_{o'} \cdot l_{o' y} \cdot \left[ T \cdot G(\mathrm{i} \omega )  \right]_o^{o'}     \Bigg] \Bigg\} ,
\end{align}
\begin{align}
&  \left( \mathcal{C}_{o o'}^z \right)^{\mathrm{spin-orb}}  \nonumber \\
& =  - \frac{1}{2}  \left(  1 + \overrightarrow{\mathcal{P}}_{o \leftrightarrow o'} \right) \Bigg\{    \mathrm{Tr}_{m, \sigma} \Big( s_{o x} \cdot T^o_{o'} \cdot l_{o' y} \cdot \rho^{o'}_o    +   s_{o' y} \cdot T_o^{o'} \cdot  l_{o x} \cdot \rho_{o'}^o     \nonumber \\
& \quad \quad \quad \quad \quad \quad \quad \quad \quad \quad  \quad  \quad +     l_{o x} \cdot T^o_{o'} \cdot s_{o' y} \cdot \rho^{o'}_o    +   l_{o' y} \cdot T_o^{o'} \cdot  s_{o x} \cdot \rho_{o'}^o        \Big)  \nonumber \\
& \quad  +     \beta \mathcal{V}^{\mathrm{spin}}_{o x} \mathcal{V}^{\mathrm{orb}}_{o' y}     + \beta   \mathcal{V}^{\mathrm{orb}}_{o x} \mathcal{V}^{\mathrm{spin}}_{o' y}     \nonumber \\
& \quad  -     \frac{1}{  \beta} \int_0^{\beta} \mathrm{d} \tau \int_0^{\beta} \mathrm{d} \tau'  \Big< \mathcal{T}_{\gamma} \Big[  \hat{\mathcal{V}}^{\mathrm{spin}}_{o x}(\tau) \, \hat{\mathcal{V}}^{\mathrm{orb}}_{o' y}(\tau')   + \hat{\mathcal{V}}^{\mathrm{orb}}_{o x}(\tau) \, \hat{\mathcal{V}}^{\mathrm{spin}}_{o' y}(\tau')          \Big] \Big>  \Bigg\} \nonumber \\
 &  \stackrel{\Gamma = 0}{=} - \frac{1}{2}  \left(  1 + \overrightarrow{\mathcal{P}}_{o \leftrightarrow o'} \right) \Bigg\{    \mathrm{Tr}_{m, \sigma} \Big( s_{o x} \cdot T^o_{o'} \cdot l_{o' y} \cdot \rho^{o'}_o    +   s_{o' y} \cdot T_o^{o'} \cdot  l_{o x} \cdot \rho_{o'}^o     \nonumber \\
& \quad \quad \quad \quad \quad \quad \quad \quad \quad \quad  \quad  \quad +     l_{o x} \cdot T^o_{o'} \cdot s_{o' y} \cdot \rho^{o'}_o    +   l_{o' y} \cdot T_o^{o'} \cdot  s_{o x} \cdot \rho_{o'}^o        \Big)  \nonumber \\
& \quad  +     \frac{1}{\beta}  \sum_{\omega} \mathrm{e}^{\mathrm{i} \omega 0^+} \mathrm{Tr}_{m, \sigma} \Bigg[ s_{o x}  \cdot \left[ G(\mathrm{i} \omega ) \cdot T \right]^o_{o'} \cdot l_{o' y} \cdot \left[ G(\mathrm{i} \omega ) \cdot T \right]_o^{o'}  \nonumber \\
& \quad  -   s_{o x}  \cdot   G(\mathrm{i} \omega )^o_{o'} \cdot l_{o' y} \cdot \left[ T \cdot G(\mathrm{i} \omega ) \cdot T \right]_o^{o'}      -  s_{o  x}  \cdot  \left[ T \cdot G(\mathrm{i} \omega ) \cdot T \right]^o_{o'} \cdot l_{o' y} \cdot  G(\mathrm{i} \omega )_o^{o'}  \nonumber \\
&   \quad + s_{o x}  \cdot \left[ T \cdot G(\mathrm{i} \omega )  \right]^o_{o'} \cdot   l_{o' y} \cdot \left[ T \cdot G(\mathrm{i} \omega )  \right]_o^{o'}    
   +      l_{o x}  \cdot \left[ G(\mathrm{i} \omega ) \cdot T \right]^o_{o'} \cdot s_{o' y} \cdot \left[ G(\mathrm{i} \omega ) \cdot T \right]_o^{o'}  \nonumber \\
&   \quad  -   l_{o x}  \cdot   G(\mathrm{i} \omega )^o_{o'} \cdot s_{o' y} \cdot \left[ T \cdot G(\mathrm{i} \omega ) \cdot T \right]_o^{o'}      -   l_{o  x}  \cdot  \left[ T \cdot G(\mathrm{i} \omega ) \cdot T \right]^o_{o'} \cdot s_{o' y} \cdot  G(\mathrm{i} \omega )_o^{o'}  \nonumber \\
&   \quad + l_{o x}  \cdot \left[ T \cdot G(\mathrm{i} \omega )  \right]^o_{o'} \cdot s_{o' y} \cdot \left[ T \cdot G(\mathrm{i} \omega )  \right]_o^{o'}   \Bigg]   \Bigg\} .
\end{align}
In the case of $o = o'$, we have
\begin{align}
& \left( \mathcal{C}_{o o}^x \right)^{\mathrm{spin}} =   \mathrm{i}     \,  \mathrm{Tr}_{\sigma}  \Big[ s_{o x}  \cdot \mathrm{Tr}_m \left( \rho^o_o \cdot A^o_o   -      A^o_o \cdot  \rho^o_o     \right)   \Big] ,  
\end{align}
\begin{align}
& \left( \mathcal{C}_{o o}^x \right)^{\mathrm{orb}} =   \mathrm{i}     \,  \mathrm{Tr}_{m}  \Big[ l_{o x}  \cdot \mathrm{Tr}_{\sigma} \left( \rho^o_o \cdot A^o_o   -      A^o_o \cdot  \rho^o_o     \right)   \Big] ,  
\end{align}
\begin{align}
& \left( \mathcal{C}_{o o}^y \right)^{\mathrm{spin}} = -   \mathrm{i}     \,  \mathrm{Tr}_{\sigma}  \Big[ s_{o y}  \cdot \mathrm{Tr}_m \left( \rho^o_o \cdot A^o_o   -      A^o_o \cdot  \rho^o_o     \right)   \Big] ,  
\end{align}
\begin{align}
& \left( \mathcal{C}_{o o}^y \right)^{\mathrm{orb}} =  - \mathrm{i}     \,  \mathrm{Tr}_{m}  \Big[ l_{o y}  \cdot \mathrm{Tr}_{\sigma} \left( \rho^o_o \cdot A^o_o   -      A^o_o \cdot  \rho^o_o     \right)   \Big] ,  
\end{align}
\begin{align}
&   \left( \mathcal{C}_{o o}^z \right)^{\mathrm{spin-spin}}    =    -   \mathrm{Tr}_{m, \sigma} \Big( s_{o x} \cdot T^o_{o} \cdot s_{o y} \cdot \rho^{o}_o    +   s_{o y} \cdot T_o^{o} \cdot  s_{o x} \cdot \rho_{o}^o   \Big) \nonumber \\
& \quad \quad \quad \quad \quad \quad  \quad -  \beta      \mathcal{V}^{\mathrm{spin}}_{o x} \mathcal{V}^{\mathrm{spin}}_{o y}      + \frac{1}{  \beta} \int_0^{\beta} \mathrm{d} \tau \int_0^{\beta} \mathrm{d} \tau'  \left< \mathcal{T}_{\gamma}  \hat{\mathcal{V}}^{\mathrm{spin}}_{o x}(\tau) \, \hat{\mathcal{V}}^{\mathrm{spin}}_{o y}(\tau')    \right>  \nonumber \\
 &  \stackrel{\Gamma = 0}{=}    -   \mathrm{Tr}_{m, \sigma} \Big( s_{o x} \cdot T^o_{o} \cdot s_{o y} \cdot \rho^{o}_o    +   s_{o y} \cdot T_o^{o} \cdot  s_{o x} \cdot \rho_{o}^o   \Big) \nonumber \\
& \quad  - \frac{1}{\beta}  \sum_{\omega} \mathrm{e}^{\mathrm{i} \omega 0^+} \mathrm{Tr}_{m, \sigma} \Bigg[ s_{o x}  \cdot \left[ G(\mathrm{i} \omega ) \cdot T \right]^o_{o} \cdot  s_{o y} \cdot \left[ G(\mathrm{i} \omega ) \cdot T \right]_o^{o}  \nonumber \\
& \quad   -  s_{o x}  \cdot   G(\mathrm{i} \omega )^o_{o} \cdot s_{o y} \cdot \left[ T \cdot G(\mathrm{i} \omega ) \cdot T \right]_o^{o}     -   s_{o x}  \cdot  \left[ T \cdot G(\mathrm{i} \omega ) \cdot T \right]^o_{o} \cdot s_{o y} \cdot  G(\mathrm{i} \omega )_o^{o}  \nonumber \\
& \quad   + s_{o x}  \cdot \left[ T \cdot G(\mathrm{i} \omega )  \right]^o_{o} \cdot s_{o  y} \cdot \left[ T \cdot G(\mathrm{i} \omega )  \right]_o^{o}   \Bigg] ,
\end{align}
\begin{align}
&   \left( \mathcal{C}_{o o}^z \right)^{\mathrm{orb-orb}}   =      -   \mathrm{Tr}_{m, \sigma}\! \left( l_{o x} \! \cdot T^o_{o} \! \cdot l_{o y} \! \cdot \rho^{o}_o    +   l_{o y} \! \cdot T_o^{o} \! \cdot  l_{o x} \! \cdot \rho_{o}^o    
-   \frac{1}{2}     \left\{ l_{o x} ; l_{o y} \right\} \! \cdot \!  \left\{ \rho ; T \right\}^o_o\right) \nonumber \\
& \quad \quad  \quad \quad \quad  \quad  - \beta   \mathcal{V}^{\mathrm{orb}}_{o x} \mathcal{V}^{\mathrm{orb}}_{o y}     + \frac{1}{  \beta} \int_0^{\beta} \mathrm{d} \tau \int_0^{\beta} \mathrm{d} \tau'  \left< \mathcal{T}_{\gamma}  \hat{\mathcal{V}}^{\mathrm{orb}}_{o x}(\tau) \, \hat{\mathcal{V}}^{\mathrm{orb}}_{o y}(\tau')    \right>  \nonumber \\
&  \stackrel{\Gamma = 0}{=}      -   \mathrm{Tr}_{m, \sigma} \left( l_{o x} \cdot T^o_{o} \cdot l_{o y} \cdot \rho^{o}_o    +   l_{o y} \cdot T_o^{o} \cdot  l_{o x} \cdot \rho_{o}^o    
-   \frac{1}{2}    \left\{ l_{o x} ; l_{o y} \right\}  \cdot   \left\{ \rho ; T \right\}^o_o\right) \nonumber \\
& \quad - \frac{1}{\beta}  \sum_{\omega} \mathrm{e}^{\mathrm{i} \omega 0^+} \mathrm{Tr}_{m, \sigma} \Bigg[ l_{o x}  \cdot \left[ G(\mathrm{i} \omega ) \cdot T \right]^o_{o} \cdot l_{o y} \cdot \left[ G(\mathrm{i} \omega ) \cdot T \right]_o^{o}  \nonumber \\
& \quad   -   l_{o x}  \cdot   G(\mathrm{i} \omega )^o_{o } \cdot  l_{o y} \cdot \left[ T \cdot G(\mathrm{i} \omega ) \cdot T \right]_o^{o }    
  -  l_{o x}  \cdot  \left[ T \cdot G(\mathrm{i} \omega ) \cdot T \right]^o_{o} \cdot l_{o y} \cdot  G(\mathrm{i} \omega )_o^{o }  \nonumber \\
& \quad   +    l_{o x}  \cdot \left[ T \cdot G(\mathrm{i} \omega )  \right]^o_{o} \cdot l_{o  y} \cdot \left[ T \cdot G(\mathrm{i} \omega )  \right]_o^{o}   \Bigg]   ,
\end{align}
\begin{align}
&  \left( \mathcal{C}_{o o}^z \right)^{\mathrm{spin-orb}}  \nonumber \\
& = -  \mathrm{Tr}_{m, \sigma} \Big( s_{o x} \cdot T^o_{o} \cdot l_{o y} \cdot \rho^{o}_o    +   s_{o y} \cdot T_o^{o} \cdot  l_{o x} \cdot \rho_{o}^o     \nonumber \\
& \quad \quad \quad \quad \quad \quad +     l_{o x} \cdot T^o_{o} \cdot s_{o y} \cdot \rho^{o}_o    +   l_{o y} \cdot T_o^{o} \cdot  s_{o x} \cdot \rho_{o}^o       \Big)  \nonumber \\
& \quad   +   \frac{1}{2}  \mathrm{Tr}_{m, \sigma} \Big[ \Big(    \left\{ s_{o x} ; l_{o y}  \right\}   + \left\{  s_{o y} ; l_{o x} \right\}       \Big) \cdot \left\{ \rho ; T \right\}^o_o\Big]    \nonumber \\
& \quad  -  \beta   \left( \mathcal{V}^{\mathrm{spin}}_{o x} \mathcal{V}^{\mathrm{orb}}_{o y}     +   \mathcal{V}^{\mathrm{orb}}_{o x} \mathcal{V}^{\mathrm{spin}}_{o y}       \right)     \nonumber \\
& \quad  + \frac{1}{  \beta} \int_0^{\beta} \mathrm{d} \tau \int_0^{\beta} \mathrm{d} \tau'  \Big< \mathcal{T}_{\gamma} \Big[  \hat{\mathcal{V}}^{\mathrm{spin}}_{o x}(\tau) \, \hat{\mathcal{V}}^{\mathrm{orb}}_{o y}(\tau')   + \hat{\mathcal{V}}^{\mathrm{orb}}_{o x}(\tau) \, \hat{\mathcal{V}}^{\mathrm{spin}}_{o y}(\tau')         \Big] \Big>   \nonumber \\
& \stackrel{\Gamma = 0}{=} -  \mathrm{Tr}_{m, \sigma} \Big( s_{o x} \cdot T^o_{o} \cdot l_{o y} \cdot \rho^{o}_o    +   s_{o y} \cdot T_o^{o} \cdot  l_{o x} \cdot \rho_{o}^o     \nonumber \\
& \quad \quad \quad \quad \quad \quad +     l_{o x} \cdot T^o_{o} \cdot s_{o y} \cdot \rho^{o}_o    +   l_{o y} \cdot T_o^{o} \cdot  s_{o x} \cdot \rho_{o}^o       \Big)  \nonumber \\
& \quad   +   \frac{1}{2}  \mathrm{Tr}_{m, \sigma} \Big[ \Big(    \left\{ s_{o x} ; l_{o y}  \right\}   + \left\{  s_{o y} ; l_{o x} \right\}       \Big) \cdot \left\{ \rho ; T \right\}^o_o\Big]    \nonumber \\
& \quad   -  \frac{1}{\beta}  \sum_{\omega} \mathrm{e}^{\mathrm{i} \omega 0^+} \mathrm{Tr}_{m, \sigma} \Bigg[ s_{o x}  \cdot \left[ G(\mathrm{i} \omega ) \cdot T \right]^o_{o'} \cdot l_{o' y} \cdot \left[ G(\mathrm{i} \omega ) \cdot T \right]_o^{o'}  \nonumber \\
& \quad    -  s_{o x}  \cdot   G(\mathrm{i} \omega )^o_{o'} \cdot l_{o' y} \cdot \left[ T \cdot G(\mathrm{i} \omega ) \cdot T \right]_o^{o'}   
  -   s_{o x}  \cdot  \left[ T \cdot G(\mathrm{i} \omega ) \cdot T \right]^o_{o'} \cdot l_{o' y} \cdot  G(\mathrm{i} \omega )_o^{o'}  \nonumber \\
& \quad   + s_{o x}  \cdot \left[ T \cdot G(\mathrm{i} \omega )  \right]^o_{o'} \cdot l_{o' y} \cdot \left[ T \cdot G(\mathrm{i} \omega )  \right]_o^{o'}   
+  l_{o x}  \cdot \left[ G(\mathrm{i} \omega ) \cdot T \right]^o_{o'} \cdot s_{o' y} \cdot \left[ G(\mathrm{i} \omega ) \cdot T \right]_o^{o'}  \nonumber \\
& \quad    -   l_{o x}  \cdot   G(\mathrm{i} \omega )^o_{o'} \cdot s_{o' y} \cdot \left[ T \cdot G(\mathrm{i} \omega ) \cdot T \right]_o^{o'}   
  -   l_{o x}  \cdot  \left[ T \cdot G(\mathrm{i} \omega ) \cdot T \right]^o_{o'} \cdot s_{o' y} \cdot  G(\mathrm{i} \omega )_o^{o'}  \nonumber \\
& \quad   + l_{o x}  \cdot \left[ T \cdot G(\mathrm{i} \omega )  \right]^o_{o'} \cdot s_{o' y} \cdot \left[ T \cdot G(\mathrm{i} \omega )  \right]_o^{o'}   \Bigg] .
\end{align}

\subsection{Exchange interactions}
\label{exch sep}

From Eqs.\eqref{J nonlocal} and \eqref{J local}, we see that $\mathcal{J}_{o o'}^{\alpha} \equiv \left( \mathcal{J}_{o o'}^{\alpha} \right)^{\mathrm{spin-spin}}   +   \left( \mathcal{J}_{o o'}^{\alpha} \right)^{\mathrm{orb-orb}} + \left( \mathcal{J}_{o o'}^{\alpha} \right)^{\mathrm{spin-orb}}$ for all $\alpha = x, y, z$. In the case of $o \neq o'$, we obtain 
\begin{align}
&  \left(   \mathcal{J}^x_{o o'}  \right)^{\mathrm{spin-spin}}    =   \mathrm{Tr}_{m, \sigma} \left( s_{o y} \cdot T^o_{o'} \cdot s_{o' y} \cdot \rho^{o'}_o    +   s_{o' y} \cdot T_o^{o'} \cdot  s_{o y} \cdot \rho_{o'}^o  \right)  \nonumber \\
& \quad  \quad \quad \quad \quad \quad \quad  + \beta \mathcal{V}^{\mathrm{spin}}_{o y} \mathcal{V}^{\mathrm{spin}}_{o' y}    - \frac{1}{\beta} \int_0^{\beta} \mathrm{d} \tau \int_0^{\beta} \mathrm{d} \tau'  \left< \mathcal{T}_{\gamma} \hat{\mathcal{V}}^{\mathrm{spin}}_{o y}(\tau) \, \hat{\mathcal{V}}^{\mathrm{spin}}_{o' y}(\tau') \right>  \nonumber \\
& \stackrel{\Gamma = 0}{=}     \mathrm{Tr}_{m, \sigma} \left( s_{o y} \cdot T^o_{o'} \cdot s_{o' y} \cdot \rho^{o'}_o    +   s_{o' y} \cdot T_o^{o'} \cdot  s_{o y} \cdot \rho_{o'}^o  \right)  \nonumber \\              
& \quad    + \frac{1}{\beta}  \sum_{\omega} \mathrm{e}^{\mathrm{i} \omega 0^+} \mathrm{Tr}_{m, \sigma} \Bigg[ s_{o y}  \cdot \left[ G(\mathrm{i} \omega ) \cdot T \right]^o_{o'} \cdot s_{o' y} \cdot \left[ G(\mathrm{i} \omega ) \cdot T \right]_o^{o'}  \nonumber \\
& \quad   -   s_{o y}  \cdot   G(\mathrm{i} \omega )^o_{o'} \cdot s_{o' y} \cdot \left[ T \cdot G(\mathrm{i} \omega ) \cdot T \right]_o^{o'}    
  -   s_{o y}  \cdot  \left[ T \cdot G(\mathrm{i} \omega ) \cdot T \right]^o_{o'} \cdot s_{o' y} \cdot  G(\mathrm{i} \omega )_o^{o'}  \nonumber \\
& \quad  + s_{o y}  \cdot \left[ T \cdot G(\mathrm{i} \omega )  \right]^o_{o'} \cdot s_{o' y} \cdot \left[ T \cdot G(\mathrm{i} \omega )  \right]_o^{o'}   \Bigg]   ,
\end{align}
\begin{align}
& \left(  \mathcal{J}^{x}_{o o'} \right)^{\mathrm{orb-orb}}  =   \mathrm{Tr}_{m, \sigma} \left( l_{o y} \cdot T^o_{o'} \cdot l_{o' y} \cdot \rho^{o'}_o    +   l_{o' y} \cdot T_o^{o'} \cdot  l_{o y} \cdot \rho_{o'}^o  \right)  \nonumber \\
&  \quad  \quad \quad \quad \quad \quad \quad+ \beta \mathcal{V}^{\mathrm{orb}}_{o y} \mathcal{V}^{\mathrm{orb}}_{o' y}   - \frac{1}{\beta} \int_0^{\beta} \mathrm{d} \tau \int_0^{\beta} \mathrm{d} \tau'  \left< \mathcal{T}_{\gamma} \hat{\mathcal{V}}^{\mathrm{orb}}_{o y}(\tau) \, \hat{\mathcal{V}}^{\mathrm{orb}}_{o' y}(\tau') \right> \nonumber \\
& \stackrel{\Gamma = 0}{=}     \mathrm{Tr}_{m, \sigma} \left( l_{o y} \cdot T^o_{o'} \cdot l_{o' y} \cdot \rho^{o'}_o    +   l_{o' y} \cdot T_o^{o'} \cdot  l_{o y} \cdot \rho_{o'}^o  \right)  \nonumber \\              
& \quad    + \frac{1}{\beta}  \sum_{\omega} \mathrm{e}^{\mathrm{i} \omega 0^+} \mathrm{Tr}_{m, \sigma} \Bigg[ l_{o y}  \cdot \left[ G(\mathrm{i} \omega ) \cdot T \right]^o_{o'} \cdot l_{o' y} \cdot \left[ G(\mathrm{i} \omega ) \cdot T \right]_o^{o'}  \nonumber \\
& \quad   -   l_{o y}  \cdot   G(\mathrm{i} \omega )^o_{o'} \cdot l_{o' y} \cdot \left[ T \cdot G(\mathrm{i} \omega ) \cdot T \right]_o^{o'}    
  -   l_{o y}  \cdot  \left[ T \cdot G(\mathrm{i} \omega ) \cdot T \right]^o_{o'} \cdot l_{o' y} \cdot  G(\mathrm{i} \omega )_o^{o'}  \nonumber \\
& \quad  + l_{o y}  \cdot \left[ T \cdot G(\mathrm{i} \omega )  \right]^o_{o'} \cdot l_{o' y} \cdot \left[ T \cdot G(\mathrm{i} \omega )  \right]_o^{o'}   \Bigg]   ,
\end{align}
\begin{align}
&    \left(  \mathcal{J}^{x}_{o o'} \right)^{\mathrm{spin-orb}}   =  \left( 1 + \overrightarrow{\mathcal{P}}_{o \leftrightarrow o'} \right) \Bigg\{ \mathrm{Tr}_{m, \sigma} \left( s_{o y} \cdot T^o_{o'} \cdot l_{o' y} \cdot \rho^{o'}_o       +    l_{o y} \cdot T^o_{o'} \cdot s_{o' y} \cdot \rho^{o'}_o       \right)  \nonumber \\
& \quad \quad \quad \quad \quad  \quad \quad +    \beta  \mathcal{V}^{\mathrm{spin}}_{o y} \mathcal{V}^{\mathrm{orb}}_{o' y}      - \frac{1}{\beta} \int_0^{\beta} \! \mathrm{d} \tau \int_0^{\beta} \! \mathrm{d} \tau'  \left< \mathcal{T}_{\gamma}     \hat{\mathcal{V}}^{\mathrm{spin}}_{o y}(\tau) \, \hat{\mathcal{V}}^{\mathrm{orb}}_{o' y}(\tau')       \right>  \Bigg\}   \nonumber \\
& \stackrel{\Gamma = 0}{=}    \left( 1 + \overrightarrow{\mathcal{P}}_{o \leftrightarrow o'} \right) \Bigg\{ \mathrm{Tr}_{m, \sigma} \left( s_{o y} \cdot T^o_{o'} \cdot l_{o' y} \cdot \rho^{o'}_o       +    l_{o y} \cdot T^o_{o'} \cdot s_{o' y} \cdot \rho^{o'}_o       \right)  \nonumber \\
&  \quad + \frac{1}{\beta}  \sum_{\omega} \mathrm{e}^{\mathrm{i} \omega 0^+} \mathrm{Tr}_{m, \sigma} \Bigg[ s_{o y}  \cdot \left[ G(\mathrm{i} \omega ) \cdot T \right]^o_{o'} \cdot l_{o' y} \cdot \left[ G(\mathrm{i} \omega ) \cdot T \right]_o^{o'}  \nonumber \\
& \quad    -   s_{o y}  \cdot   G(\mathrm{i} \omega )^o_{o'} \cdot l_{o' y} \cdot \left[ T \cdot G(\mathrm{i} \omega ) \cdot T \right]_o^{o'}    
  -   s_{o y}  \cdot  \left[ T \cdot G(\mathrm{i} \omega ) \cdot T \right]^o_{o'} \cdot l_{o' y} \cdot  G(\mathrm{i} \omega )_o^{o'}  \nonumber \\
& \quad   +   s_{o y}  \cdot \left[ T \cdot G(\mathrm{i} \omega )  \right]^o_{o'} \cdot l_{o' y} \cdot \left[ T \cdot G(\mathrm{i} \omega )  \right]_o^{o'}   \Bigg] ,
\end{align}

\begin{align}
&  \left(   \mathcal{J}^y_{o o'}  \right)^{\mathrm{spin-spin}}    =   \mathrm{Tr}_{m, \sigma} \left( s_{o x} \cdot T^o_{o'} \cdot s_{o' x} \cdot \rho^{o'}_o    +   s_{o' x} \cdot T_o^{o'} \cdot  s_{o x} \cdot \rho_{o'}^o  \right)  \nonumber \\
& \quad  \quad \quad \quad \quad \quad \quad  + \beta \mathcal{V}^{\mathrm{spin}}_{o x} \mathcal{V}^{\mathrm{spin}}_{o' x}    - \frac{1}{\beta} \int_0^{\beta} \mathrm{d} \tau \int_0^{\beta} \mathrm{d} \tau'  \left< \mathcal{T}_{\gamma} \hat{\mathcal{V}}^{\mathrm{spin}}_{o x}(\tau) \, \hat{\mathcal{V}}^{\mathrm{spin}}_{o' x}(\tau') \right>  \nonumber \\
& \stackrel{\Gamma = 0}{=}     \mathrm{Tr}_{m, \sigma} \left( s_{o x} \cdot T^o_{o'} \cdot s_{o' x} \cdot \rho^{o'}_o    +   s_{o' x} \cdot T_o^{o'} \cdot  s_{o x} \cdot \rho_{o'}^o  \right)  \nonumber \\              
& \quad    + \frac{1}{\beta}  \sum_{\omega} \mathrm{e}^{\mathrm{i} \omega 0^+} \mathrm{Tr}_{m, \sigma} \Bigg[ s_{o x}  \cdot \left[ G(\mathrm{i} \omega ) \cdot T \right]^o_{o'} \cdot s_{o' x} \cdot \left[ G(\mathrm{i} \omega ) \cdot T \right]_o^{o'}  \nonumber \\
& \quad   -   s_{o x}  \cdot   G(\mathrm{i} \omega )^o_{o'} \cdot s_{o' x} \cdot \left[ T \cdot G(\mathrm{i} \omega ) \cdot T \right]_o^{o'}    
  -   s_{o x}  \cdot  \left[ T \cdot G(\mathrm{i} \omega ) \cdot T \right]^o_{o'} \cdot s_{o' x} \cdot  G(\mathrm{i} \omega )_o^{o'}  \nonumber \\
& \quad  + s_{o x}  \cdot \left[ T \cdot G(\mathrm{i} \omega )  \right]^o_{o'} \cdot s_{o' x} \cdot \left[ T \cdot G(\mathrm{i} \omega )  \right]_o^{o'}   \Bigg]   ,
\end{align}
\begin{align}
& \left(  \mathcal{J}^{y}_{o o'} \right)^{\mathrm{orb-orb}}  =   \mathrm{Tr}_{m, \sigma} \left( l_{o x} \cdot T^o_{o'} \cdot l_{o' x} \cdot \rho^{o'}_o    +   l_{o' x} \cdot T_o^{o'} \cdot  l_{o x} \cdot \rho_{o'}^o  \right)  \nonumber \\
&  \quad  \quad \quad \quad \quad \quad \quad+ \beta \mathcal{V}^{\mathrm{orb}}_{o x} \mathcal{V}^{\mathrm{orb}}_{o' x}   - \frac{1}{\beta} \int_0^{\beta} \mathrm{d} \tau \int_0^{\beta} \mathrm{d} \tau'  \left< \mathcal{T}_{\gamma} \hat{\mathcal{V}}^{\mathrm{orb}}_{o x}(\tau) \, \hat{\mathcal{V}}^{\mathrm{orb}}_{o' x}(\tau') \right> \nonumber \\
& \stackrel{\Gamma = 0}{=}     \mathrm{Tr}_{m, \sigma} \left( l_{o x} \cdot T^o_{o'} \cdot l_{o' x} \cdot \rho^{o'}_o    +   l_{o' x} \cdot T_o^{o'} \cdot  l_{o x} \cdot \rho_{o'}^o  \right)  \nonumber \\              
& \quad    + \frac{1}{\beta}  \sum_{\omega} \mathrm{e}^{\mathrm{i} \omega 0^+} \mathrm{Tr}_{m, \sigma} \Bigg[ l_{o x}  \cdot \left[ G(\mathrm{i} \omega ) \cdot T \right]^o_{o'} \cdot l_{o' x} \cdot \left[ G(\mathrm{i} \omega ) \cdot T \right]_o^{o'}  \nonumber \\
& \quad   -   l_{o x}  \cdot   G(\mathrm{i} \omega )^o_{o'} \cdot l_{o' x} \cdot \left[ T \cdot G(\mathrm{i} \omega ) \cdot T \right]_o^{o'}    
  -   l_{o x}  \cdot  \left[ T \cdot G(\mathrm{i} \omega ) \cdot T \right]^o_{o'} \cdot l_{o' x} \cdot  G(\mathrm{i} \omega )_o^{o'}  \nonumber \\
& \quad  + l_{o x}  \cdot \left[ T \cdot G(\mathrm{i} \omega )  \right]^o_{o'} \cdot l_{o' x} \cdot \left[ T \cdot G(\mathrm{i} \omega )  \right]_o^{o'}   \Bigg]   ,
\end{align}
\begin{align}
&    \left(  \mathcal{J}^{y}_{o o'} \right)^{\mathrm{spin-orb}}   =  \left( 1 + \overrightarrow{\mathcal{P}}_{o \leftrightarrow o'} \right) \Bigg\{ \mathrm{Tr}_{m, \sigma} \left( s_{o x} \cdot T^o_{o'} \cdot l_{o' x} \cdot \rho^{o'}_o       +    l_{o x} \cdot T^o_{o'} \cdot s_{o' x} \cdot \rho^{o'}_o       \right)  \nonumber \\
& \quad \quad \quad \quad \quad  \quad \quad +    \beta  \mathcal{V}^{\mathrm{spin}}_{o x} \mathcal{V}^{\mathrm{orb}}_{o' x}      - \frac{1}{\beta} \int_0^{\beta} \! \mathrm{d} \tau \int_0^{\beta} \! \mathrm{d} \tau'  \left< \mathcal{T}_{\gamma}     \hat{\mathcal{V}}^{\mathrm{spin}}_{o x}(\tau) \, \hat{\mathcal{V}}^{\mathrm{orb}}_{o' x}(\tau')       \right>  \Bigg\}   \nonumber \\
& \stackrel{\Gamma = 0}{=}    \left( 1 + \overrightarrow{\mathcal{P}}_{o \leftrightarrow o'} \right) \Bigg\{ \mathrm{Tr}_{m, \sigma} \left( s_{o x} \cdot T^o_{o'} \cdot l_{o' x} \cdot \rho^{o'}_o       +    l_{o x} \cdot T^o_{o'} \cdot s_{o' x} \cdot \rho^{o'}_o       \right)  \nonumber \\
&  \quad + \frac{1}{\beta}  \sum_{\omega} \mathrm{e}^{\mathrm{i} \omega 0^+} \mathrm{Tr}_{m, \sigma} \Bigg[ s_{o x}  \cdot \left[ G(\mathrm{i} \omega ) \cdot T \right]^o_{o'} \cdot l_{o' x} \cdot \left[ G(\mathrm{i} \omega ) \cdot T \right]_o^{o'}  \nonumber \\
& \quad    -   s_{o x}  \cdot   G(\mathrm{i} \omega )^o_{o'} \cdot l_{o' x} \cdot \left[ T \cdot G(\mathrm{i} \omega ) \cdot T \right]_o^{o'}    
  -   s_{o x}  \cdot  \left[ T \cdot G(\mathrm{i} \omega ) \cdot T \right]^o_{o'} \cdot l_{o' x} \cdot  G(\mathrm{i} \omega )_o^{o'}  \nonumber \\
& \quad   +   s_{o x}  \cdot \left[ T \cdot G(\mathrm{i} \omega )  \right]^o_{o'} \cdot l_{o' x} \cdot \left[ T \cdot G(\mathrm{i} \omega )  \right]_o^{o'}   \Bigg] ,
\end{align}
and the terms related to $\mathcal{J}^{z}_{o o'}$ are just obtained as the averages of the respective terms related to $\mathcal{J}^{x}_{o o'}$ and $\mathcal{J}^{y}_{o o'}$, according to the relation $\mathcal{J}^{z}_{o o'} = \left( \mathcal{J}^{x}_{o o'} + \mathcal{J}^{y}_{o o'} \right) / 2$.

\subsection{Magnetic field}

To separate the magnetic field as $\boldsymbol{\mathcal{B}}_o \equiv \boldsymbol{\mathcal{B}}_o^{\mathrm{spin}} +  \boldsymbol{\mathcal{B}}_o^{\mathrm{orb}}$, from Eq.\eqref{B} we notice that
\begin{align}
\boldsymbol{\mathcal{B}}_i = \mu_{\mathrm{B}} g_i   \boldsymbol{B}_i   \left< \hat{\boldsymbol{S}}_i \right> \cdot \boldsymbol{u}^z_i = \mu_{\mathrm{B}} g_i   \boldsymbol{B}_i  \,  \mathrm{Tr}_M  \left( \rho^i_i \boldsymbol{S}_i \right) \cdot \boldsymbol{u}^z_i .
\end{align}
Substituting $i \equiv (o S)$, where we recall that the orbital degree of freedom $o = (a, n , l)$, we notice that $\boldsymbol{B}_i \rightarrow \boldsymbol{B}_o \rightarrow \boldsymbol{B}_a$, since the external magnetic field depends only on position, and $\boldsymbol{u}^z_i \rightarrow \boldsymbol{u}^z_o$, since by hypothesis the separation into spin and orbital parts is done under the assumption that the dynamical vectors $\boldsymbol{e}_i$ (and therefore also their initial values $\boldsymbol{u}^z_i$) depend only on the orbital degrees of freedom . We then obtain
\begin{align}
\boldsymbol{\mathcal{B}}_o & = \mu_{\mathrm{B}}  \boldsymbol{B}_a \sum_S   g_{o S}  \, \mathrm{Tr}_M  \left( \rho^{o S}_{o S} \boldsymbol{S}_{o S} \right) \cdot \boldsymbol{u}^z_o \equiv  \mu_{\mathrm{B}}  \boldsymbol{B}_a       \, \mathrm{Tr}_{M, S}  \left( g_{o } \rho^{o }_{o } \boldsymbol{S}_{o } \right) \cdot \boldsymbol{u}^z_o \nonumber \\
& = \mu_{\mathrm{B}}  \boldsymbol{B}_a       \, \mathrm{Tr}_{m, \sigma}  \left( g_{o } \rho^{o }_{o } \boldsymbol{S}_{o } \right) \cdot \boldsymbol{u}^z_o  \equiv  \boldsymbol{\mathcal{B}}_o^{\mathrm{spin}} +  \boldsymbol{\mathcal{B}}_o^{\mathrm{orb}},
\end{align}
where 
\begin{align}
& \boldsymbol{\mathcal{B}}_o^{\mathrm{spin}} \equiv g_{1/2}  \mu_{\mathrm{B}}  \boldsymbol{B}_a       \left[ \mathrm{Tr}_{ \sigma}  \left(  \boldsymbol{s}_{o } \mathrm{Tr}_{m } \rho^{o }_{o } \right)  \cdot \boldsymbol{u}^z_o \right]   , \nonumber \\
& \boldsymbol{\mathcal{B}}_o^{\mathrm{orb}} \equiv   g_l \mu_{\mathrm{B}}  \boldsymbol{B}_a      \left[ \mathrm{Tr}_{m}  \left(  \boldsymbol{l}_{o } \mathrm{Tr}_{ \sigma}  \rho^{o }_{o }  \right) \cdot \boldsymbol{u}^z_o   \right]  ,
\label{B sep}
\end{align}
where $g_{1/2}$ and $g_l$ are the intrinsic-spin and orbital $g$-factors, respectively.

\subsection{Local exchange interactions (diagonal anisotropy)}

From Eqs.\eqref{J local} we obtain
\begin{align}
&  \mathcal{J}_{o o}^{x} -  \mathcal{J}_{o o}^z =  \mathcal{B}^{z}_{o} +  \widetilde{\mathcal{M}}^{o y}_{o y}    + \frac{1}{2} \sum_{o' \neq o} \left(    \mathcal{J}^{x}_{o o'} +  \mathcal{J}^{y}_{o o'}  \right)  ,  \nonumber \\
&  \mathcal{J}_{o o}^{y} -  \mathcal{J}_{o o}^z =  \mathcal{B}^{z}_{o} +  \widetilde{\mathcal{M}}^{o x}_{o x}    + \frac{1}{2} \sum_{o' \neq o} \left(    \mathcal{J}^{x}_{o o'} +  \mathcal{J}^{y}_{o o'}  \right)  ,
\end{align}
where the parameters $\mathcal{B}_o^z$, $\mathcal{J}^{x}_{o o'}$ and $\mathcal{J}^{y}_{o o'}$ for $o \neq o'$ were determined in the previous paragraphs. The additional terms, for $\alpha = x$ or $y$, are written as
\begin{align*}
\widetilde{\mathcal{M}}^{o \alpha}_{o \alpha} = \left( \widetilde{\mathcal{M}}^{o \alpha}_{o \alpha} \right)^{\mathrm{spin-spin}}  + \left( \widetilde{\mathcal{M}}^{o \alpha}_{o \alpha} \right)^{\mathrm{orb-orb}} + \left( \widetilde{\mathcal{M}}^{o \alpha}_{o \alpha} \right)^{\mathrm{spin-orb}} ,  
\end{align*}
where
\begin{align}
& \left( \widetilde{\mathcal{M}}^{o \alpha}_{o \alpha} \right)^{\mathrm{spin-spin}}   \equiv   2 \mathrm{Tr}_{m, \sigma} \left( s_{o \alpha} \cdot T^o_{o} \cdot s_{o \alpha} \cdot \rho^{o}_o       \right)     -   \frac{1}{4}  \mathrm{Tr}_{m, \sigma}    \left\{ \rho ; T \right\}^o_o   \nonumber \\
& \quad \quad\quad\quad\quad\quad\quad \quad  + \beta \left( \mathcal{V}^{\mathrm{spin}}_{o \alpha} \right)^2    - \frac{1}{\beta} \int_0^{\beta} \mathrm{d} \tau \int_0^{\beta} \mathrm{d} \tau'  \left< \mathcal{T}_{\gamma} \hat{\mathcal{V}}^{\mathrm{spin}}_{o \alpha}(\tau) \, \hat{\mathcal{V}}^{\mathrm{spin}}_{o  \alpha}(\tau') \right>  \nonumber \\
&  \stackrel{\Gamma = 0}{=}  2 \mathrm{Tr}_{m, \sigma} \left( s_{o \alpha} \cdot T^o_{o} \cdot s_{o \alpha} \cdot \rho^{o}_o       \right)     -   \frac{1}{4}  \mathrm{Tr}_{m, \sigma}    \left\{ \rho ; T \right\}^o_o   \nonumber \\
& \quad +  \frac{1}{\beta}  \sum_{\omega} \mathrm{e}^{\mathrm{i} \omega 0^+} \mathrm{Tr}_{m, \sigma} \Big\{ s_{o \alpha}  \cdot \left[ G(\mathrm{i} \omega ) \cdot T \right]^o_{o} \cdot s_{o \alpha} \cdot \left[ G(\mathrm{i} \omega ) \cdot T \right]_o^{o}  \nonumber \\
& \quad   -  2 s_{o \alpha}  \cdot   G(\mathrm{i} \omega )^o_{o} \cdot s_{o \alpha} \cdot \left[ T \cdot G(\mathrm{i} \omega ) \cdot T \right]_o^{o}   + s_{o \alpha}  \cdot \left[ T \cdot G(\mathrm{i} \omega )  \right]^o_{o} \cdot s_{o \alpha} \cdot \left[ T \cdot G(\mathrm{i} \omega )  \right]_o^{o}   \Big\} ,
\label{loc ss}
\end{align}
\begin{align}
& \left( \widetilde{\mathcal{M}}^{o \alpha}_{o \alpha} \right)^{\mathrm{orb-orb}}  \equiv   2 \mathrm{Tr}_{m, \sigma} \left( l_{o \alpha} \cdot T^o_{o} \cdot l_{o \alpha} \cdot \rho^{o}_o      \right)       -      \mathrm{Tr}_{m } \Big(   l_{o \alpha} \cdot l_{o \alpha }       \cdot \mathrm{Tr}_{ \sigma} \left\{ \rho ; T \right\}^o_o\Big)   \nonumber \\
& \quad \quad\quad\quad\quad\quad\quad  \, \,   + \beta \left( \mathcal{V}^{\mathrm{orb}}_{o \alpha} \right)^2   - \frac{1}{\beta} \int_0^{\beta} \mathrm{d} \tau \int_0^{\beta} \mathrm{d} \tau'  \left< \mathcal{T}_{\gamma} \hat{\mathcal{V}}^{\mathrm{orb}}_{o \alpha}(\tau) \, \hat{\mathcal{V}}^{\mathrm{orb}}_{o \alpha}(\tau') \right> \nonumber \\
&  \stackrel{\Gamma = 0}{=}   2 \mathrm{Tr}_{m, \sigma} \left( l_{o \alpha} \cdot T^o_{o} \cdot l_{o \alpha} \cdot \rho^{o}_o      \right)       -      \mathrm{Tr}_{m } \Big(   l_{o \alpha} \cdot l_{o \alpha }       \cdot \mathrm{Tr}_{ \sigma} \left\{ \rho ; T \right\}^o_o\Big)   \nonumber \\ \nonumber \\
& \quad \, \, + \frac{1}{\beta}  \sum_{\omega} \mathrm{e}^{\mathrm{i} \omega 0^+} \mathrm{Tr}_{m, \sigma} \Big\{ l_{o \alpha}  \cdot \left[ G(\mathrm{i} \omega ) \cdot T \right]^o_{o} \cdot l_{o \alpha} \cdot \left[ G(\mathrm{i} \omega ) \cdot T \right]_o^{o}  \nonumber \\
& \quad  \, \,    -  2 l_{o \alpha}  \cdot   G(\mathrm{i} \omega )^o_{o} \cdot l_{o \alpha} \cdot \left[ T \cdot G(\mathrm{i} \omega ) \cdot T \right]_o^{o}   
 + l_{o \alpha}  \cdot \left[ T \cdot G(\mathrm{i} \omega )  \right]^o_{o} \cdot l_{o \alpha} \cdot \left[ T \cdot G(\mathrm{i} \omega )  \right]_o^{o}   \Big\} ,
 \label{loc oo}
\end{align}
\begin{align}
& \left( \widetilde{\mathcal{M}}^{o \alpha}_{o \alpha} \right)^{\mathrm{spin-orb}}  
 \equiv    2  \mathrm{Tr}_{m, \sigma} \left( s_{o \alpha} \cdot T^o_{o} \cdot l_{o \alpha} \cdot \rho^{o}_o        +     l_{o \alpha} \cdot T^o_{o} \cdot s_{o \alpha} \cdot \rho^{o}_o      \right)  \nonumber \\
& \quad \quad\quad\quad\quad\quad\quad  \quad  -     \mathrm{Tr}_{m, \sigma} \Big( \left\{ s_{o \alpha} ; l_{o \alpha}      \right\} \cdot \left\{ \rho ; T \right\}^o_o\Big)    \nonumber \\
& \quad \quad\quad\quad\quad\quad\quad  \quad  + 2  \beta   \mathcal{V}^{\mathrm{spin}}_{o \alpha} \mathcal{V}^{\mathrm{orb}}_{o \alpha}       - \frac{2}{\beta} \int_0^{\beta} \mathrm{d} \tau \int_0^{\beta} \mathrm{d} \tau'  \left< \mathcal{T}_{\gamma}   \hat{\mathcal{V}}^{\mathrm{spin}}_{o \alpha}(\tau) \, \hat{\mathcal{V}}^{\mathrm{orb}}_{o \alpha}(\tau')     \right>  \nonumber \\
&  \stackrel{\Gamma = 0}{=} 2  \mathrm{Tr}_{m, \sigma} \left( s_{o \alpha} \cdot T^o_{o} \cdot l_{o \alpha} \cdot \rho^{o}_o        +     l_{o \alpha} \cdot T^o_{o} \cdot s_{o \alpha} \cdot \rho^{o}_o      \right)   
  -     \mathrm{Tr}_{m, \sigma} \Big( \left\{ s_{o \alpha} ; l_{o \alpha}      \right\} \cdot \left\{ \rho ; T \right\}^o_o\Big)    \nonumber \\
& \quad +  \frac{1}{\beta}  \sum_{\omega} \mathrm{e}^{\mathrm{i} \omega 0^+} \mathrm{Tr}_{m, \sigma} \Big\{ s_{o \alpha}  \cdot \left[ G(\mathrm{i} \omega ) \cdot T \right]^o_{o} \cdot l_{o \alpha} \cdot \left[ G(\mathrm{i} \omega ) \cdot T \right]_o^{o}  \nonumber \\
& \quad      -   s_{o \alpha}  \cdot   G(\mathrm{i} \omega )^o_{o} \cdot l_{o \alpha} \cdot \left[ T \cdot G(\mathrm{i} \omega ) \cdot T \right]_o^{o}    
   -   s_{o \alpha}  \cdot  \left[ T \cdot G(\mathrm{i} \omega ) \cdot T \right]^o_{o} \cdot l_{o \alpha} \cdot  G(\mathrm{i} \omega )_o^{o}  \nonumber \\
& \quad     + s_{o \alpha}  \cdot \left[ T \cdot G(\mathrm{i} \omega )  \right]^o_{o} \cdot l_{o \alpha} \cdot \left[ T \cdot G(\mathrm{i} \omega )  \right]_o^{o}   \Big\} .
\label{loc so}
\end{align}
The local exchange interaction parameters can then be written as 
\begin{align*}
\mathcal{J}_{o o}^{\alpha} -  \mathcal{J}_{o o}^z  \equiv  \left( \mathcal{J}_{o o}^{\alpha} -  \mathcal{J}_{o o}^z \right)^{\mathrm{spin-spin}} +
\left( \mathcal{J}_{o o}^{\alpha} -  \mathcal{J}_{o o}^z \right)^{\mathrm{orb-orb}} + \left( \mathcal{J}_{o o}^{\alpha} -  \mathcal{J}_{o o}^z \right)^{\mathrm{spin-orb}}  ,
\end{align*}
where
\begin{align}
\left( \mathcal{J}_{o o}^{\alpha} -  \mathcal{J}_{o o}^z \right)^{\mathrm{spin-spin}} = \left( \mathcal{B}^{z}_{o} \right)^{\mathrm{spin}} \!  + \! \left( \widetilde{\mathcal{M}}^{o \bar{\alpha}}_{o \bar{\alpha}}  \right)^{\mathrm{spin-spin}} \! + \frac{1}{2} \sum_{o' \neq o} \left(    \mathcal{J}^{x}_{o o'} +  \mathcal{J}^{y}_{o o'}  \right)^{\mathrm{spin-spin}} ,
\end{align}
\begin{align}
\left( \mathcal{J}_{o o}^{\alpha} -  \mathcal{J}_{o o}^z \right)^{\mathrm{orb-orb}} = \left( \mathcal{B}^{z}_{o} \right)^{\mathrm{orb}}  +  \left( \widetilde{\mathcal{M}}^{o \bar{\alpha}}_{o \bar{\alpha}}  \right)^{\mathrm{orb-orb}}   + \frac{1}{2} \sum_{o' \neq o} \left(    \mathcal{J}^{x}_{o o'} +  \mathcal{J}^{y}_{o o'}  \right)^{\mathrm{orb-orb}} ,
\end{align}
\begin{align}
\left( \mathcal{J}_{o o}^{\alpha} -  \mathcal{J}_{o o}^z \right)^{\mathrm{spin-orb}} =    \left( \widetilde{\mathcal{M}}^{o \bar{\alpha}}_{o \bar{\alpha}}  \right)^{\mathrm{spin-orb}}   + \frac{1}{2} \sum_{o' \neq o} \left(    \mathcal{J}^{x}_{o o'} +  \mathcal{J}^{y}_{o o'}  \right)^{\mathrm{spin-orb}} ,
\end{align}
where $\left( \alpha, \bar{\alpha} \right) = \left( x, y \right)$ or $\left( y, x \right)$, and the various terms are given by Eqs.\eqref{B sep}, \eqref{loc ss}, \eqref{loc oo}, \eqref{loc so}, and in Section \ref{exch sep}.

\section{Conclusion}
\label{sec: Conclusion}

To conclude, in this work we have established the mapping between a relativistic electronic system with rotationally invariant interactions onto an effective \emph{classical} spin model, via the equivalence of their thermodynamic potentials under rotations of the local total magnetic moments up to the second order in the rotation angles, when the spin configuration of the electrons is symmetry-broken and out of equilibrium. The parameters of the effective spin model were obtained as functionals of the single- and two-electron Green's functions of the electronic system. We have removed two approximations which were adopted in previous works on non-relativistic systems \cite{Katsnelson00, Katsnelson02, Secchi13}, namely: (1) here we take into account the vertices of the two-electron Green's functions, (2) here we include the non-local components of the self-energies. Besides, we have extended the theory in order to completely account for relativistic effects, determining the complete relativistic exchange tensors in a unified framework. For two components ($x$ and $y$) of the Dzyaloshinskii-Moriya vectors, which had already been determined previously \cite{Katsnelson10}, we have recovered the known results and extended them to the non-collinear case; moreover, here we have determined also the third component ($z$), together with the completely new terms describing anisotropic exchange and other out-of-diagonal symmetric terms of the exchange tensors. In the particular case of spin-$1/2$ (single-band Hubbard model) we have recovered the known expressions for the isotropic exchange parameters both in the general case of non-local self-energy \cite{Secchi13} and in the particular case of local self-energy \cite{Katsnelson00, Katsnelson02}. Having included also an external magnetic field, we have shown how it determines a renormalization of the exchange tensor via linear and non-linear contributions (the details can be found in \ref{app: Analysis}). Finally, we have shown how to study separately the orbital and spin-$1/2$ contributions to magnetism, as well as a combined ``spin-orbital'' contribution which cannot be decoupled.

We remark that our theory should be used to predict spin dynamics in a given phase of the electronic system, but it cannot be used to predict the phases of the system themselves. In fact, the application of the theory for computations requires fixing the initial spin configuration in a definite out-of-equilibrium phase. The subsequent classical spin dynamics is then determined by the magnetic interactions given by our theory.

Building on this work, we foresee three main possible paths for further theoretical investigation: (1) study of the response of the thermodynamic potential to higher orders in the rotation angles, (2) extension of the effective spin model to include higher-order spin-spin interactions (the quadratic model considered here is enough for the second-order response in the angles of rotation, but may not be enough for higher orders in the angles), (3) inclusion of time-dependent external electromagnetic fields, which up to now was done only for a non-relativistic system \cite{Secchi13}. This latter extension would be desirable in order to realistically describe the manipulation of magnetism and the ultrafast spin dynamics induced by sub-picosecond laser pulses.

\section*{Acknowledgements}

We acknowledge useful scientific discussions with Vladimir V. Mazurenko, Johan Mentink, Martin Eckstein and Alexander Chudnovskiy. This work is supported by the European Union Seventh Framework Programme under grant agreement No. 281043 (FEMTOSPIN), and by Deutsche Forschungsgemeinschaft under grant SFB-668.

\appendix

\section{Considerations on the rotational invariance of the interaction Hamiltonian}
\label{app: Invariance}

The whole treatment has been based on the assumption that the interaction term is rotationally invariant. We now discuss this issue more in detail. We start by assuming that the interaction is local (on-site), in the spirit of the multi-orbital Hubbard model. However, on a single site it can mix states belonging to different shells and having different angular momenta. The interaction Hamiltonian is
\begin{align}
\hat{H}_V^{\phi} \equiv & \frac{1}{2}   \sum_{1, 2, 3, 4}    \hat{\phi}^{\dagger}_{i_1, M_1}  \hat{\phi}^{\dagger}_{i_2, M_2}    V_{ \lbrace i_3, M_3 \rbrace, \lbrace i_4, M_4 \rbrace}^{\lbrace i_1, M_1 \rbrace, \lbrace i_2, M_2\rbrace} \hat{\phi}^{i_3, M_3}   \hat{\phi}^{i_4, M_4} ,
\end{align}
where
\begin{align}
   V_{ \lbrace i_3, M_3 \rbrace, \lbrace i_4, M_4 \rbrace}^{\lbrace i_1, M_1 \rbrace, \lbrace i_2, M_2\rbrace}   & \equiv \int \text{d} v(\boldsymbol{x}) \int \text{d} v(\boldsymbol{x}')  \, \phi^*_{i_1, M_1}(\boldsymbol{x}) \, \phi^*_{i_2, M_2}(\boldsymbol{x}') \nonumber \\
& \quad \quad \times  V(\boldsymbol{x} - \boldsymbol{x}') \,  \phi^{i_3, M_3}(\boldsymbol{x}') \, \phi^{i_4, M_4}(\boldsymbol{x}) .    
\end{align}
The on-site interaction is \emph{supposed} to be rotationally invariant, i.e., $\hat{H}_V^{\phi} = \hat{H}_V^{\psi}$. To check the conditions under which this condition is fulfilled, we perform the rotation of the fermionic fields according to Eq.\eqref{transformation}, obtaining
\begin{align}
  \hat{H}_V^{\phi}   = &  \frac{1}{2}   \sum_{1, 2, 3, 4}   \hat{\psi}^{\dagger}_{i_1, M_1} \hat{\psi}^{\dagger}_{i_2, M_2} \hat{\psi}^{i_3, M_3}  \hat{\psi}^{i_4, M_4}   \nonumber \\
& \cdot   \sum_{M_5, M_6, M_7, M_8} \!  R^{\dagger}(i_1)_{M_5}^{M_1}   \,  R^{\dagger}(i_2)_{M_6}^{M_2}   \, V_{ \lbrace i_3, M_7 \rbrace, \lbrace i_4, M_8 \rbrace}^{\lbrace i_1, M_5 \rbrace, \lbrace i_2, M_6\rbrace}     R(i_3)_{M_3}^{M_7}  \,  R(i_4)_{M_4}^{M_8}   .
\end{align}
Suppose that the interaction is intra-atomic ($a_1 = a_2 = a_3 = a_4 \equiv a$), as in the Hubbard model, and that 
\begin{align}
V_{a, \lbrace n_3, l_3, S_3, M_7 \rbrace, \lbrace n_4, l_4, S_4, M_8 \rbrace}^{\lbrace n_1, l_1, S_1, M_5 \rbrace, \lbrace n_2, l_2, S_2, M_6\rbrace} = \delta_{M_7}^{M_6} \delta_{M_8}^{M_5} \delta_{S_3}^{S_2} \delta_{S_4}^{S_1} V_{a, \lbrace n_3, l_3, S_3 \rbrace, \lbrace n_4, l_4, S_4 \rbrace}^{\lbrace n_1, l_1, S_1 \rbrace, \lbrace n_2, l_2, S_2 \rbrace} , 
\label{assumption inv 1}
\end{align}
then we obtain:
\begin{align}
\hat{H}_V^{\phi}  = & \frac{1}{2} \sum_a \sum_{\lbrace n \rbrace} \sum_{ \lbrace l \rbrace} \sum_{\lbrace S \rbrace} \sum_{ \lbrace M \rbrace } \hat{\psi}^{\dagger}_{a, n_1, l_1, S_1, M_1} \hat{\psi}^{\dagger}_{a, n_2, l_2, S_2, M_2}      \delta_{S_3}^{S_2} \delta_{S_4}^{S_1}    \nonumber \\
& \cdot V_{a, \lbrace n_3, l_3, S_3 \rbrace, \lbrace n_4, l_4, S_4 \rbrace}^{\lbrace n_1, l_1, S_1 \rbrace, \lbrace n_2, l_2, S_2 \rbrace} \Bigg[ \sum_{M_5}  R^{\dagger}(a, n_1, l_1, S_1)_{M_5}^{M_1}     R(a, n_4, l_4, S_1)_{M_4}^{M_5}       \nonumber \\
&    \cdot \sum_{M_6}     R^{\dagger}(a, n_2, l_2, S_2)_{M_6}^{M_2} R(a, n_3, l_3, S_2)_{M_3}^{M_6} \Bigg] \hat{\psi}^{a, n_3, l_3, S_3, M_3}  \hat{\psi}^{a, n_4, l_4, S_4, M_4} ,
\end{align}
where $\sum_{\lbrace n \rbrace} \equiv \sum_{n_1, n_2, n_3, n_4}$, etcetera. To get rotational invariance of the interaction Hamiltonian we have now two possibilities: either 1) we take the rotations to be only site-dependent (i.e., not resolved with respect to the shells and orbital angular momenta), so that the rotation matrices are independent of $n$ and $l$, or 2) we further assume that the interaction parameter is 
\begin{align}
\propto \delta^{n_1}_{n_4} \delta^{n_2}_{n_3} \delta^{l_1}_{l_4} \delta^{l_2}_{l_3},
\label{assumption inv 2}
\end{align}
which implies that the local Coulomb interaction is spherically symmetric. In both cases, we can perform the summations over $M_5$ and $M_6$, obtaining:
\begin{align}
\hat{H}_V^{\phi}  = & \frac{1}{2} \sum_a \sum_{\lbrace n  \rbrace} \sum_{ \lbrace l  \rbrace} \sum_{\lbrace S  \rbrace} \sum_{ \lbrace M  \rbrace }  \hat{\psi}^{\dagger}_{a, n_1, l_1, S_1, M_1} \hat{\psi}^{\dagger}_{a, n_2, l_2, S_2, M_2}   \nonumber \\
& \cdot   \delta^{M_1}_{M_4}  \delta^{M_2}_{M_3}      \delta_{S_3}^{S_2} \delta_{S_4}^{S_1} V_{a, \lbrace n_3, l_3, S_3 \rbrace, \lbrace n_4, l_4, S_4 \rbrace}^{\lbrace n_1, l_1, S_1 \rbrace, \lbrace n_2, l_2, S_2 \rbrace}    \hat{\psi}^{a, n_3, l_3, S_3, M_3}  \hat{\psi}^{a, n_4, l_4, S_4, M_4} \nonumber \\
= & \frac{1}{2} \sum_a \sum_{n_1, l_1, S_1, M_1} \sum_{n_2, l_2, S_2, M_2} \sum_{n_3, l_3} \sum_{n_4, l_4} \hat{\psi}^{\dagger}_{a, n_1, l_1, S_1, M_1} \hat{\psi}^{\dagger}_{a, n_2, l_2, S_2, M_2}   \nonumber \\
& \cdot    V_{a, \lbrace n_3, l_3, S_2 \rbrace, \lbrace n_4, l_4, S_1 \rbrace}^{\lbrace n_1, l_1, S_1 \rbrace, \lbrace n_2, l_2, S_2 \rbrace}    \hat{\psi}^{a, n_3, l_3, S_2, M_2}  \hat{\psi}^{a, n_4, l_4, S_1, M_1},
\end{align}
which is invariant. Therefore, the assumption \eqref{assumption inv 1} guarantees the invariance of the interaction Hamiltonian under rotations of the magnetic moments \emph{site-resolved} but \emph{not shell-resolved}, while the additional assumption \eqref{assumption inv 2} allows for rotational invariance under \emph{shell-resolved} rotations.

\section{Analysis of the matrix $\widetilde{\mathcal{M}}^{i \alpha}_{i' \alpha'}$}
\label{app: Analysis}

In our theory, the matrix $\widetilde{\mathcal{M}}^{i \alpha}_{i' \alpha'}$ [cfr. Eq.\eqref{tilde M brevity}] is one of the key quantities in terms of which the magnetic parameters are expressed. We now take a closer look at the structure of this matrix, starting with the term $\mathcal{W}^{i \alpha}_{i' \alpha'}$ [cfr. Eq.\eqref{A tilde ii'}].

We see that we can decompose Eq.\eqref{A tilde ii'} in order to separate the parts of $\mathcal{W}^{i \alpha}_{i' \alpha'}$ involving the local single-particle Hamiltonian $T_i^i$ from those involving the non-local part. We put
\begin{align}
\mathcal{W}^{i \alpha}_{i' \alpha'}  \equiv \left(  \mathcal{W}^{i \alpha}_{i' \alpha'}  \right)^{\mathrm{loc}} + \left(  \mathcal{W}^{i \alpha}_{i' \alpha'}  \right)^{\mathrm{loc / nloc}} + \left(  \mathcal{W}^{i \alpha}_{i' \alpha'}  \right)^{\mathrm{nloc}} ,
 \label{decomp A}
\end{align}
where the three parts originate from terms in the summation over $(j, j')$ in Eq.\eqref{A tilde ii'} having different specific values of $j$ and $j'$ in relation with $i$ and $i'$. 
Namely, the local parts are the terms with $j = i$ and $j' = i'$,
\begin{align}
 \left(  \mathcal{W}^{i \alpha}_{i' \alpha'}  \right)^{\mathrm{loc}}   & \equiv      \sum_{M_1 M_2 M_3 M_4}  \!     \left[ S_{i \alpha}  , T^{i}_{i}    \right]^{M_2}_{M_1} \widetilde{\chi}^{(i M_1) (i' M_3)  }_{ (i M_2) (i' M_4)}  \left[ S_{i' \alpha'}  , T^{i'}_{i'}    \right]^{M_4}_{M_3}        ;
    \label{A tilde ii' local}
\end{align}
then, there are terms involving both local and non-local components of the single-particle Hamiltonian, corresponding to $(j = i, j' \neq i')$ or $(j \neq i, j' = i')$,
\begin{align}
& \left( \mathcal{W}^{i \alpha}_{i' \alpha'}  \right)^{\mathrm{loc / nloc}}     \equiv       \sum_{M_1 M_2 M_3 M_4}  \Bigg\{ \nonumber \\
 &   \quad \sum_{j \neq i} \left[ \left( S_{i \alpha}  \cdot T^{i}_{j}   \right)^{M_2}_{M_1}     \widetilde{\chi}^{(j M_1) (i' M_3) }_{ (i M_2)  (i' M_4)}         -        \left( T^{j}_{i} \cdot   S_{i \alpha} \right)^{M_2}_{M_1}    \widetilde{\chi}^{(i M_1)  (i' M_3) }_{ (j M_2)  (i' M_4)}       \right]  \left[ S_{i' \alpha'}  , T^{i'}_{i'}   \right]^{M_4}_{M_3}   \nonumber \\
 &    + \!   \sum_{j' \neq i'} \! \left[ S_{i \alpha}  , T^{i}_{i}  \right]^{M_2}_{M_1}  \left[     \widetilde{\chi}^{(i M_1) (j' M_3) }_{ (i M_2)  (i' M_4)}  \! \left( S_{i' \alpha'}  \cdot T^{i'}_{j'}   \right)^{M_4}_{M_3}      -    \widetilde{\chi}^{(i M_1) (i' M_3)  }_{ (i M_2) (j' M_4)}      \!          \left( T^{j'}_{i'} \! \cdot   S_{i' \alpha'} \right)^{M_4}_{M_3}           \right] \!\! \Bigg\}  ; 
    \label{A tilde ii' nonlocal 1}
\end{align}
finally, the completely non-local term corresponds to $(j \neq i, j' \neq i')$,
\begin{align}
  \left(  \mathcal{W}^{i \alpha}_{i' \alpha'}  \right)^{\mathrm{nloc}}   \equiv  &   \sum_{j \neq i} \sum_{ j' \neq i'}   \sum_{M_1 M_2 M_3 M_4}     \Bigg[ \left( S_{i \alpha}  \cdot T^{i}_{j}   \right)^{M_2}_{M_1}   \left( S_{i' \alpha'}  \cdot T^{i'}_{j'}   \right)^{M_4}_{M_3} \widetilde{\chi}^{(j M_1) (j' M_3) }_{ (i M_2)  (i' M_4)}          \nonumber \\
    & \quad \quad \quad \quad \quad \quad -           \left( S_{i \alpha}  \cdot T^{i}_{j}   \right)^{M_2}_{M_1}     \left( T^{j'}_{i'} \cdot   S_{i' \alpha'} \right)^{M_4}_{M_3}  \widetilde{\chi}^{(j M_1) (i' M_3)  }_{ (i M_2) (j' M_4)}         \nonumber \\
    & \quad \quad \quad \quad  \quad \quad   -        \left( T^{j}_{i} \cdot   S_{i \alpha} \right)^{M_2}_{M_1}   \left( S_{i' \alpha'}  \cdot T^{i'}_{j'}   \right)^{M_4}_{M_3} \widetilde{\chi}^{(i M_1)  (j' M_3) }_{ (j M_2)  (i' M_4)}     \nonumber \\
    & \quad \quad \quad \quad \quad \quad +      \  \left( T^{j}_{i} \cdot   S_{i \alpha} \right)^{M_2}_{M_1}      \left( T^{j'}_{i'} \cdot   S_{i' \alpha'} \right)^{M_4}_{M_3} \widetilde{\chi}^{(i M_1) (i' M_3)  }_{ (j M_2) (j' M_4)}     \Bigg] .
    \label{A tilde ii' nonlocal 2}
\end{align}
The terms $\left(  \mathcal{W}^{i \alpha}_{i' \alpha'}  \right)^{\mathrm{loc}} $ and $\left( \mathcal{W}^{i \alpha}_{i' \alpha'}  \right)^{\mathrm{loc / nloc}} $ are completely relativistic terms, since they vanish when $\left[ S_{i \alpha}  , T^{i}_{i}    \right] = 0$, i.e., when there is no external magnetic field and no local anisotropy. The term $\left(  \mathcal{W}^{i \alpha}_{i' \alpha'}  \right)^{\mathrm{nloc}}$, on the other hand, survives also in the non-relativistic regime. It should be noted that in our theory, in the general relativistic case, all the terms of the exchange tensor depend on the magnetic field and the local anisotropy not only via the intrinsic dependence of the Green's functions, but also via terms which are explicitly linear and even quadratic in such parameters, as it is evident by looking at the terms of Eqs.\eqref{A tilde ii' local}, \eqref{A tilde ii' nonlocal 1}, \eqref{A tilde ii' nonlocal 2}, as well as by considering the terms arising from $\mathcal{M}^{i \alpha}_{i' \alpha'}        + \beta  \mathcal{V}_{i \alpha} \mathcal{V}_{i' \alpha'}$ in Eq.\eqref{tilde M brevity}.

To get some insights into the structure of the parameters, it is instructive to consider the case $(i \alpha) = (i' \alpha')$, which is relevant for Eqs.\eqref{third set}. Using Eq.\eqref{local commutator}, we can write
\begin{align}
 \left(  \mathcal{W}^{i \alpha}_{i \alpha}  \right)^{\mathrm{loc}}     \equiv  &    \sum_{M_1 M_2 M_3 M_4}  \!     \left[ S_{i \alpha}  , A^{i}_{i}    \right]^{M_2}_{M_1} \widetilde{\chi}^{(i M_1) (i M_3)  }_{ (i M_2) (i M_4)}  \left[ S_{i \alpha}  , A^{i}_{i}    \right]^{M_4}_{M_3} \nonumber \\
 &        + 2 \mathrm{i} \mu_{\mathrm{B}} g_i   \sum_{M_1 M_2 M_3 M_4}  \!     \left[ \left( \boldsymbol{B}_i \times \boldsymbol{S}_i \right) \cdot \boldsymbol{u}_i^{\alpha}      \right]^{M_2}_{M_1} \widetilde{\chi}^{(i M_1) (i M_3)  }_{ (i M_2) (i M_4)}  \left[ S_{i \alpha}  , A^{i}_{i}    \right]^{M_4}_{M_3} \nonumber \\
 &  - \mu_{\mathrm{B}}^2 g_i^2 \! \sum_{M_1 M_2 M_3 M_4}  \!     \left[ \left( \boldsymbol{B}_i \times \boldsymbol{S}_i \right) \cdot \boldsymbol{u}_i^{\alpha}    \right]^{M_2}_{M_1} \widetilde{\chi}^{(i M_1) (i M_3)  }_{ (i M_2) (i M_4)}  \left[ \left( \boldsymbol{B}_i \times \boldsymbol{S}_i \right) \cdot \boldsymbol{u}_i^{\alpha}   \right]^{M_4}_{M_3} ,
\end{align}
\begin{align}
 \left(  \mathcal{W}^{i \alpha}_{i \alpha}  \right)^{\mathrm{loc / nloc}}     \equiv  &  \, \,  2   \sum_{M_1 M_2 M_3 M_4} \left( \left[ S_{i \alpha}  , A^{i}_{i}  \right]^{M_2}_{M_1}    +  \mathrm{i} \mu_{\mathrm{B}} g_i \left[ \left( \boldsymbol{B}_i \times \boldsymbol{S}_i \right) \cdot \boldsymbol{u}_i^{\alpha} \right]^{M_2}_{M_1}  \right)  \nonumber \\
 & \quad \cdot     \sum_{j \neq i}    \Bigg[     \widetilde{\chi}^{(i M_1) (j M_3) }_{ (i M_2)  (i M_4)}  \! \left( S_{i \alpha}  \cdot T^{i}_{j}   \right)^{M_4}_{M_3}      -    \widetilde{\chi}^{(i M_1) (i M_3)  }_{ (i M_2) (j M_4)}      \!          \left( T^{j}_{i} \! \cdot   S_{i \alpha} \right)^{M_4}_{M_3}           \Bigg]    . 
\end{align}
We then obtain, for the quantities relevant to Eqs.\eqref{third set},
\begin{align}
 \widetilde{\mathcal{M}}^{i \alpha}_{i \alpha} = &  \left. \widetilde{\mathcal{M}}_{i \alpha}^{i \alpha} \right|_{\boldsymbol{B}_i = \boldsymbol{0}}   - \mu_{\mathrm{B}} g_i \left(  \boldsymbol{B}_i \cdot \left< \hat{\boldsymbol{S}}_i \right> - B_{i}^{\alpha} \left< \hat{S}_{i \alpha} \right> \right)    \nonumber \\
 &       - 2 \mathrm{i} \mu_{\mathrm{B}} g_i   \sum_{M_1 M_2 M_3 M_4}  \!     \left[ \left( \boldsymbol{B}_i \times \boldsymbol{S}_i \right) \cdot \boldsymbol{u}_i^{\alpha}      \right]^{M_2}_{M_1} \widetilde{\chi}^{(i M_1) (i M_3)  }_{ (i M_2) (i M_4)}  \left[ S_{i \alpha}  , A^{i}_{i}    \right]^{M_4}_{M_3} \nonumber \\
 &  + \mu_{\mathrm{B}}^2 g_i^2 \! \sum_{M_1 M_2 M_3 M_4}  \!     \left[ \left( \boldsymbol{B}_i \times \boldsymbol{S}_i \right) \cdot \boldsymbol{u}_i^{\alpha}    \right]^{M_2}_{M_1} \widetilde{\chi}^{(i M_1) (i M_3)  }_{ (i M_2) (i M_4)}  \left[ \left( \boldsymbol{B}_i \times \boldsymbol{S}_i \right) \cdot \boldsymbol{u}_i^{\alpha}   \right]^{M_4}_{M_3} \nonumber \\
 &     -  2  \mathrm{i} \mu_{\mathrm{B}} g_i  \sum_{M_1 M_2 M_3 M_4}      \left[ \left( \boldsymbol{B}_i \times \boldsymbol{S}_i \right) \cdot \boldsymbol{u}_i^{\alpha} \right]^{M_2}_{M_1}     \nonumber \\
 & \quad \cdot     \sum_{j \neq i}    \Bigg[     \widetilde{\chi}^{(i M_1) (j M_3) }_{ (i M_2)  (i M_4)}  \! \left( S_{i \alpha}  \cdot T^{i}_{j}   \right)^{M_4}_{M_3}      -    \widetilde{\chi}^{(i M_1) (i M_3)  }_{ (i M_2) (j M_4)}      \!          \left( T^{j}_{i} \! \cdot   S_{i \alpha} \right)^{M_4}_{M_3}           \Bigg] \nonumber \\
 &   +  \beta  \mu_{\mathrm{B}}^2 g^2_i \left[ \left( \boldsymbol{B}_i \times \left< \hat{\boldsymbol{S}}_i \right> \right)  \cdot \boldsymbol{u}^{\alpha}_i \right]^2 + 2 \beta \mu_{\mathrm{B}} g_i \left. \mathcal{V}_{i \alpha} \right|_{\boldsymbol{B}_i = \boldsymbol{0}} \left( \boldsymbol{B}_i \times \left< \hat{\boldsymbol{S}}_i \right> \right)  \cdot \boldsymbol{u}^{\alpha}_i  , 
 \label{tilde M i i alpha alpha}
\end{align}
where $\left. \widetilde{\mathcal{M}}_{i \alpha}^{i \alpha} \right|_{\boldsymbol{B}_i = \boldsymbol{0}}$, we recall, is not independent on $\boldsymbol{B}_i$, but it is a term which does not vanish when $\boldsymbol{B}_i = \boldsymbol{0}$. A simplification of the parts which depend explicitly on the local relativistic terms can be achieved after decomposing the two-particle Green's functions as in Eq.\eqref{2p GF},
\begin{align}
 \chi^{1, 3}_{2 , 4}(\tau, \tau^+, \tau', \tau'^+) & = \left( \chi^0 \right)^{1, 3}_{2 , 4}(\tau, \tau^+, \tau', \tau'^+) + \left( \chi^{\Gamma} \right)^{1, 3}_{2 , 4}(\tau, \tau^+, \tau', \tau'^+) ,
\end{align}
where $\left( \chi^0 \right)^{1, 3}_{2 , 4}(\tau, \tau^+, \tau', \tau'^+) = -    \rho^1_2 \, \rho^3_4  - G^1_4(\tau - \tau' - \varepsilon ) \, \,  G^3_2(\tau' - \tau - \varepsilon)$ and $\chi^{\Gamma}$ contains the vertex corrections. Using the Matsubara-frequency representation and Eq.\eqref{integration Matsubara}, we can then reduce Eq.\eqref{tilde M i i alpha alpha} to
\begin{align}
 \widetilde{\mathcal{M}}^{i \alpha}_{i \alpha} = &  \left. \widetilde{\mathcal{M}}_{i \alpha}^{i \alpha} \right|_{\boldsymbol{B}_i = \boldsymbol{0}}   - \mu_{\mathrm{B}} g_i \left(  \boldsymbol{B}_i \cdot \left< \hat{\boldsymbol{S}}_i \right> - B_{i}^{\alpha} \left< \hat{S}_{i \alpha} \right> \right)    \nonumber \\
  & - \frac{\mu_{\mathrm{B}}^2 g_i^2}{\beta}   \mathrm{Tr}_M  \! \left\{    \left[ \left( \boldsymbol{B}_i \times \boldsymbol{S}_i \right) \cdot \boldsymbol{u}_i^{\alpha}    \right] \cdot \! \sum_{\omega} \mathrm{e}^{\mathrm{i} \omega 0^+}  \! G^{ i    }_{    i  }(\mathrm{i} \omega) \! \cdot \! \left[ \left( \boldsymbol{B}_i \times \boldsymbol{S}_i \right) \cdot \boldsymbol{u}_i^{\alpha}   \right] \! \cdot     G^{  i    }_{  i   }(\mathrm{i}\omega) \! \right\} \nonumber \\
 &  + \mu_{\mathrm{B}}^2 g_i^2 \! \sum_{M_1 M_2 M_3 M_4}  \!     \left[ \left( \boldsymbol{B}_i \times \boldsymbol{S}_i \right) \cdot \boldsymbol{u}_i^{\alpha}    \right]^{M_2}_{M_1} \left( \widetilde{\chi}^{\Gamma} \right)^{(i M_1) (i M_3)  }_{ (i M_2) (i M_4)}  \left[ \left( \boldsymbol{B}_i \times \boldsymbol{S}_i \right) \cdot \boldsymbol{u}_i^{\alpha}   \right]^{M_4}_{M_3} \nonumber \\
 &     -  2  \mathrm{i} \mu_{\mathrm{B}} g_i  \sum_{M_1 M_2 M_3 M_4}             \sum_{j  }    \Bigg[   \left(   \widetilde{\chi}^{\Gamma} \right)^{(i M_1) (j M_3) }_{ (i M_2)  (i M_4)}  \! \left( S_{i \alpha}  \cdot \left( T^{i}_{j}  \right)_{\boldsymbol{B}_i = \boldsymbol{0}}  \right)^{M_4}_{M_3}  \nonumber \\
 & \quad \quad \quad  \quad   \quad   -    \left(   \widetilde{\chi}^{\Gamma} \right)^{(i M_1) (i M_3)  }_{ (i M_2) (j M_4)}      \!          \left( \left( T^{j}_{i} \right)_{\boldsymbol{B}_i = \boldsymbol{0}} \! \cdot   S_{i \alpha} \right)^{M_4}_{M_3}           \Bigg]  \left[ \left( \boldsymbol{B}_i \times \boldsymbol{S}_i \right) \cdot \boldsymbol{u}_i^{\alpha} \right]^{M_2}_{M_1}   \nonumber \\
 &     +   \frac{2  \mathrm{i}    \mu_{\mathrm{B}} g_i }{\beta}  \, \mathrm{Tr}_M   \Bigg\{ \!    \left[ \left( \boldsymbol{B}_i \times \boldsymbol{S}_i \right) \cdot \boldsymbol{u}_i^{\alpha} \right]     \cdot \! \sum_{\omega} \mathrm{e}^{\mathrm{i} \omega 0^+}     \Big[     G^{  i  }_{   i  }(\mathrm{i} \omega) \cdot S_{i \alpha}  \cdot \left[   T_{\boldsymbol{B}_i = \boldsymbol{0}}   \cdot   G(\mathrm{i} \omega) \right]^i_i    \nonumber \\
 & \quad \quad \quad   \quad \quad \quad \quad \quad \quad  -  \left[  G(\mathrm{i} \omega)  \cdot   T_{\boldsymbol{B}_i = \boldsymbol{0}}  \right]^{i}_{i}   \cdot   S_{i \alpha} \cdot   G^{  i    }_{  i    }(\mathrm{i} \omega)     \Big] \Bigg\} .
 \label{tilde M i i alpha alpha reduced}
\end{align}
In addition to the first term, which survives in the non-relativistic regime (and also includes anisotropy contributions), and to the second term in the first line, which as discussed should be identified with a component of the effective magnetic field, Eq.\eqref{tilde M i i alpha alpha reduced} explicitly shows how the relativistic exchange parameters have a non-trivial dependence on the magnetic field $\boldsymbol{B}_i$.

\end{document}